\documentclass[fleqn,usenatbib]{mnras}

\usepackage{newtxtext,newtxmath}

\usepackage[T1]{fontenc}

\DeclareRobustCommand{\VAN}[3]{#2}
\let\VANthebibliography\thebibliography
\def\thebibliography{\DeclareRobustCommand{\VAN}[3]{##3}\VANthebibliography}

\usepackage{graphicx}	
\usepackage{amsmath}	
\usepackage{todonotes}
\usepackage{multirow}
\usepackage{colortbl}
\usepackage{subcaption}
\usepackage{soul}
\usepackage{caption}

\newcommand{\psrpoppy}{{\sc PsrPopPy}}
\newcommand{\psrpop}{{\sc Psr\textsc{POP}}}
\newcommand{\illustris}{{\sc Illustris\textsc{TNG}}}
\newcommand{\sevn}{{\sc sevn}}

\newcommand{\orcidicon}[1]{\href{https://orcid.org/#1}{\includegraphics[width=11pt]{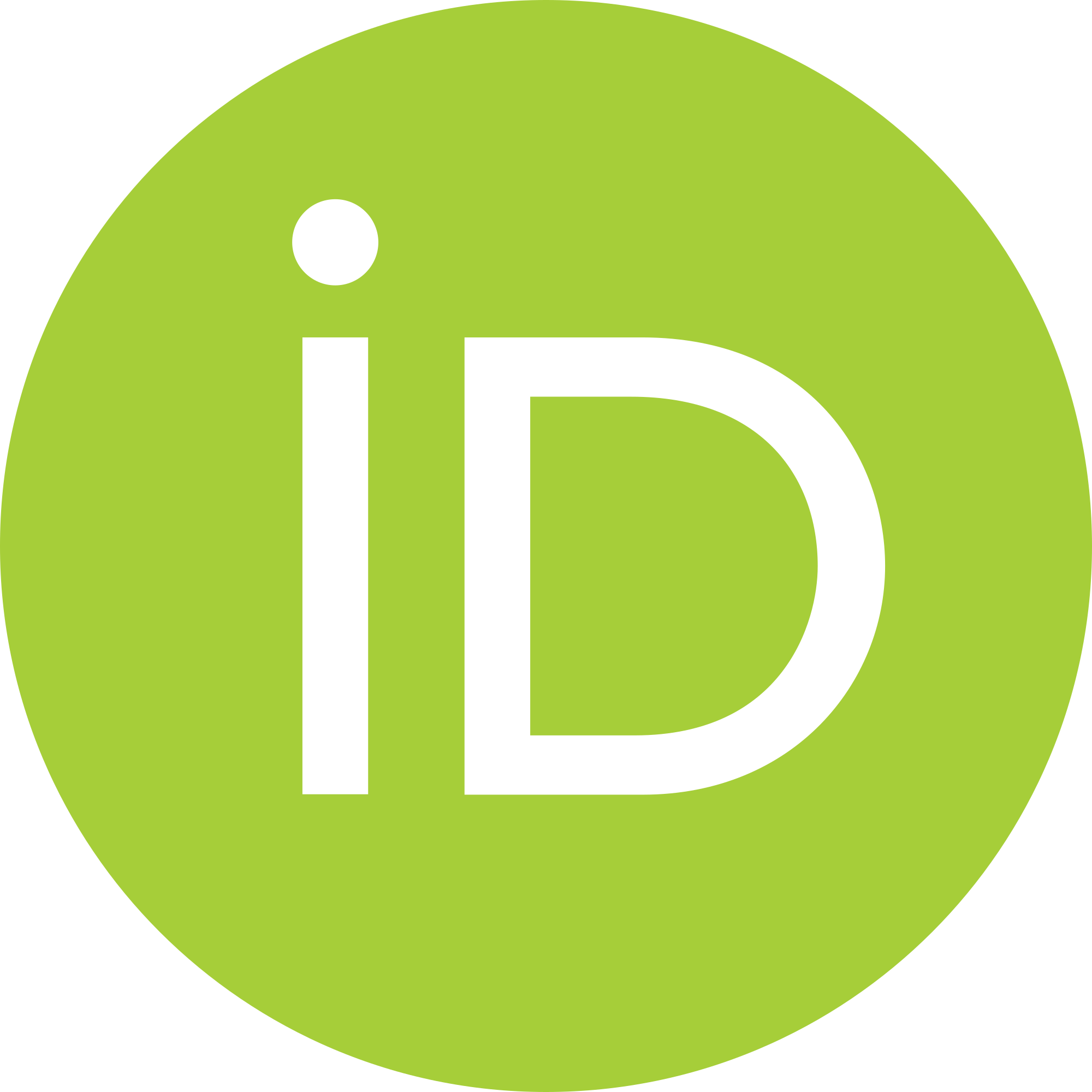}}}

\newcommand{\orcid}[1]{\href{https://orcid.org/#1}{\protect\orcidicon{#1}}}

\title[BNSs in the Milky Way]{Binary neutron star populations in the Milky Way}

\author[C. Sgalletta et al.]{
Cecilia Sgalletta\orcid{0009-0003-7951-4820}$^{1,2}$\thanks{E-mail: cecilia.sgalletta@sissa.it},
Giuliano Iorio\orcid{0000-0003-0293-503X}$^{2,3}$\thanks{E-mail: giuliano.iorio@unipd.it},
Michela Mapelli\orcid{0000-0001-8799-2548}$^{2,3,4}$\thanks{E-mail: michela.mapelli@unipd.it},
M. Celeste Artale\orcid{0000-0003-0570-785X}$^{2,3,5}$,
Lumen Boco\orcid{0000-0003-3127-922X}$^{1}$, 
\newauthor{}
Debatri Chattopadhyay\orcid{0000-0001-5867-5033}$^{6}$,
Andrea Lapi\orcid{0000-0001-9728-8132}$^{1}$, 
Andrea Possenti$^{7}$,
Stefano Rinaldi\orcid{0000-0001-5799-4155}$^{8,9}$, 
Mario Spera\orcid{0000-0003-0930-6930}$^{1, 10}$
\\
$^{1}$SISSA, via Bonomea 365, I--34136 Trieste, Italy\\
$^{2}$INFN-Padova, Via Marzolo 8, I--35131 Padova, Italy\\
$^{3}$Dipartimento di Fisica e Astronomia Galileo Galilei, Universit\`a di  Padova, Vicolo dell'Osservatorio 3, I--35122 Padova, Italy\\
$^{4}$INAF-Padova, Vicolo dell'Osservatorio 5, I--35122 Padova, Italy\\
$^{5}$ Departamento de Ciencias Fisicas, Universidad Andres Bello, Fernandez Concha 700, Las Condes, Santiago, Chile \\
$^{6}$ Gravity Exploration Institute, School of Physics and Astronomy, Cardiff University, Cardiff CF24 3AA, UK \\
$^{7}$INAF–Osservatorio Astronomico di Cagliari, Via della Scienza 5, I-09047
Selargius, CA, Italy\\
$^{8}$Dipartimento di Fisica “E. Fermi”, Universit\`a di Pisa, I-56127 Pisa, Italy \\
$^{9}$INFN, Sezione di Pisa, I-56127 Pisa, Italy \\
$^{10}$National Institute for Nuclear Physics - INFN, Sezione di Trieste, I-34127 Trieste, Italy
}

\date{Accepted XXX. Received YYY; in original form ZZZ}

\pubyear{2023}

\begin{document}
\label{firstpage}
\pagerange{\pageref{firstpage}--\pageref{lastpage}}
\maketitle

\begin{abstract}
Galactic binary neutron stars (BNSs) are a unique laboratory to probe the evolution of BNSs and their progenitors. Here, we use a new version of the population synthesis code \sevn{} to evolve the population of Galactic BNSs, by modeling the spin up and down of pulsars self-consistently. We analyze the merger rate $\mathcal{R}_{\rm MW}$, orbital period $P_{\rm orb}$, eccentricity $e$, spin period $P$, and spin period derivative $\dot{P}$ of the BNS population. Values of the common envelope parameter $\alpha=1 - 3$ and an accurate model of the Milky Way star formation history best reproduce the BNS merger rate in our Galaxy ($\mathcal{R}_{\rm MW}\approx{}30$~Myr$^{-1}$). We apply radio-selection effects to our simulated BNSs and compare them to the observed population. Using a Dirichlet process Gaussian mixture method, we evaluate the four-dimensional likelihood in the $(P_{\rm orb}, e, P, \dot{P})$ space, by comparing our radio-selected simulated pulsars against Galactic BNSs. Our analysis favours an uniform initial distribution for both the magnetic field ($10^{10-13}$ G) and the spin period ($10-100$ ms). The implementation of radio selection effects is critical to match not only the spin period and period derivative, but also the orbital period and eccentricity of Galactic BNSs. According to our fiducial model, the Square Kilometre Array will detect $\sim 20$ new BNSs in the Milky Way.  

\end{abstract}




\section{Introduction}

Radio pulsars are highly magnetized, rapidly spinning neutron stars (NSs) that emit beams of electromagnetic radiation, making them some of the most precise cosmic clocks known to science \citep{hewish1968,pacini1968}. They can provide us with  a wealth of information. Their spin period and spin period derivative can be measured with high accuracy. 
The timing precision of radio pulsars allows for tests of general relativity in the strong-field regime, which would otherwise be unfeasible with terrestrial laboratories. The only binary pulsar system known to date, PSR J0737-3039A/B \citep{burgay2003, lyne2004}, 
is a unique laboratory for tests of gravity theories and for the study of highly condensed matter \citep[e.g.,][]{kramer2006, lattimer2021}. Moreover, the pulsar timing array \citep[PTA,  a set of pulsars which is analysed to search for correlated signatures in the pulse arrival times][]{hobbs2010, hobbs2013, manchester2013,lentati2015,arzoumanian2018,miles2023} 
makes it possible to investigate the sources of low-frequency gravitational waves (GWs), such as super-massive black hole mergers \citep{Arzoumanian2020}. 

The formation and evolution properties of pulsars are still matter of debate. Earlier efforts attempted to infer the birth distribution of pulsar properties from their observed Galactic population \citep{ostriker1969, fauchergiguere2006}, the main long-standing issue being that of magnetic field decay \citep{fauchergiguere2006}. Based on a theoretical argument, magnetic field decay might generate from Ohmic dissipation, caused by the formation of electric currents on the NS crust \citep{bhattacharya1992, konar1997, konar1999}. This effect is probably more relevant, however, during accretion \citep{kiel2008}. A number of magnetic field decay timescales has been explored to date \citep{gonthier2004, fauchergiguere2006, kiel2008, oslowski2011, chattopadhyay2020, chattopadhyay2021}. The understanding of pulsars birth properties might help to shed light on the supernova (SN) explosion mechanism, because we expect that the newly-born NS properties are tightly correlated with the pre-SN star \citep{fauchergiguere2006, kapil2023}. 

The presence of a radio pulsar in a binary system allows us to measure the orbital properties (orbital period, eccentricity, and masses) to a high level of accuracy \citep{kaspi1994, stairs2002, kramer2006, lorimer2008, ozel2016, kramer2021}. Thanks to this property of radio pulsars, we have detected about a dozen of binary  NSs (BNSs) in the Milky Way (MW), i.e. binary systems in which both components are NSs \citep{hulse1975a, burgay2003, lyne2004, martinez2015, ozel2016, tauris2017}.
BNSs are loud sources of GWs \citep[][]{burgay2003, abbott2017a, abbottGW190425, pol2019, radice2020,  pol2020, thrane2020, mandel2022, spera2022}, their merger powers short gamma-ray bursts \citep[][]{abbott2017b, goldstein2017, murase2018, colombo2022, perna2022} and kilonovae \citep[][]{tanvir2013, smartt2017, troja2017, kasen2017, metzger2017, metzger2019, nedora2023}, and is a fundamental source of chemical enrichment by r-process elements \citep[][]{eichler1989, hotokezaka2018, cote2019, kobayashi2022, fujibayashi2023, combi2023}.

The formation of a BNS from the evolution of a massive binary star is still controversial \citep[e.g.,][]{tauris2017,vignagomez2018, kruckow2018, chruslinska2018,giacobbo2018b,  belczynski2018, coenrad2019, andrews2019, chattopadhyay2020, vignagomez2020, belczynski2020, olejak2021, riley2022, broekgaarden2022, olejak2022, iorio2022}. 
According to the standard scenario, the two progenitor stars undergo at least one Roche-lobe overflow, after which the first SN takes place, leading to the formation of the first NS. If the binary system is tight enough to avoid disruption by the natal kick, the evolution of the companion star then initiates a common envelope (CE) phase, which results in the ejection of the envelope and the formation of a tight binary system composed of a NS and a naked He star. If such system avoids disruption even during the second SN, a BNS forms, which might then harden by GW emission until it reaches coalescence \citep[][and references therein]{tauris2017}.  An  alternative scenario consists in an early CE phase between the two progenitor stars, leading to the formation of a binary system composed of two naked cores onto a very tight orbit. Such binary system might avoid ionization during the two SN explosions and lead to the formation of a BNS \citep{brown1995, dewi2006, justham2011, vignagomez2018, vignagomez2020, broekgaarden2021a, iorio2022}. 

Two of the most important unknowns of binary star evolution are the physics of CE \citep{ivanova2013, klencki2021,roepke2022} and the natal kick \citep{woosley1987, janka1994, lai2001, hobbs2005, odoherty2023}. CE  \citep{Paczynski1976, webbink10985}  is usually described with a simple energy formalism, in which we assume that a fraction $\alpha$ of the kinetic orbital energy of the two cores is transferred to the envelope and helps unbinding it \citep[see e.g.][for alternative models]{nelemans2000, hirai2022, distefano2023}. The distribution of natal kicks is often modelled with a Maxwellian curve with one-dimensional root-mean square $\sigma=265$ km~s$^{-1}$, based on the proper motions of 73 young Galactic radio pulsars \citep{hobbs2005}. However, core-collapse SN models suggest that the kick might be much lower in presence of a stripped or ultra-stripped SN, i.e. a SN triggered by a naked He or CO core \citep{tauris2015, tauris2017, bray2016, bray2018, giacobbo2020}. Moreover, electron-capture SNe, especially in binary systems, might be associated with low kicks \citep{gessner2018, giacobbo2019, muller2019, stevenson2022}.

The aim of this work is to characterize the population of Galactic BNSs, by using a population synthesis code  coupled with a MW model. We evolve the orbital binary properties, the spin period  and  magnetic field of pulsars. With this framework, we explore the merger rates, the orbital properties and the population of radio pulsars. We consider many different assumptions for the MW model, the magnetic field, and the binary population parameters. We compare  our models with current observations of Galactic BNSs. 

To this purpose, we use the state-of-the-art population synthesis code \sevn{} \citep{spera2017, spera2019,mapelli2020, iorio2022} to implement single and binary stellar evolution. We adopt four  MW models,  employing semi-empirical prescriptions \citep{chiappini1997, courteau2014, pezzulli2015, grisoni2017, boco2019} and cosmological simulations \citep{schaye2015, eagle2017, nelson2019, nelson2019tng50}. 
We explore several parameters, such as $\alpha$, the birth distribution of spin periods and surface magnetic fields of pulsars, and the magnetic field decay timescale.
We test our results against the observed Galactic pulsar population.

\section{Methods} \label{sec:methods}
We study the Galactic BNS  population by coupling various MW models with BNS catalogues obtained with the \textsc{sevn} population synthesis code. Here below, we present the details of our models.

\subsection{\textsc{sevn}}\label{sec:sevn}

The stellar evolution for N-body (\sevn{}) code is a state-of-the-art binary population synthesis code that implements  single stellar evolution by interpolating pre-computed stellar tracks on the fly, and binary processes through analytic and semi-analytic models \citep{spera2017,spera2019, mapelli2020}. 
We use  the latest version of the \sevn{} code, described in \cite{iorio2022}\footnote{\sevn{} is publicly available at \url{https://gitlab.com/sevncodes/sevn.git}. The version used in this work is the release {\it Sgalletta23}: \url{https://gitlab.com/sevncodes/sevn/-/releases/sgalletta23}.}. 
In the following, we will give a general overview of the main features of \sevn{} focusing on the processes most relevant for this work. 
We refer to \cite{iorio2022} for a detailed description of the code. 
The first distinctive mark of \sevn{} is the  way it handles single stellar evolution: the interpolation of look-up tables makes \sevn{}  fast and versatile. In fact, it is possible to change stellar evolution models by simply substituting the input stellar track tables.  
The stellar tracks adopted in this work have been evolved with the \textsc{parsec} code \footnote{We adopt the tables labelled \emph{SEVNtracks\textunderscore parsec \textunderscore ov05 \textunderscore AGB} for the H-stars and \emph{SEVNtracks \textunderscore parsec \textunderscore pureHe36} for the naked He stars. }  \citep[][]{bressan2012, costa2019,  costa2021, nguyen2022}.

\textsc{sevn} includes several prescriptions for core-collapse SN explosions. In this work, we investigate the impact on our results of three of them: the \emph{rapid} and \emph{delayed} models by \cite{fryer2012} and the \emph{rapid-Gauss} model.
The rapid and the delayed models are both based on a convection-enhanced neutrino-driven mechanism for the SN explosion, though the revival
of the shock wave happens within the first $250 \text{ms}$ after the collapse in the rapid model, whereas such timescale can be much longer for the delayed model.  
The two prescriptions predict the masses of compact remnants as the sum of the mass of the proto-compact object and the amount of fallback material.
The third model we adopted is based on the rapid explosion mechanism but draws the NS masses from a Gaussian distribution peaked at $1.33$ M$_\odot$ with standard deviation $0.09$ M$_{\odot}$, resulting from a fit to the Galactic BNS masses \citep{ozel2012, ozel2016}. We decided to introduce this model because both the rapid and delayed prescriptions fail to reproduce the observed NS mass distribution \citep{vignagomez2018}. In the delayed, rapid, and rapid-Gauss models, we do not allow for NS masses $<1.1$ M$_\odot$.  
\sevn{} assumes that if a compact remnant has mass $\in[1.1,\,{} 3)$~M$_\odot$ it is a NS, whereas if it has a mass $\ge{}3$ M$_\odot$ it is a black hole. 
The NS radius is fixed at $11\,{} \text{km}$ \citep{ozel2012, Bogdanov2016, bauswein2017, abbott2018}.

\textsc{sevn} includes several formalisms to model natal kicks. 
Here, we draw black hole natal kicks following \citet[][]{giacobbo2020}:
\begin{equation}
V_{\rm kick}=f_{\rm H05}\,{}\frac{\langle{}M_{\rm NS}\rangle{}}{M_{\rm rem}}\,{}\frac{M_{\rm ej}}{\langle{}M_{\rm ej}\rangle},
\end{equation}    
 where $\langle{}M_{\rm NS}\rangle{}$ and $\langle{}M_{\rm ej}\rangle$ are the average NS mass and ejecta mass from single stellar evolution, respectively, while $M_{\rm rem}$ and $M_{\rm ej}$ are the compact object mass and the ejecta mass \citep{giacobbo2020}. The term $f_{\rm H05}$ is a random number drawn from a Maxwellian distribution with  one-dimensional root mean square $\sigma_\mathrm{kick}=265 \ \mathrm{km}\,{} \mathrm{s}^{-1}$, coming from a fit to the proper motions of 73 young pulsars ($<3$ Myr) in the MW \citep{hobbs2005}.   In this formalism, stripped  and ultra-stripped  SNe  result in lower kicks with respect to the other explosions, owing to the lower amount of ejected mass $M_{\rm ej}$ \citep{bray2016,bray2018}. In addition to the natal kick, we also calculate a Blaauw kick \citep{blaauw1961} resulting from the instantaneous mass loss in a binary system triggered by a SN explosion. We use the same formalism as described in Appendix~A of \cite{hurley2002}.

\subsection{Binary evolution}

In the following, we focus on the binary evolution processes implemented in \sevn{} that are relevant for this work, especially Roche-lobe overflow (RLO) and the CE phase \citep{iorio2022}. 
The Roche lobe of a star in a binary system defines the region of space within which matter is gravitationally bound to the star itself. Therefore, when a star fills its Roche lobe, matter flows to the companion object under its gravitational attraction. This process is known as RLO. 
RLO thus involves variations in the mass ratio, in the masses and radii of the two stars and in the semi-major axis of the system. At each timestep, \sevn{} evaluates the Roche-lobe radius of the two stars in the binary system using the analytical expression derived in \cite{eggleton1983}:
\begin{equation}\label{eq:RocheLobeRadius}
\frac{R_{L}}{a} = \frac{0.49 \,{}q^{2/3}}{0.6 q^{2/3} + \ln(1+q^{1/3})},
\end{equation}
where $q$ is the mass ratio and $a$ is the semi-major axis. If either star radii satisfy the condition $r \geq R_{L}$, a RLO episode initiates and mass falls from the donor (the star filling its Roche lobe) to the accretor.
Depending on the response of the Roche-lobe and donor radius to mass stripping, mass transfer can be stable or unstable. 
To assess the stability of mass transfer, \textsc{sevn} adopts a formalism common to many population synthesis codes (see e.g. \citealt{hurley2002, hobbs2005}). The mass ratio of the donor $d$ star to the accretor $a$ star, $q$, is compared to a critical value $q_{\rm c}$,
dependent on the stellar evolutionary phase of the donor star: if $q> q_{\rm c}$ the mass transfer is unstable on a dynamical timescale. Here, we use the values of $q_{\rm c}$ adopted in the fiducial simulations of \cite[][see their Table~3]{iorio2022}.  
In this model, mass transfer is always stable if the donor star is in the main sequence or in the Hertzprung gap evolutionary phase.
In the case of stable mass transfer, the mass loss rate in \sevn{} scales as \citep{hurley2002} 
\begin{equation}
    \dot{M}_{\rm d} = -F(M_{\rm d})\left(\ln{\frac{R_{\rm d}}{R_{\rm L,d}}}\right)^3 \text{M}_\odot \,{}\text{yr}^{-1},
\end{equation}
where $F(M_{\rm d})$ is a normalization factor, $R_{\rm d}$ is the radius of the donor and ${R_{\rm L,d}}$ is the Roche lobe of the donor star. 
\textsc{sevn} allows for non conservative mass transfer; therefore, the mass lost by the donor can be bigger than the amount accreted by the companion. The mass accreted is modeled as follows:
\begin{equation}
\dot{M}_\mathrm{a} = \left\{
 \begin{array}{ll}
 \min{(\dot{M}_{\rm Edd},\,{}-f_\mathrm{MT}\,{}\dot{M}_\mathrm{d})} &\mathrm{if}\,{}{\rm the}\,{} \mathrm{accretor}\,{} \mathrm{is}\,{} \mathrm{a}\,{} \mathrm{compact}\,{} \mathrm{object} \\
  - f_\mathrm{MT}\,{} \dot{M}_\mathrm{d}&\mathrm{otherwise}, 
  \end{array}
  \right.
\label{eq:maccrlo}
\end{equation}
 where $\dot{M}_{\rm Edd}$ is the Eddington rate and $f_{\rm MT} \in [0,1]$ is the mass accretion efficiency. We set $f_{\rm MT}=0.5$ in our simulations \citep{Bouffanais2021, iorio2022}.

When mass transfer is unstable the outcome can  be either a stellar merger or a CE. 
During the CE phase, the cores of the two stars orbit each other within a shared envelope. 
\sevn{} parametrizes the CE phase using the $\alpha$ formalism \citep{webbink10985,tout1997}. 
The envelope’s binding energy at the onset of the CE is evaluated as:
\begin{equation}
    E_{\rm bind, i} = - G\left(\frac{M_1 \,{}M_{\rm env, 1}}{\lambda_{1} R_1} + \frac{M_2\,{} M_{\rm env, 2}}{\lambda_{2} R_2} \right),
\end{equation}

where $M$ is the mass of the star, $M_{\rm env}$ is the mass of the envelope, 
and $R$ is the star’s radius. The subscripts $1$ and $2$ refer to the primary and
secondary star, respectively. Finally, $\lambda$ takes into account the structural properties of the envelope of each star, which is calculated following the model by \cite{claeys2014}, as described in Appendix A of \cite{iorio2022}.
The variation of the orbital energy is 
\begin{equation}
    \Delta E_{\rm orb} =  \frac{G\,{} M_{\rm c,1}\,{} M_{\rm c,2}}{2}\left( a_{\rm f}^{-1} - a_{\rm i}^{-1}\right),
\end{equation}
where $M_{\rm c,1}$ and $M_{\rm c,2}$ are the core masses of the two stars and $a_{\rm i}$ and $a_{\rm f}$ are the semi-major axis at the onset and at the end of the CE, respectively. \sevn{} infers the values of $a_{\rm f}$  through the condition $E_{\rm bind, i} = \alpha\,{}\Delta E_{\rm orb}$. The $\alpha$ parameter represents the efficiency of energy transport from the binary orbit to the envelope. According to its original definition, $\alpha$ should take values between 0 and 1, but recently values of $\alpha>1$ have been explored, to account for the missing physics in this very simplified formalism \citep[e.g.,][]{fragos2019}.
If the core radii of the two stars are both smaller than their Roche lobes  at the end of CE (Eq.~\ref{eq:RocheLobeRadius}), then the envelope is ejected. Otherwise, the two stars merge during the CE phase.

\subsection{NS properties}

To study the evolution of the physical properties of NSs with \sevn{}, at the time of formation we assign to each NS a spin, a magnetic field and an angle between the rotation and the magnetic axis, $\alpha_{\rm B}$. For all of our NSs, we generate ${\cos{\alpha_{\rm B}}}$ from an uniform distribution between $0$ and $1$. The evolution of each NS depends on its interaction with the companion star. If the  NS is not accreting matter from the companion, the evolution is the same as if the NS were isolated, i.e. the NS spin and magnetic field decrease with time (\emph{spin down}). 
In contrast, if the companion star fills its Roche lobe, the infalling matter can transfer angular momentum to the NS, causing the latter to spin up. When the latter process takes place the resulting NS is said to be \emph{recycled}. In fact, observed recycled pulsars are usually characterized by short spin periods and relatively low magnetic fields \citep{lorimer2008, lorimer2011, ozel2016}.

\subsubsection{Spin down}

If the evolution of the pulsar proceeds unperturbed, the pulsar can be seen as a rotating magnet. We model the magnetic field of the pulsar with a dipole and  compute the loss of energy accordingly. In fact, as the pulsar rotates it emits electromagnetic radiation, losing rotational energy, that is the pulsar spins down.
The angular frequency of the pulsar changes according to \cite{goldreich1969}:
\begin{equation}\label{eq:spindown}
    \dot{\Omega} = -\frac{8\,{}\pi\,{} B^2 \,{}R^6\,{} \sin{\alpha_{\rm B}}^2\,{} \Omega^3}{3\,{} \mu_0\,{} c^3\,{} I},
\end{equation}
where $\Omega$ is the angular frequency,  
$B$ is the surface magnetic field, $R$ is the radius, $\alpha_{\rm B}$ is the angle between the rotational axis and the magnetic axis, $I$ is the moment of inertia of the pulsar, $c$ is the speed of light, and $\mu_0$ is the vacuum magnetic permeability. 
The time derivative of the angular frequency $\dot{\Omega}$ is linked to $\Omega$ through the relation $\dot{\Omega} \propto \Omega^n$, where $n$ is the magnetic braking index. 
Observational evidence points toward a value of $n$ lying in the range $2.5-3.5$ \citep{manchester2005}.
In addition, we assume that the magnetic field decays with time because of Ohmic dissipation. 
The finite resistivity in the NS crust, caused by electron scatterings, converts the magnetic energy into heat \citep{goldreich1992, konar1997, urpin1997,konar1999}. Furthermore, observations seem to support the decay of the magnetic field: younger pulsars show in general stronger magnetic fields with respect to older ones \citep{ostriker1969}.
In our models, we evolve the magnetic field following an exponential decay \citep{kiel2008, oslowski2011, chattopadhyay2020, chattopadhyay2021}:
\begin{equation}\label{eq:magneticdecay}
    B=(B_0 - B_{\rm min})\,{} e^{-t/\tau_{\rm d}} + B_{\rm min}, 
\end{equation}
where $B_0$ is the initial surface magnetic field, $B_{\rm min}$ is the minimum surface magnetic field strength and $\tau_{\rm d}$ is the magnetic field decay timescale. 
We assume that the pulsar magnetic field stops decreasing when it reaches the value $B_{\rm min}$. Throughout this work, we adopt $B_{\rm min}= 10^8 \text{G}$, as suggested by \cite{zhang2006}. 
The magnetic field decay timescale $\tau_{\rm d}$ is a free parameter of the model. Its value  is controversial: ranging from $2-5 \,{}\text{Myr}$ \citep[e.g.,][]{oslowski2011} to $2000 \,{}\text{Myr}$ \citep[e.g.,][]{kiel2008}. \cite{fauchergiguere2006} carry out their analysis considering no magnetic field decay at all. 

Following \cite{chattopadhyay2020}, we get an analytic expression for the evolution of the angular frequency of the pulsar:

\begin{align}\label{eq:spinevolve}
\begin{split}
    \frac{1}{\Omega_{\rm f}^2} = & \frac{1}{\Omega_{\rm i}^2} + \frac{16\,{}\pi\,{} R^6\, Pé\sin^2{\alpha_{\rm B}}}{3\,{} \mu_0 \,{}c^3 \,{}I} \times \\
    & \times \left[B^2_{\rm min}\,{}\Delta t - \tau_{\rm d}\,{} B_{\rm min} (B_{\rm f} - B_{\rm i}) - \frac{\tau_{\rm d}}{2} (B_{\rm f}^2 - B_{\rm i}^2)\right],    
\end{split}
\end{align}
where $\Omega_{\rm i}$ and $\Omega_{\rm f}$ are the initial and final angular frequencies, $B_{\rm i}$ and $B_{\rm f}$ are the initial and final magnetic fields, and $\Delta t$ is the time elapsed between the final and the initial states.

The spin period $P$ and the spin down rate $\dot{P}$ can then be obtained from $\Omega$ and $\dot{\Omega}$ through the following relations:
\begin{align}
    P &= \frac{2\,{} \pi}{ \Omega} \\
    \dot{P} &= - \frac{\dot{\Omega}\,{} P}{\Omega}.
\end{align}
We update the values of the NS spin periods and magnetic fields in \textsc{sevn} after each time-step according to Eqs.~\ref{eq:magneticdecay} and \ref{eq:spinevolve}. We can trace the pulsar evolution in the $P-\dot{P}$ diagram.
Because of spin down, pulsars evolve towards larger spin periods and lower magnetic fields. In the $P-\dot{P}$ plane, this trend results in a diagonal shift, towards the lower right corner of the plot \citep[see e.g. Figure 2 in][]{chattopadhyay2020}. 
The timescale of this evolution depends on the choice of $\tau_{\rm d}$: for shorter values of $\tau_{\rm d}$ the traversal of the $P-\dot{P}$ plane is faster.

\subsubsection{Spin up}

Matter exchange processes in binary stars, such as  those happening during a RLO, can significantly affect the evolution of NS spin and magnetic field. For instance, part of the angular momentum of the exchanged material, can be transferred to the NS, that, consequently, spins up. In \textsc{sevn}, we have implemented  the spin-up process of NSs during RLO mass transfer following the same prescriptions as in \cite{kiel2008} and \cite{chattopadhyay2020}.
The rate of change in the angular momentum of the pulsar $\dot{J}_{\rm acc}$ scales linearly with the amount of accreted mass, that is:
\begin{equation}\label{eq:jdot}
    \dot{J}_{\rm acc} =\epsilon{}\,{} V_{\rm diff} \,{} R_{\rm A}^2 \dot{M}_{\rm NS},
\end{equation}

where $\epsilon$ is an efficiency factor, set at $1$ in our models, $\dot{M}_{\rm NS}$ is the mass accretion rate on the NS, and $R_{\rm A}$ is the magnetic radius. We define $R_{\rm A} = R_{\rm Alfven}/2$ as in \cite{chattopadhyay2020}. 
The Alfven radius $R_{\rm Alfven}$ is the radius at which the magnetic pressure equals the ram pressure:
\begin{equation}\label{alfven}
R_{\rm Alfven} = \left(\frac{2\,{}\pi^2}{G\,{}\mu_0^2}\right)^{1/7}\times\left(\frac{R^6}{\dot{M}_{\rm NS}\,{}M_{\rm NS}^{1/2}}\right)^{1/7}\times B^{4/7}.
\end{equation}

$V_{\rm diff}$ is the difference between the Keplerian angular velocity at the magnetic radius $\Omega_{\rm K}|_{R_{\rm A}}$ and the co-rotation angular velocity $\Omega_{\rm co}$:
\begin{equation}
    V_{\rm diff} = \Omega_{\rm K}|_{R_{\rm A}}-\Omega_{\rm co}.
\end{equation}

We can see from Eq.~\ref{eq:jdot}, that the condition for the NS to spin up is $V_{\rm diff}>0$. In this case, the variation in angular momentum of the NS is positive and its spin increases: as matter reaches the magnetic radius, the magnetic pressure dominates and matter follows the magnetic field lines  and  is accreted on the NS polar caps. In contrast, if $V_{\rm diff}<0$, then the NS acts as a propeller \citep{kiel2008}: the velocity of the magnetic field lines at the magnetic radius is higher than the local Keplerian velocity ($\Omega_{\rm co}>\Omega_{\rm K}|_{R_{\rm A}}$) and matter is blown away from the NS \citep{illarionov1975}. In \textsc{sevn}, we have implemented the propeller effect as well.

We also assume that the magnetic field decays exponentially with the amount of mass accreted. Observations of the low magnetic fields of binary and millisecond pulsars suggest accretion-induced field decay \citep{konar1997}. The magnetic field evolves according to the following equation during RLO in our models:
\begin{equation}\label{eq:magneticdecayrlo}
    B = (B_0 - B_{\rm min})\times \exp{\left(-\frac{\Delta M_{\rm NS}}{\Delta M_{\rm d}}\right)} + B_{\rm min}.
\end{equation}

$\Delta M_{\rm NS}$ is the amount of mass accreted by the NS  and $\Delta M_{\rm d}$ is the magnetic field decay mass scale, another free-parameter of the model.  
We fixed $\Delta M_{\rm d}$ to the fiducial value $\Delta M_{\rm d} = 0.025$~M$_\odot$, which corresponds to the optimal value obtained in \cite{chattopadhyay2020}. We further discuss the values adopted in our simulations in Sec. \ref{sec:models}.

At each time-step, \textsc{sevn} evaluates if matter accretes onto the NS via RLO. If not, the variations to the spin and magnetic field of the NS are only due to spin down (Eqs.~\ref{eq:spindown} and \ref{eq:magneticdecay}). Otherwise, the algorithm first evaluates $V_{\rm diff}$. If $V_{\rm diff}>0$, 
\textsc{sevn} calculates also the spin up from Eqs.~\ref{eq:jdot} and \ref{eq:magneticdecayrlo}:
\begin{equation}\label{omegastep}
    \Omega_{\rm i+1} = \Omega_{\rm i} + \frac{\Delta J_{\rm acc}}{I},
\end{equation}
where $\Delta J_{\rm acc} = V_{\rm diff}\,{}R_{\rm A}^2 \Delta M_{\rm NS}$.
At the end of the time-step, \sevn{} updates both the spin and the magnetic field accordingly. If $V_{\rm diff}<0$, no matter accretes during that time-step and spin and magnetic fields are updated only accounting for spin down. This accurate treatment of spin up and down benefits from the adaptive time-step formalism of \sevn{} \citep{iorio2022}.

According to \cite{chattopadhyay2020}, a CE phase can trigger further spin up of the NS, as suggested by \cite{macleod2015}. Here, we decide not to include this possible spin up by CE, because it is quite controversial \citep{oslowski2011, chamandy2018}.

\subsection{Initial conditions for the \sevn{} binary catalogues}

We sample the masses of the primary stars ($M_1$) from a Kroupa's initial mass function \citep{kroupa2001}, in a range between $5$ and $150 \text{M}_\odot$:
\begin{equation}
\mathcal{F}(M) \propto M_{1}^{-2.3}
\end{equation}

We draw the secondary star's mass ($M_2$), the initial binary orbital period ($P_{\rm orb}$), and the eccentricity following the distributions derived by  \cite{sana2012}, which are based on observations of massive binary stars in young clusters:
\begin{equation}
\mathcal{F}(q) \propto q^{-0.1}
\end{equation}

where $q=M_2/M_1\in [q_{\rm min}, 1]$, $q_{\rm min} = \max \left( \frac{2.2 \text{M}_\odot}{M_{1}}, 0.1 \right)$, so that the secondary star mass distribution is cut at $2.2$M$_\odot$,
\begin{equation}
\mathcal{F}(P_{\rm orb}) \propto (\log P_{\rm orb})^{-0.55}
\end{equation}
with $0.15 \leq \log (P_{\rm orb}/{\rm d}) \leq 5.5$,  
and
\begin{equation}
\mathcal{F}(e) \propto e^{-0.42}
\end{equation}
with $0\leq e \leq{} 0.9$. 
For each parameter (described in section \ref{sec:models}), we simulate with \sevn{} 11 sub-sets varying the metallicity: $Z=0.0002$, $0.0004$, $0.0008$, $0.0012$, $0.0016$, $0.002$, $0.004$, $0.008$, $0.012$, $0.016$, $0.02$. Each \sevn{} sub-set evolves $10^6$ binary systems. Therefore, for each model we ran a total of $1.1\times 10^7$ binaries. 

\subsection{Milky Way (MW) models} \label{sec:mwmodel}

In our work, we couple the catalogues of binary NS  systems from \textsc{sevn} with a MW-like galaxy. We choose four Galaxy models 
that have total stellar mass and current star formation rate (SFR) as close as possible to the measured values of the MW \citep{artale2019}. Therefore, we require a total stellar mass $( 5.43 \pm 0.57 ) \times{} 10^{10}$ M$_\odot$  \citep{mcmillan2017} and a current SFR $\sim 1.65$~M$_\odot$~yr$^{-1}$ \citep{licquia2015}. 
The use of different Galaxy models allows us to test the robustness of our results against the specific features of the chosen MW models. Here below, we describe the four models in detail. We summarize them in Table~\ref{tab:mwmodels}.

\subsubsection{Constant  SFR (Const model)}

The first model for the MW assumes a constant SFR, fixed at 1.65~M$_{\odot}$~yr$^{-1}$. All binary systems in this model have  solar metallicity, $Z=0.0142$ \citep{asplund2009, vignagomez2018, chattopadhyay2020}. The integration of the SFR through cosmic time leads to a total stellar mass $M_{\star} \sim 1.9 \times 10^{10}$ M$_{\odot}$, well below the observed value. 
This MW model has been used by most of the previous works \citep[e.g.,][]{vignagomez2018,shao2018, chattopadhyay2020}, adopting a fixed SFR at 1~M$_{\odot}$~yr$^{-1}$. 

\subsubsection{The \textsc{eagle} model}
The \textsc{eagle} project \footnote{\url{https://icc.dur.ac.uk/Eagle/}} \citep{schaye2015, eagle2017} comprises a set of cosmological hydrodynamical simulations with comoving box sizes of $25$, $50$ and $100$~Mpc, evolved from redshift $z = 20$ to $z\sim 0$. The simulations are run with the \textsc{gadget}-3 code \citep{springel2005}. A full description of the sub-grid baryonic processes included can be found in \cite{schaye2015}. 
The cosmological framework adopted is the $\Lambda$CDM cosmological model, with cosmological parameters: $\Omega_{\Lambda,0}=0.693$, $\Omega_{\rm M,0}=0.307$, $\Omega_{\rm b,0}=0.048$, $\sigma_8=0.8288$, $n_s=0.9611$, and $h=0.6777$ \citep{planck2014}.
We selected a MW-like galaxy from the \textsc{eagle} run labeled {\sc L025N0752}, characterized by a periodic box of side $25\,{}\text{Mpc}$ comoving. This is the run with the highest resolution in the \textsc{eagle} suite: the baryonic particle and the dark matter mass resolutions are $2.26 \times 10^5 M_\odot$ and $1.21\times 10^6 M_\odot$,  respectively. 
By imposing the requirements outlined in Section \ref{sec:mwmodel}, we selected a best-fitting galaxy identified by the IDs of the halo, Gnr~$=23$ and  the sub-halo, Sgrn~$=0$. This galaxy is characterized by a total stellar mass $3.65 \times 10^{10} M_\odot$ and a current SFR $1.38$~
M$_{\odot}$~yr$^{-1}$. 

\subsubsection{The \illustris{} model}

The \illustris{} project \footnote{\url{https://www.tng-project.org/}} consists of three simulation volumes, with comoving box size of $50$, $100$ and $300$~Mpc, from  redshift $z = 127$ to $z\sim 0$. The simulations have been run with the moving mesh code \textsc{arepo} \citep{springel2010}.A full description of the baryonic processes included in this simulation can be found in \cite{nelson2019} and \cite{pillepich2019}.
A $\Lambda$CDM cosmological model is assumed, with cosmological parameters from \cite{ade2016}: $\Omega_{\Lambda,0}=0.6911$, $\Omega_{\rm M,0}=0.3089$, $\Omega_{\rm b,0}=0.0486$, $\sigma_8=0.8159$, $n_s=0.9667$ and $h=0.6774$.
We select our MW-like galaxy from the {\sc TNG50} run (with $50$~Mpc comoving length, \citealt{nelson2019tng50}). 
The best candidate galaxy for our work is identified by Gnr~$=244$ and Sgrn~$=547844$, with a total stellar mass $5.10 \times 10^{10}M_{\odot}$ and current SFR  $1.675$~M$_\odot$~yr$^{-1}$. 
The galaxy is included in the MW-type catalog released by \citet{pillepich2023} from the TNG50 simulation.

\subsubsection{The empirical (Emp) model}

We also include a MW-like model built from empirical relations. We adopt an exponentially decaying SFR function:
\begin{equation}\label{eq:empiricalsfr}
    {\rm SFR}(\tau) = A \exp{(-\tau/\tau_{\Psi})}
\end{equation}
where $A$ is  a proportionality constant, $\tau$ is the internal Galactic age, defined as the time $t_{\rm f}$ elapsed since the beginning of significant star formation activity in the Galaxy, and $\tau_{\Psi}$ is the characteristic timescale for the suppression of SFR \citep{grisoni2017, bovy2017, boco2019}. 
This model is commonly adopted to describe low-redshift, disc-dominated galaxies \citep{chiappini1997, courteau2014, pezzulli2015, grisoni2017}.
Integrating the SFR over the cosmic time yields the total stellar mass of the galaxy:
\begin{equation}\label{eq:starmassmw}
    M_\star (t_0) = \int_{t_{\rm f}}^{t_0} \,{}{\rm SFR}(t)\,{} {\rm d}t,
\end{equation}
where $t_0$ is the cosmic time today.
For our MW-like galaxy, we set $\tau_{\Psi} = 7\,{}$Gyr, as the SFR decay timescale \citep{bovy2017}. 
To derive the two unknowns in Equations \ref{eq:empiricalsfr} and \ref{eq:starmassmw}, $A$ and $t_{\rm f}$, we imposed a total stellar mass $M_\star (t_0)= 5 \times 10^{10}$~M$_\odot$ and a current star formation rate SFR$(t_0) = 1.65$~M$_\odot$~yr$^{-1}$. 

As for the metallicity evolution, we assume the fundamental metallicity relation \citep[FMR,][]{mannucci2010, mannucci2011}. The FMR has been obtained empirically through observations of the local galaxies of the Sloan Digital Sky Survey,  and its robustness has been confirmed for galaxies up to $z\sim3.5$ \citep{mannucci2010, hunt2016}. The FMR links $M_\star$, SFR and the gas metallicity $Z$, whereas it is almost independent of redshift \citep{chruslinska2021, boco2021, santoliquido2022}.  
Following this approach, the Galaxy metallicity is evaluated through Eq.~2 of \cite{mannucci2011}:
\begin{equation}\label{eq:fmr}
    12 + \log ({\rm O}/{\rm H})=
    \begin{cases}
    \begin{split}
        8.90+ & 0.37\,{}m -0.14\,{}s-0.19\,{}m^2 \\
        &+0.12\,{}m\,{}s -0.054\,{}s^2 
    \end{split}
    & \text{for}\,{} \mu_{0.32}\geq9.5 \\
    8.93 + 0.51 \,{}(\mu_{0.32} - 10) & \text{for}\,{} \mu_{0.32}<9.5  
    \end{cases}
\end{equation}

with
      $\mu_{0.32} = \log M_\star - 0.32\,{}  \log {\rm SFR}$,
       $m=\log{M_\star} -10$, and 
      $s=\log{\rm SFR}$.
We build our empirical model by  evaluating the SFR (Eq.~\ref{eq:empiricalsfr}) and the metallicity at a series of time-steps, ranging from $\sim 12$~Gyr ago until today. At each time-step, we assume that the metallicity follows a log-normal distribution within the Galaxy, with mean equal to  the value predicted by the FMR (Eq.~\ref{eq:fmr}) and variance $\sigma_{\log{Z}} = 0.1$~dex.

\subsection{Populating the MW  with BNSs}

To populate the  MW galaxy models with the catalogues of BNSs evolved with \textsc{sevn}, we follow the same  procedure as outlined in \cite{mapelli2017} and \cite{mapelli2018}. For each simulation set, we read the eccentricity $e$ and the semi-major axis $a$ of the binary, the times of the SN explosions $t_1$ and $t_2$, the NS masses $m_1$ and $m_2$, the spin periods $P_1$ and $P_2$ and magnetic fields $B_1$ and $B_2$ at the formation of the second compact object.
Along with this, we save the total simulated initial stellar mass $M_{\textsc{sevn}}$ for each of the 11 metallicity sub-sets. 

For the \textsc{eagle} and \illustris{} models, we read every stellar particle from the snapshot of the Galaxy at redshift $z=0$. 
In particular, we need the stellar particle mass $m_\star$, its formation redshift $z_\star$, and its metallicity at formation $Z_\star$. 
For each stellar particle, we select the \textsc{sevn} catalogue with the metallicity closest to $Z_\star$, and  associate to it a number of BNSs proportional to the ratio $m_\star / M_{\textsc{sevn}}$:
\begin{equation}\label{eq:nbns}
    n_{\rm BNS} = N_{\textsc{sevn}} \frac{m_\star}{ M_{\textsc{sevn}}} f_{\rm corr}\,{} f_{\rm bin},
\end{equation}
where $N_{\textsc{sevn}}$ is the number of BNSs within the selected catalogue, $f_{\rm corr}$, and $f_{\rm bin}$ are correction factors. The factor $f_{\rm corr}=0.285$ takes into account that we consider only systems with primary star mass $\geq 5 M_\odot$. By integrating the Kroupa IMF between $0$ and $5$~M$_\odot$, we obtain the relative weight of stars with a mass $\geq 5$~M$_\odot$: this yields $f_{\rm corr} = 0.285$. The factor  $f_{\rm bin}$ corrects for the binary fraction, i.e. takes into account that we evolve only binary systems. From \cite{sana2012}, $f_{\rm bin} = 0.4$, which is equivalent to saying that $40\%$ of stellar mass lie in binary systems. We randomly choose, with a Monte Carlo-like approach, $n_{\rm BNS}$ BNSs from the \textsc{sevn} catalogue. 

For the empirical model, we do not have stellar particles, but we can derive the amount of stellar mass formed by the Galaxy $\Delta M_\star$ at each time-step. Knowing the metallicity distribution at the same time-step, we can derive how much mass has been accreted $\Delta m_\star |_{Z}$, for each of the \textsc{sevn} metallicities. The sum over all the 11 metallicities is:
\begin{equation}
    \Delta m_{\star} = \sum_{i=1}^{11} \Delta m_\star|_{Z_i} 
\end{equation}
So, for each metallicity we evaluate $n_{\rm BNS}$ simply by substituting the value of  $\Delta M_\star|_{Z_i}$ in Eq.~\ref{eq:nbns}. 

We convert the formation redshift $z_\star$ of each stellar particle in lookback time:
\begin{equation}
t_{\rm lb} (z) = \frac{1}{H_0}\int_0^z \frac{dz'}{(1+z')\sqrt{\Omega_{\rm M,0}\,{} (z'+1)^3 
+\Omega_{\Lambda{},0}}}\label{eq:tlook}
\end{equation}
 choosing the cosmological parameters consistently with the adopted cosmological simulation (Section~\ref{sec:mwmodel}). Then, the formation time of a BNS within the Galaxy is \footnote{
 $t_1$ and $t_2$ mark the SN explosion times of the primary and secondary progenitor stars, respectively. An episode of mass transfer during the evolution of the system might revert the initial mass ratio, thus leading to $t_1 > t_2$.}
\begin{equation}
t_{\rm BNS} = t_\star - \max{(t_1,\,{} t_2)} ,
\end{equation}
where $t_\star$ is the lookback time of the formation of the stellar particle corresponding to the redshift $z_\star$. From now on, all the times will be expressed in lookback times. 

We assume that the subsequent evolution is driven only by the emission of gravitational wave radiation. Integrating the differential equations for gravitational wave radiation, derived in \cite{peters1964}, we evolve the eccentricity and the semi-major axis:

\begin{eqnarray}
\frac{{\rm d}a}{{\rm d}t} = - \frac{64}{5}\frac{G^3\,{} m_1 \,{}m_2 (m_1+m_2)}{c^5 \,{}a^3\,{} (1-e^2)^{7/2}}\left(1+\frac{73}{24}\,{}e^2 + \frac{37}{96}\,{}e^4\right), \nonumber\\
\frac{{\rm d}e}{{\rm d}t} = - \frac{304}{15}\frac{G^3 \,{}m_1\,{} m_2 \,{}(m_1+m_2)}{c^5 \,{}a^3\,{} (1-e^2)^{5/2}}\left(1+\frac{121}{304}\,{}e^2 \right).
\end{eqnarray}

For the integration, we employ an Euler method with an adaptive time-step: the time-steps are smaller if the system is rapidly evolving (i.e. close to the merger); if the system is instead on a loose-wide orbit, the integration time-steps are larger. This allows us to have a fast algorithm  without losing in precision. 
We evolve each system from $t_{\rm BNS}$ until today, i.e. $t_{\rm lb}=0$. If, during the evolution $a = 3\,{}r_{\rm S}$, where $r_{\rm S}$ is the Schwartzschild radius, we assume that the system has merged and do not evolve it anymore.

In parallel, we evolve the NS spin and magnetic field. In particular, we update the spin and magnetic field according to Eqs.~\ref{eq:magneticdecay} and \ref{eq:spinevolve}, with 
\begin{equation}
    \Delta t = t_{\rm BNS} - t_{\rm BNS, final},
\end{equation}
where $t_{\rm BNS, final} = t_{\rm merge}$ if the system has merged, or $t_{\rm BNS, final}=0$ (i.e., the present time) otherwise. We repeat the same procedure for each parameter set and for each Galaxy model keeping track of both the set of merged BNSs and the population of BNSs today in the MW.

\subsection{Simulation set up} \label{sec:models}


\begin{table} \centering
\begin{tabular}{lcr}
\hline
MW model &  $M_\star \,{}(10^{10}$ M$_{\odot})$   &   SFR (M$_{\odot}$ yr$^{-1})$ \\ \hline

\textsc{eagle}  &   $3.7$   & $1.38$   \\
\illustris{}   &   $5.1$   & $1.67$   \\
Empirical (Emp)   &   $5.0$   & $1.65$   \\
Constant  (Const)   &   $1.9$   & $1.65$   \\ \hline
\end{tabular}
\caption{Properties of the MW models. Columns 1 and 2 refer to the total stellar mass $M_\star$ and current SFR of the considered Galaxy models.}
\label{tab:mwmodels}
\end{table}

\begin{table*}
    \centering
    \begin{tabular}{c c c c c}
         \hline
         Initial Distribution & $B_{\rm birth}$ range (G) & $B_{\rm birth}$ distribution & $P_{\rm birth}$ range (ms) & $P_{\rm birth}$ distribution \\ \hline
         U & ($10^{10} - 10^{13}$) & Uniform & ($10-100$) & Uniform \\
         FL & ($10^{10} - 10^{13}$) & Flat in log & ($10-100$) & Uniform \\
         FG &  --  & \cite{fauchergiguere2006} & -- & \cite{fauchergiguere2006} \\ \hline
    \end{tabular}
    \caption{Initial distributions adopted in this work for the spins and magnetic fields.}
    \label{tab:initdistr}
\end{table*}


\begin{table*}
    \centering
    \begin{tabular}{>{\columncolor[gray]{0.9}}lccc cc   >{\columncolor[gray]{0.9}}lccc}
        \hline
    \rowcolor{white}
         & $\alpha$ & Init. Distr. & $\tau_{\rm d}$ (Gyr) &&& & $\alpha$ & Init. Distr. & $\tau_{\rm d}$ (Gyr) \\ \hline
        Ua0.5t0.1Emp & 0.5 & U & 0.1 &&&  Ua3t0.1Emp &  3 & U & 0.1 \\
        Ua0.5t0.5Emp & 0.5 & U & 0.5 &&& Ua3t0.5Emp & 3 & U & 0.5 \\
        Ua0.5t1Emp & 0.5 & U & 1 &&&  \textbf{Ua3t1Emp} &  \textbf{3} & \textbf{U} & \textbf{1}\\
        Ua0.5t2Emp & 0.5 & U & 2 &&& Ua3t2Emp &  3 & U & 2\\
        &&&&&&&&& \\
        FLa0.5t0.1Emp  & 0.5 & FL & 0.1 &&& FLa3t0.1Emp  & 3 & FL & 0.1 \\ 
        FLa0.5t0.5Emp  & 0.5 & FL & 0.5 &&& FLa3t0.5Emp  & 3 & FL & 0.5 \\
        FLa0.5t1Emp  & 0.5 & FL & 1 &&& FLa3t1Emp & 3 & FL & 1 \\
        FLa0.5t2Emp  & 0.5 & FL & 2 &&& FLa3t2Emp  & 3 & FL & 2 \\ 
        &&&&&&&&& \\
        FGa0.5t0.1Emp & 0.5 & FG & 0.1 &&& FGa3t0.1Emp & 3 & FG & 0.1 \\
        FGa0.5t0.5Emp & 0.5 & FG & 0.5 &&&  FGa3t0.5Emp  & 3 & FG & 0.5 \\
        FGa0.5t1Emp & 0.5 & FG & 1 &&& FGa3t1Emp & 3 & FG & 1 \\
        FGa0.5t2Emp & 0.5 & FG & 2 &&& FGa3t2Emp & 3 & FG & 2 \\ 
        &&&&&&&&& \\
        
        Ua1t0.1Emp & 1 & U & 0.1 &&& Ua5t0.1Emp & 5 & U & 0.1 \\
        Ua1t0.5Emp & 1 & U & 0.5 &&& Ua5t0.5Emp & 5 & U & 0.5 \\
        Ua1t1Emp & 1 & U & 1 &&&  Ua5t1Emp & 5 & U & 1 \\
        Ua1t2Emp & 1 & U & 2 &&& Ua5t2Emp & 5 & U & 2 \\
        &&&&&&&&& \\
        FLa1t0.1Emp & 1 & FL & 0.1 &&& FLa5t0.1Emp  & 5 & FL & 0.1 \\ 
        FLa1t0.5Emp & 1 & FL & 0.5 &&& FLa5t0.5Emp & 5 & FL & 0.5 \\
        FLa1t1Emp & 1 & FL & 1 &&& FLa5t1Emp & 5 & FL & 1 \\
        FLa1t2Emp & 1 & FL & 2 &&& FLa5t2Emp & 5 & FL & 2 \\ 
        &&&&&&&&& \\
        FGa1t0.1Emp & 1 & FG & 0.1 &&& FGa5t0.1Emp & 5 & FG & 0.1 \\
        FGa1t0.5Emp & 1 & FG & 0.5 &&& FGa5t0.5Emp & 5 & FG & 0.5 \\
        FGa1t1Emp & 1 & FG & 1 &&& FGa5t1Emp & 5 & FG & 1 \\
        FGa1t2Emp & 1 & FG & 2 &&& FGa5t2Emp & 5 & FG & 2\\ 

        \hline
    \end{tabular}
    \caption{ Summary of the simulations. The initial distributions for spins and magnetic fields of NSs (U, FL and FG) are summarized in Table \ref{tab:initdistr}. All models listed here have been run within the Empirical MW model. We have explored the same grid of parameters with the other MW models ({\sc eagle}, \illustris{}, and Const). All models assume the rapid-Gauss prescription for the SN explosion. Our fiducial model is highlighted  in bold text in the Table.} 
    \label{tab:models}
\end{table*}

We varied several parameters in our model: the SN explosion mechanism prescription (rapid, delayed, rapid-gauss), the CE efficiency parameter $\alpha$ ($\alpha=0.5,$ 1, 3, and 5), the initial spin and magnetic-field distributions, and the magnetic field decay timescale $\tau_{\rm d}$. 
All the models adopt the rapid-gauss prescription to reproduce the masses of the NSs. In the Appendix \ref{sec:masses}, we test for comparison also the rapid and the delayed models 
from \cite{fryer2012}. 
We investigate different values of the parameter $\alpha$. In particular we set $\alpha=0.5$, $1$, $3$, and $5$.
Both spins and magnetic fields are drawn from initial distributions. We need to make this assumption because the link between the properties of the pre-SN star and those of the NS is highly uncertain. 
Given the small sample of observed pulsars, the initial distributions are still quite uncertain. There is no agreement in the literature on the favoured initial parameter distributions, see for example \cite{fauchergiguere2006, oslowski2011}. 
In our work we consider the initial distributions as free parameters and  test some of the most common models in the literature, as summarized in Table~\ref{tab:initdistr}:
\begin{enumerate}
    \item \emph{uniform} (hereafter, U): the spin period and the magnetic field are drawn uniformly in the range $[10, 100]$~ms and $[10^{10}, 10^{13}]$~G, respectively.
    \item \emph{flat-in-log} (hereafter, FL): the spin periods are drawn uniformly between $10$ and $100$~ms, the magnetic fields are distributed according to a flat-in-log distribution in the range $[10^{10}, 10^{13}]$~G. 
    \item \emph{Faucher-Giguère} (hereafter, FG): spin periods and magnetic fields follow the distributions presented in \cite{fauchergiguere2006}. The spin periods follow a normal distribution with mean $ \langle{P}\rangle=300$~ms and variance $\sigma_P = 150$~ms, while the magnetic fields are drawn from a log-normal distribution with mean $\langle\log (B/G)\rangle=12.65$ and variance $\sigma_{\langle \log B \rangle} = 0.55$.
\end{enumerate}
 
Moreover, we consider different values for the magnetic field decay timescale $\tau_{\rm d}$ 
(Eq.~\ref{eq:magneticdecay}): $\tau_{\rm d}=0.1$, $0.5$, $1$ and $2$~Gyr.
We test our results on the different MW-like galaxies as well (Section \ref{sec:mwmodel}). We adopt a uniform terminology for our models. Every model is identified with a string built as follows:
`\{distr\}a\{alpha\}t\{tau\}\{MW\}',
 where within the brackets we substitute the actual value (or acronym) assumed by that specific parameter in the model. So that, in the place of `distr' we substitute the acronym for the initial spins and magnetic fields distributions (see Table \ref{tab:initdistr}); `alpha' corresponds to the value $\alpha$ for CE efficiency, `tau' represents the value $\tau_{\rm d}$ of the magnetic field decay timescale, and lastly `MW' the MW-like galaxy chosen. We use the following abbreviations for the MW models: \textsc{eagle}, \illustris{}, Emp and Const. 
For example,  model `Ua3t1Emp' adopts the uniform distribution of initial spin and magnetic field, $\alpha=3$, $\tau_{\rm d}=1$~Gyr, and the empirical MW model. 
The models are summarized in Table \ref{tab:models}. We choose `Ua3t1Emp' as our fiducial model.

\subsection{Selection effects} \label{sec:seleffects}

To compare our modelled BNSs with observations, we need to account for selection effects. For instance, as a pulsar crosses the death line (see section \ref{sec:deathlines}) it ceases to emit in the radio. Furthermore, pulsars have quite narrow beaming opening angles, thus we can only detect the pulsars whose beam intersects the line of sight. 
Radio selection effects also depend on the sky location of the source. The flux density scales as the inverse of the distance squared, which means that, for a given intrinsic luminosity, observations are biased towards close pulsars.
Moreover, scattering by free electrons in the interstellar medium  smears the pulsars’ signal, lowering the signal-to-noise ratio \citep{cordes2002}. Other effects that may affect the observations are pulse nulling and intermittency, i.e. when the pulsed emission ceases for many pulse periods and quasi-periodic on/off cycles, respectively \citep{lyne2010}. On top of this, the Doppler shifting of the period caused by orbital motion also smears the pulsar signal \citep{andersen2018, balakrishnan2022}.
We use the python implementation of \psrpop{} \citep{lorimer2011},  \psrpoppy{} \footnote{\url{https://github.com/samb8s/PsrPopPy}} \citep{bates2014},  to account for some of the mentioned selection effects.
Specifically, we model the death lines and beam geometry, the dependence on sky location, interstellar radio scintillation, pulsar luminosity and binary selection effects.

\subsubsection{Death lines} \label{sec:deathlines}

Pulsars cease to emit in the radio when the magnetic field is not strong enough for the production of electron-positron pairs. The death lines, empirical relations in the $P-\dot{P}$ plane, mark the locus of points beyond which the pulsars stop emitting: if a pulsar crosses one of these lines in the $P-\dot{P} \,{}$ plane, it turns off.
We adopt the death lines from \cite{rudak1994}:
\begin{align}\label{eq:deathlines}
    \log\dot{P} &= 3.29 \,{}\log P - 16.55 \nonumber{}\\
    \log\dot{P} &= 0.92 \,{} \log P - 18.65.
\end{align}

Moreover, to avoid the piling up of pulsars at the death lines, we also add a cut-off on the efficiency of radio emission as in \cite{szary2014}. The radio efficiency $\xi_{\rm R}$ is defined as:
\begin{equation}
    \xi_{\rm R}=\frac{L}{\dot{E}},
\end{equation}
where $L$ is the pulsar radio luminosity and $\dot{E} = 4 \,\pi^2\,{} I\,{} P^{-3}\dot{P}$ is the pulsar spin down power. Following the model by \cite{szary2014}, if the radio efficiency exceeds a certain threshold ($\xi_{\rm R} > \xi_{\rm R, max}$), the pulsar ceases to emit. We set the threshold to $\xi_{\rm R, max}=0.01$, as in \cite{chattopadhyay2020, chattopadhyay2021}.
Following these prescriptions, \psrpoppy{} classifies a pulsar as dead if either it has crossed the death lines (Eq.~\ref{eq:deathlines}) or if $\xi_{\rm R} > \xi_{\rm R, max}$.

\subsubsection{Beaming fraction}

Pulsar radio emission is concentrated on collimated beams with finite width, so that pulsars sweep out only a limited area of the sky. As a consequence, we can detect only a fraction of the whole pulsar population, i.e. those whose beam crosses the observer's line of sight.
The beaming fraction $f_{\rm beam}$ represents the fraction of pulsars beaming towards us. Previous studies agree that $f_{\rm beam}$ is period dependent. We adopt the prescription proposed in \cite{tauris1998}, an empirical relation obtained by fitting slow-rotating pulsars ($P\gtrsim 100$ ms),
\begin{equation}\label{eq:beamingfraction}
    f_{\rm beam} = 0.09\left(\log{P} - 1\right)^2 + 0.03 \qquad 0\leq f_{\rm beam} \leq 1, 
\end{equation}
where $P$ is the spin period of the pulsar in seconds. Equation \ref{eq:beamingfraction} shows that the beaming fraction is higher for pulsars spinning faster. Therefore such objects are also more likely to be detected.
\psrpoppy{} takes into account the beaming effects using a rejection sampling method: each pulsar can be detected with a  probability $f_{\rm beam}$, computed according to Eq. \ref{eq:beamingfraction}.

\subsubsection{Survey sensitivity} \label{subsec:surveysens}

We use the radiometer Equation \citep{dewey1985, lorimer2004} to evaluate the minimum flux that a source must have to be detected 

\begin{equation} \label{eq:radiometereq}
    S_{\rm min}= \beta\,{} \frac{(S/N_{\rm min})\,{}(T_{\rm rec} + T_{\rm sky})}{G_{\rm A} \sqrt{n_{\rm pol}\,{} t_{\rm int}\,{} \Delta \nu}} \sqrt{ \frac{W}{P - W}}, 
\end{equation}

where $S/N_{\rm min}$ is the minimum signal-to-noise ratio, $\beta$ takes into account losses in sensitivity due to sampling antenna and digitalisation noise, $n_{\rm pol}$ is the number of polarizations, $T_{\rm rec}$ and $T_{\rm sky}$ are the receiver and sky temperatures, $G_{\rm A}$ is the antenna gain, $\Delta \nu$ is the observing bandwidth, $t_{\rm int}$ is the integration time, $W$ is the detected pulse width, and $P$ is the pulse period. 
Most of the latter quantities are survey dependent (e.g. receiver temperature, observing bandwidth, integration time). \psrpoppy{} evaluates the radiometer equation for the Parkes Multibeam Pulsar Survey \citep[PMSURV]{manchester2001}, the Swinburne Multibeam Pulsar Survey \citep[SWINMB]{edwards2001}, and the High Time Resolution Universe Pulsar Survey \citep[HTRUP]{keith2010}.
For each survey, $S/N_{\rm min}$ is fixed to the value set by the survey itself. Moreover, the observed pulsars sample considered in our analyses includes only the objects detected by the surveys listed above.
The effects caused by the propagation of the pulsed signal through the interstellar medium enters Eq.~\ref{eq:radiometereq} via $W$. In fact, collisions with free electrons cause a broadening of the received signal. Also, if $W \gtrsim P$ the signal is no longer detectable as the pulse is smeared into the background \citep{lorimer2008, lorimer2011}. 

\psrpoppy{} assesses the observed pulse widths from the following relation \citep{burgay2003}:
\begin{equation}
    W^2 = W_{\rm i}^2 + \tau_{\rm samp}^2 + \left( \tau_{\rm samp} \frac{\rm DM}{\rm DM_0}\right)^2 + \tau_{\rm scatt}^2
\end{equation}
Here $W_{\rm i}$ is the intrinsic pulse width, $\tau_{\rm samp}$ is the sampling timescale, $\tau_{\rm scatt}$ is the mean scattering timescale. ${\rm DM}$ and ${\rm DM}_0$ are the dispersion measures in the direction of the pulsar and the diagonal dispersion of the survey respectively. For simplicity we fix the duty cycle $W_{\rm i} / P = 0.05 \%$ for all pulsars \citep[e.g.][]{lyne1988}.

As the above quantities depend on the sky position of the source, we use the built-in functions of \psrpoppy{} to plant our evolved pulsars in the Galaxy. 
\psrpoppy{} employs the distribution obtained in \cite{yusifov2004} to assign a radial position to each pulsar. 
Once assessed the position of the pulsar, the sky noise temperature $T_{\rm sky}$ is evaluated accordingly by fitting the \cite{haslam1981} Table and rescaling for the correct frequency range \citep{szary2014}. \psrpoppy{} evaluates the dispersion measure DM by integrating the electron density $n_{\rm e}$ over the line of sight. For $n_{\rm e}$ we adopt the NE2001 model \citep{cordes2004}. Finally, DM is used to evaluate $\tau_{\rm scatt}$ by applying the prescriptions described in  \cite{bhat2004}.

The last missing ingredient is the luminosity. We adopt the luminosity function calculated in \cite{fauchergiguere2006},
\begin{equation}
    \log L = \log L_0 + \alpha_{\rm F06}\,{} \log P + \beta_{\rm F06}\,{} \log (\dot{P}/10^{-15}) + \delta_L, 
\end{equation}

with $L_0 = 0.18    \,{}\text{mJy}\,{} \text{kpc}^2$, $\delta_L$ is drawn from a normal distribution with $\sigma_{\delta_L} = 0.8$, $\alpha_{\rm F06} = -1.5$ and $\beta_{\rm F06} = 0.5$. 

For each pulsar in the simulated Galactic population, we evaluate both $S_{\rm min}$ and the flux 
    $F = {L}/{(4\,{}\pi\,{} \mathcal{D}^2)},$ 
where $\mathcal{D}$ is the pulsar distance.
A source is 
detectable if $F>S_{\rm min}$. We repeat this procedure for each survey, finally combining all the simulated-detected pulsars into a single sample.

\subsubsection{Selection effects of binary systems}

A comprehensive procedure to account for binary-pulsar  selection effects is still unavailable.
The Doppler shifting of the period causes a smearing of the signal and thus a reduction in the signal-to-noise ratio \citep{balakrishnan2022}. This effect is stronger for binary systems with shorter periods and for shorter integration times. 
The Doppler shifting also depends on the eccentricity of the system. \cite{bagchi2013} show that the reduction in the signal-to-noise ratio decreases for higher eccentricity. \cite{chattopadhyay2021} derive a fitting formula from the  results of \cite{bagchi2013}. They assume a NS with a mass of $1.4M_\odot$, $1000$~s duration of observation, and $60^\circ$ inclination angle; then, fitting via linear regression for eccentricities $e=0.1, 0.5$ and $0.8$, they find a detection cut-off
\begin{equation}
    P_{\rm orb}/d \geq m \times P/s + c,
\end{equation}

with
\begin{align}
    m &= m_{\rm m} \,{} e + c_{\rm m} \\
    c &= m_{\rm c} \,{} e + c_{\rm c}, 
\end{align}
where $m_{\rm m} = -8.90$, $c_{\rm m}=-27.68$, $m_{\rm c}=-3.40$ and $c_{\rm c}=5.72$. Thus, pulsars in shorter orbital period binaries are also the most difficult to be detected. 
We implement  the aforementioned fit  \citep{chattopadhyay2021} in \psrpoppy{}. 

\subsubsection{Final setup}

Using the formalism discussed above we apply the radio selection effects on the final population of BNSs in the MW. We consider only the binaries survived in the MW until the present day and run \psrpoppy{} on these catalogues. The code returns the sample of pulsars, picked from the initial catalogue, that are detectable by the selected surveys.

We ran \psrpoppy{} multiple times on the total population, in order to filter out stochastic fluctuations. In this way, we obtain multiple realisations of the radio-selected pulsars. This procedure has a further advantage: as a bootstrap technique, it augments the statistic of our final sample.

We repeat this process for each of our models. For each simulation, we compare simulated and observed pulsars, according to three surveys (the Parkes Multibeam Pulsar Survey, Swinburne Multibeam Pulsar Survey, and High Time Resolution Universe Pulsar Survey, Section~\ref{sec:seleffects}). We list the properties of the pulsars considered in this study in Table~\ref{tab:pulsars} \citep{manchester2005}.

\subsection{Statistical analysis} \label{sec:statisticalanalysis} 

We briefly summarize here the main points of the statistical framework used to compare the adopted models. In the following, we will denote the properties of the $N_D$ observed pulsars (orbital period $P_\mathrm{orb}$, eccentricity $e$, spin period $P$ and rate of change of the spin period $\dot P$) with $D = \{P_{\mathrm{orb}}, e, P, \dot{P}\}$ , whereas $\theta_i$ denotes all the hyper-parameters varied through the models (i.e., the SN explosion mechanism, the CE parameter $\alpha$, the initial spin and magnetic-field distributions, the magnetic-field decay $\tau_{\rm d}$, and the MW model) and, by extension, the $i-$th model itself.

To quantify the relative performance of two competing models in describing the available data, we will compute the Bayes' factor $B_{12}$\footnote{In general, to do model selection one should compute the odds ratio $O_{12} = B_{12} P(\theta_1)/P(\theta_2)$. However, under the assumption that the two models are a priori equally likely, the odds ratio reduces to the Bayes' factor.}:
\begin{equation}
    B_{12} = \frac{P(D, N_D|\theta_1)}{P(D, N_D|\theta_2)}\,.
\end{equation}
In general, the likelihood $P(D|\theta)$ requires to evaluate the probability of the available data conditioned on the model. In this case, however, we do not have a functional relationship between the astrophysical hyper-parameters $\theta$ and the pulsar's observed parameters.

On the other hand, we have catalogues of simulated observed pulsars, as described in Section~\ref{sec:seleffects}. Each of these catalogues, denoted with $\xi(\theta)$, is composed of $N_{\xi}$ pulsars, representative of the underlying observed distribution on the four pulsar properties. These realisations can be used to reconstruct this distribution and assign a probability for the data $D$.

The likelihood therefore becomes
\begin{multline}
P(D, N_D|\theta) = \int P(D, N_D|\xi(\theta),N_\xi)\,{}P(\xi(\theta), N_\xi|\theta)\,{}{\rm d}\xi \,{}{\rm d}N_\xi =\\= \int P(N_D|N_\xi)\,{}P(D|\xi(\theta))\,{}P(\xi(\theta), N_\xi|\theta)\,{}{\rm d}\xi \,{}{\rm d}N_\xi\,.
\end{multline}
The first term,  $P(N_D|N_\xi)$, is an inhomogenous Poisson process:
\begin{equation}
     P(N_D|N_\xi) = \frac{N_\xi^{N_D}e^{-N_\xi}}{N_D!}\,.
\end{equation}
The probability $P(D|\xi(\theta))$ is modelled using a Dirichlet process Gaussian mixture model (DPGMM, e.g., 
\citealt{Rinaldi2022}, and references therein). The DPGMM can be used to approximate arbitrary probability densities given a set of samples drawn from the unknown distribution. 
In particular, 
\begin{equation}
    P(D|\xi(\theta)) = \int P(D|\lambda)\,{}P(\lambda|\xi(\theta))\,{} {\rm d}\lambda\,,
\end{equation}
where $\lambda$ denotes the parameters of the DPGMM.

Both integrals can be calculated via Monte Carlo approximation, since we are able to sample $\xi(\theta)$ and $N_\xi$ using the methods outlined above. Samples for $\lambda$, conditioned on a specific realisation of $\xi(\theta)$, are drawn using \textsc{figaro}\footnote{\textsc{figaro} is publicly available at \url{https://github.com/sterinaldi/figaro}}.  

With the Monte Carlo approximation, the full likelihood reads
\begin{multline}\label{MC_likelihood}
    P(D,N_D|\theta) \simeq \\\frac{1}{M}\sum_{\xi_j(\theta)}^{\xi_M} \left(\frac{ \left(N_{\xi_j}^{N_D}e^{-N_{\xi_j}}\right)/N_D!}{K} \sum_{\lambda_k|\xi_j(\theta)}^{\lambda_K}\prod_i P(D_i|\lambda_k)\right)\,,
\end{multline}
where $K$ denotes the number of draws for $\lambda$ and $M$ the number of realisations for $\xi(\theta)$, and we made also use of the fact that the observations $D$ are independent.

Using Eq.~\eqref{MC_likelihood}, we can compute the likelihood for each model and, consequently, discriminate between models. If $B_{12} > 1$, the model $\theta_1$ is favoured over model $\theta_2$.
In the following, we will compute the Bayes' factor for each model over our fiducial `Ua3t1Emp'.

\begin{table*}\centering
\resizebox{\textwidth}{!}{\begin{tabular}{lccccccccccc}
\hline
Radio pulsar  & Type  &   $P$ &   $\dot{P}$   &   $B$ &   $P_{\rm orb}$   &   $e$ &   $M_{\rm psr}$   &   $M_{\rm comp}$  &   Dist.   & $t_{\rm GW}$ &  Survey \\
 & & (ms)  & ($10^{-18}$) &   ($10^9$ G)  &   (d)  &   &   ($\text{M}_\odot$) &   ($\text{M}_\odot$) &   (kpc)   &   (Myr) &    \\ \hline

J0737-3039 A$^\text{a}$  &   recycled    &   $22.7$    &   $1.76$    &   $2.0$ &   $0.102$   &   $0.088$   &   $1.338$ &   $1.249$   &   $1.15$    &   $86$ & PMSURV, HTRU \\

J0737-3039 B$^\text{a}$  &   recycled    &   $277.3$    &   $892.0$    &   $1590$ &   $0.102$   &   $0.088$   &   $1.249$ &   $1.338$   &   $1.15$    &   $86$  & PMSURV\\

J1753-2240$^\text{b}$    & recycled    &   $95.1$    &   $0.970$   &   $2.7$ &   $13.638$  &   $0.304$   &   -   & -   &   $3.46$    &   $>t_{\rm H}$ & PMSURV, HTRU\\
J1755-2550$^\text{c}$   & young   &   $315.2$    &   $2430$    &   $270$ &   $9.696$   &   $0.089$   &   - &   $>0.40$   &   $10.3$    &   $>t_{\rm H}$ & HTRU\\
J1756-2251$^\text{d}$  & recycled    &   $28.5$    &   $1.02$    &   $1.7$ &   $0.320$   &   $0.181$   &   $1.341$ &   $1.230$   &   $0.73$    &   $1660$ & PMSURV, HTRU\\
J1757-1854$^\text{e}$    & recycled    &   $21.5$    &   $2.63$    &   $7.6$ &   $0.184$   &   $0.606$   &   $1.338$ &   $1.395$   &   $7.40$    &   $78$ & HTRU\\
J1811-1736$^\text{f}$   & recycled    &   $104.2$    &   $0.901$    &   $3.0$ &   $18.779$   &   $0.828$   &   $<1.64$ &   $>0.93$   &   $5.93$    &   $>t_{\rm H}$ & PMSURV, HTRU\\
B1913+16$^\text{g}$   &   recycled    &   $59.0$    &   $8.63$    &   $7.0$ &   $0.323$   &   $0.617$   &   $1.440$ &   $1.389$   &   $9.80$    &   $301$ & PMSURV\\
\hline
\end{tabular}}
\caption{Sample of radio pulsars considered in this work. Column 1: pulsar name; column 2: spin $P$; column 3: spin derivative $\dot{P}$; column 4: magnetic field $B$; column 5: orbital period $P_{\rm orb}$; column 6: eccentricity $e$; column 7: pulsar mass $M_{\rm psr}$; column 8: companion mass $M_{\rm comp}$ (here we consider only pulsars which have another NS as companion); column 9: distance from the Sun; column 10: $t_{\rm GW}$ is the merging time of the binary systems, the BNSs with $t_{\rm GW}$ greater than the Huble time ($t_{\rm H}$) are shown with $> t_{\rm H}$. The last column (Column 12) shows in which surveys each pulsar has been discovered, the nomenclature is the same as in Sec. \protect\ref{sec:seleffects}. References: $^\text{a}$ \protect\cite{kramer2006, breton2008} , $^\text{b}$ \protect\cite{keith2009}, $^\text{c}$ \protect\cite{ng2015, ng2018} , $^\text{d}$ \protect\cite{faulkner2004}, $^\text{e}$ \protect\cite{cameron2018}, $^\text{f}$ \protect\cite{corongiu2007}, $^\text{g}$ \protect\cite{hulse1975a,weisberg2016}.} 
\label{tab:pulsars}
\end{table*}

\section{Results} \label{sec:results}
\subsection{Merger rates}

\begin{figure}
    \centering
    \includegraphics[width=.5\textwidth]{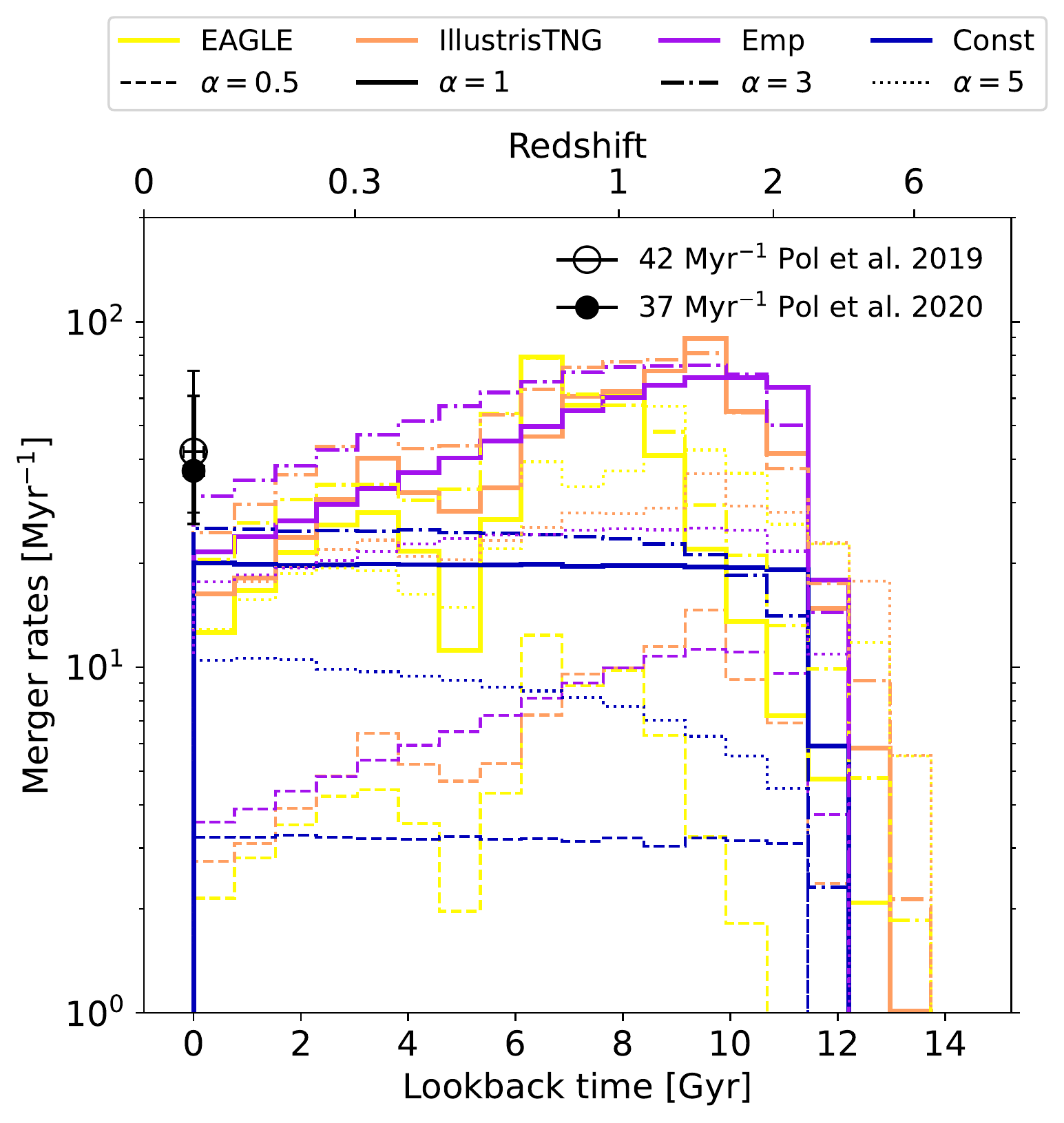}
    \caption{BNS merger rate in the MW as a function of the lookback time for different  Galaxy models and  values of the CE parameter $\alpha$. Each colour identifies a Galaxy model. Yellow: \textsc{eagle}; pink: \illustris{};
    purple: Emp; blue: Const. Each line-style is associated with a value of $\alpha$. Dashed line: $\alpha=0.5$;  solid line: $\alpha=1$; dash-dotted line: $\alpha=3$; dotted line: $\alpha=5$. 
    The circles show the BNS merger rate in the MW inferred from observations: the unfilled circle shows the value by \protect\cite{pol2019}, $\mathcal{R}_{\rm MW}=42^{+30}_{-14}\,{}\text{Myr}^{-1}$; the black filled circle shows the updated value by  \protect\cite{pol2020}, $\mathcal{R}_{\rm MW}=37^{+24}_{-11}\,{}\text{Myr}^{-1}$. All simulations shown in this Figure  assume $\tau_{\rm d}=1~\text{Gyr}$ and the U distribution for  initial spins and magnetic fields. We do not expect NS spins and magnetic fields to affect the merger rate.}
    \label{fig:mergerrates}
\end{figure}

\begin{table}
    \centering
    \begin{tabular}{c cccc}
        \multicolumn{5}{c}{ \textsc{Merger Rates} ($\text{Myr}^{-1}$)} \\ \hline
        &  \multicolumn{4}{c}{$\alpha$} \\ \hline
        & 0.5   & 1   & 3 & 5 \\ 
        MW model & &  & &  \\ \hline
        \textsc{eagle} & 2.2 & 12.6 & 20.5 & 12.9 \\
        \illustris{} & 2.7 & 16.3 & 24.5 & 16.2 \\
        Emp & 3.6 & 21.5 & 31.3 & 17.7 \\
        Const & 3.2 & 20.0 & 25.2 & 10.5 \\ \hline
\end{tabular}
\caption{BNS merger rate in the MW at present time for different  Galaxy models and  values of the CE parameter $\alpha$. All simulations shown in this Figure  assume $\tau_{\rm d}=1~\text{Gyr}$ and the U distribution for  initial spins and magnetic fields.}
\label{tab:mergerrates}
\end{table}

Figure \ref{fig:mergerrates} shows the BNS merger rate history for the Galaxy models and for the values of the CE parameter $\alpha$ adopted in this work. Table \ref{tab:mergerrates} summarizes the BNS merger rates at present time predicted by our models.
For comparison, we also show the BNS merger rate of the MW  inferred by \citet[][$\mathcal{R}_{\rm MW}=42^{+30}_{-14}\text{Myr}^{-1}$]{pol2019} and \citet[][$\mathcal{R}_{\rm MW}=37^{+24}_{-11}\text{Myr}^{-1}$]{pol2020}.

The parameter $\alpha$ has a large impact on the BNS merger rate. This result agrees with previous works, showing that  almost all BNS mergers form via CE \citep{tauris2017,giacobbo2018,kruckow2018,vignagomez2018,mapelli2018,mapelli2019,mandel2022,iorio2022}. 
In general, a larger value of $\alpha$ means that energy is transferred more efficiently to the envelope, facilitating its expulsion. 
In our models, the highest BNS merger rates are produced for $\alpha=3$. Different values of $\alpha$, both higher and lower, 
yield values of $\mathcal{R}_{\rm MW}$ which are lower than the one  inferred from observations by more than one standard deviation.

We find that the merger rate in the local Universe strongly correlates with the SFR of the Galaxy model \citep{artale2019}.
As such, the Galaxy models that better reproduce the observed $\mathcal{R}_{\rm MW}$ are also the ones with the current SFR closer to the one of the MW. 
In particular, the Emp model with $\alpha=3$ best reproduces the observed BNS merger rates (see Table \ref{tab:mergerrates}). 
For this reason, we choose the Emp model as our fiducial model. In contrast, the \textsc{eagle} model underestimates the local BNS merger rate of the MW. 
In general, the merger rate history follows the same evolution with redshift as the SFR of the host galaxy, since most BNSs merge shortly after their formation.  
Our findings are in agreement with \cite{artale2019}, who observe a tight correlation between the BNS merger rate and both the mass and  the SFR of the host galaxy \citep[see][for more details on this correlation]{artale2019,artale2020a,artale2020b,chattopadhyay2021}.

\subsection{Orbital period -- eccentricity}

\begin{figure*}
	\centering
	\includegraphics[width=0.95\textwidth]{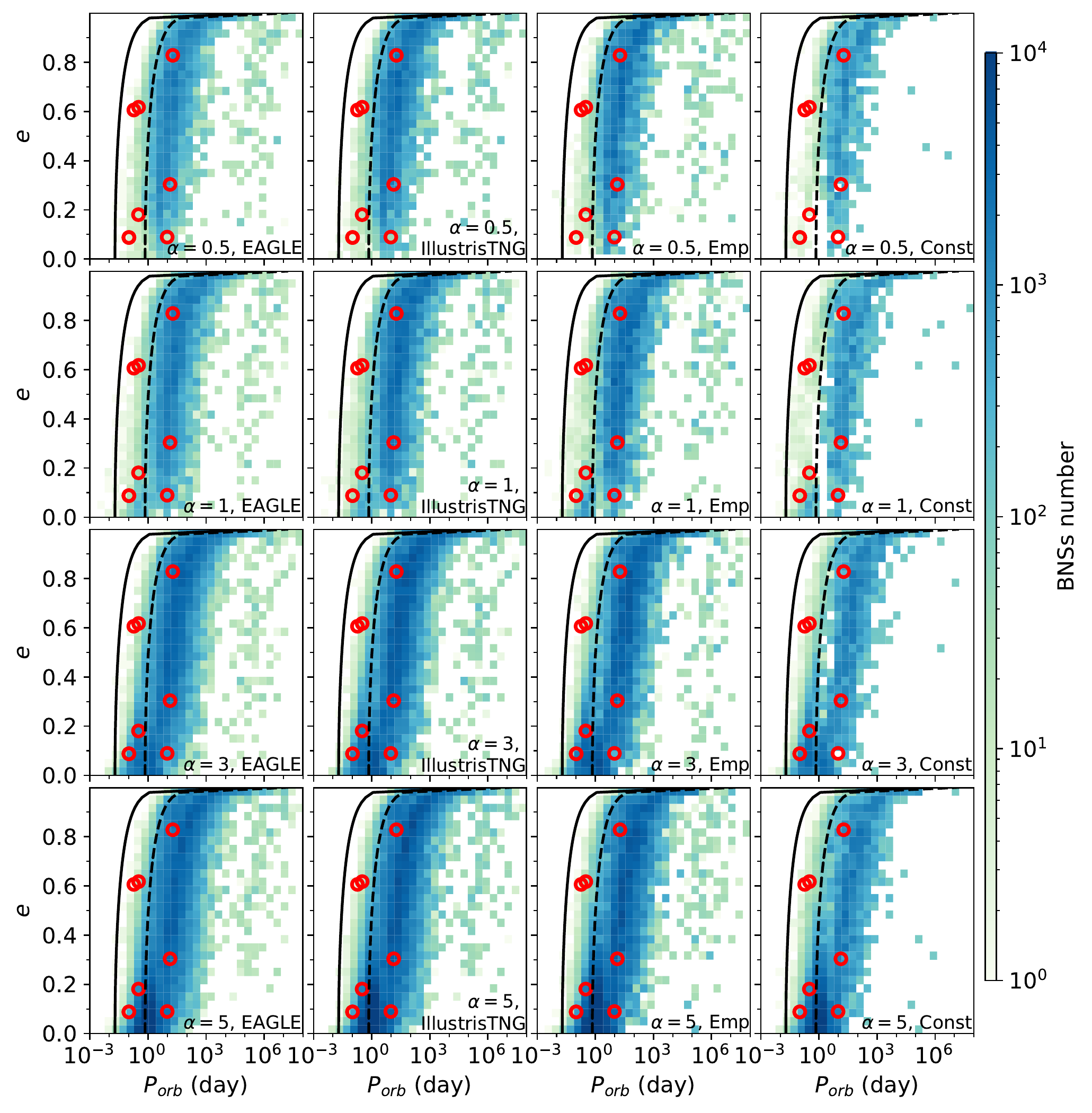}
	\caption[Just for LOF]{Distribution of the simulated BNSs  at the present time in the $P_{\rm orb}-e$ plane. Each column shows the results for a different  Galaxy model, from left to right: \textsc{eagle}, \illustris{}, Emp and Const. The rows show different $\alpha$ values, from top to bottom: $\alpha=0.5$, $1$, $3$, and $5$. All the models shown here assume $\tau_{\rm d}=1$~Gyr, and model U for the initial spin and magnetic field. The red circles show the observed population of BNSs selected for this study. The solid black line shows values of constant $t_{\rm GW}=1$~Myr, while the dashed black line corresponds to $t_{\rm GW}$ equal to the Hubble time. We obtained the lines of constant $t_{\rm GW}$  with equation D6 in  \cite{iorio2022} \protect\footnotemark[9], assuming a mass of $1.35\,{}\text{M}_\odot$ for the NSs.}
	\label{fig:porbecc}
\end{figure*}


Figure \ref{fig:porbecc} shows the distributions of our simulated BNSs at lookback time $t_{\rm lb} = 0$ (today) in the orbital period (P$_{\rm orb}$)-- eccentricity ($e$) plane. We show the results varying $\alpha$ and the host Galaxy model.
These plots do not include systems that have merged throughout the history of the MW, but only those that survived until today. All the simulations in Figure~\ref{fig:porbecc} have $\tau_{\rm d}=1$~Gyr and an uniform (U) distribution for the initial spins and magnetic fields. Indeed, these parameters do not produce significant variations in the final distribution of the P$_{\rm orb}-e$ plane, as expected. The markers show the observed Galactic pulsars in BNSs systems  (Table \ref{tab:pulsars}). 

Different models in Fig.~\ref{fig:porbecc} share similar trends. It is possible to distinguish a main branch characterized by short orbital periods ($\sim{1-10^{3}}$~d) and a second one, much less populated, with orbital periods at $\sim 10^{5-6}$~d. At shorter periods, the distributions are bound by the  GW merger timescale ($t_{\rm merge}\lesssim 1~\text{Myr} $).  

The parameter $\alpha$ has a strong impact on the distribution of BNSs in the P$_{\rm orb}-e$ plane. 
In particular,  the simulations with $\alpha=5$ predict about twice as many BNSs as the $\alpha=3$ model, and roughly five times more binaries compared to the $\alpha=0.5$ model.  
This is expected as higher $\alpha$ values are associated with a more efficient expulsion of the CE, therefore more systems are able to survive the CE phase. In contrast, as we decrease $\alpha$, the probability that a system prematurely merges during the CE increases \citep{iorio2022}. 
These plots display another interesting feature: low-$\alpha$ simulations ($\alpha=0.5$, $1$)  lack  BNSs with low eccentricity, which are instead present in the $\alpha=3$ and $5$ models, favouring more eccentric binaries. 

The host Galaxy model also plays a role on the final BNS distribution. In particular, we found that the most important parameter  
in this case is the total stellar mass $M_\star$ of the mock Galaxy. The procedure we adopt to populate a galaxy implies a direct correlation between $M_\star$ and the number of BNSs (see Eq.~\ref{eq:nbns}). For this reason the Const MW model produces a much lower number of final BNS systems.

Our results qualitatively agree with the observed BNS distribution of orbital periods and eccentricities. A more quantitative comparison is not  straightforward owing to  observational biases. In fact, eccentric binaries are 
more difficult to detect \citep{tauris2017}, having stronger Doppler effects \citep{chattopadhyay2021}.

\subsection{$P-\dot{P}$ plane}

\begin{figure*}
	\centering
	\includegraphics[width=1.0\textwidth]{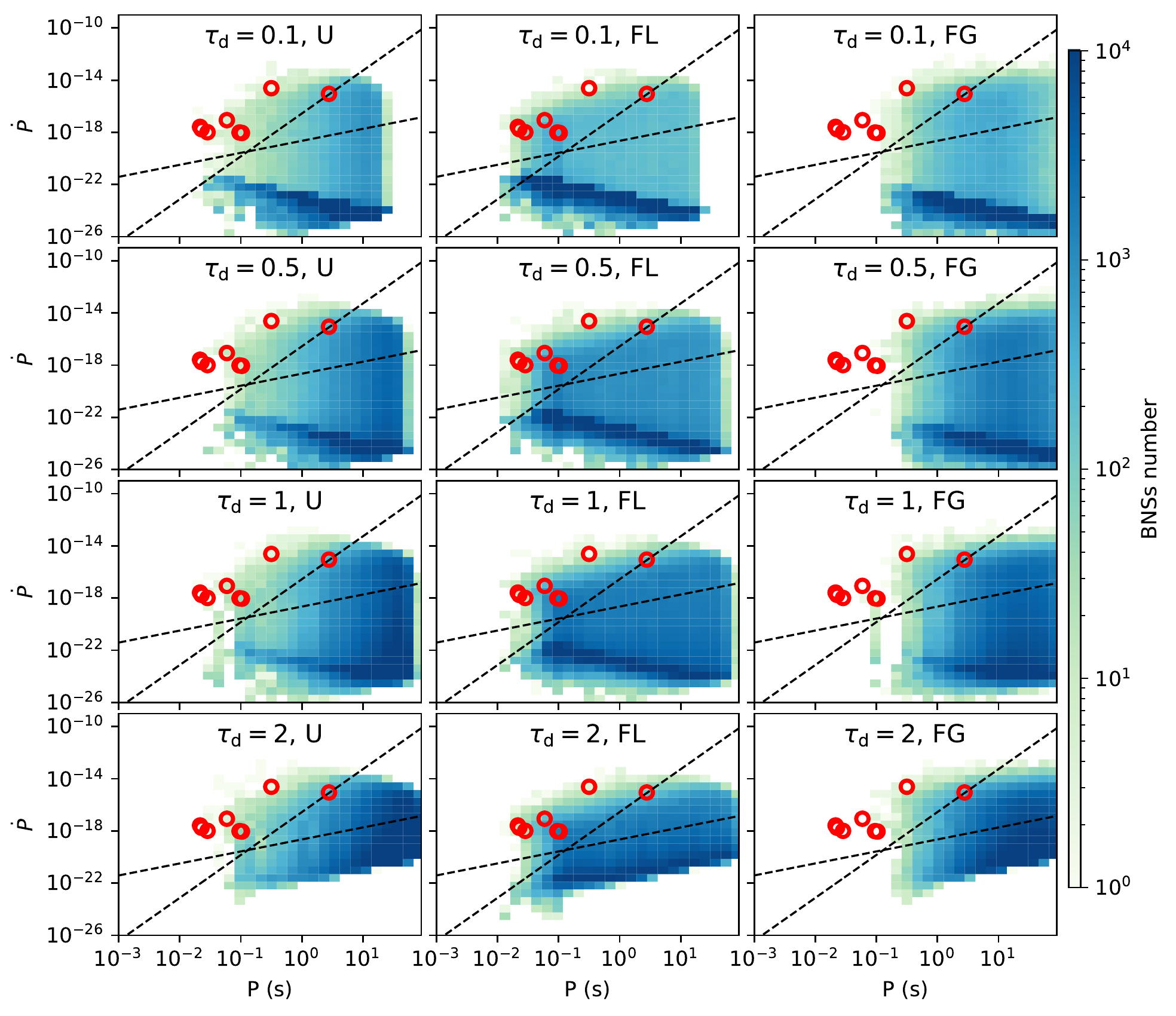}
	\caption{Distribution of the simulated BNSs at the present time in the $P-\dot{P}$ plane. Each column shows the results for a different model of the initial spin and magnetic field, from left to right: uniform (U), flat-in-log (FL), Faucher-Giguère (FG). The rows  assume different values of $\tau_{\rm d}$, from left to right: $\tau_{\rm d}=0.1$, $0.5$, $1$, $2$~Gyr. All the runs assume $\alpha=3$ and the Emp MW model. The markers show the observed population of pulsars (Table~\ref{tab:pulsars}). The dashed black lines show the  death lines defined in Eq.~\ref{eq:deathlines}.}
	\label{fig:ppdot}
\end{figure*}

Figure \ref{fig:ppdot} shows the distribution of our simulated BNSs at the present time  in the $P-\dot{P}$ plane. We show the results varying $\tau_{\rm d}$ and the model for the initial spins and magnetic fields. Figure \ref{fig:ppdot} clearly shows the importance of the initial spin and magnetic field distribution on the final pulsar population. 
Although the number of pulsars in each model varies with  $\alpha$ and with the MW model, the shape of the distribution in the $P-\dot{P}$ plane is not significantly affected. 
Most of the detected BNSs in the MW data are characterized by a spin period $\lesssim 100\,{} \text{ms}$. Most of the pulsars in the data-set are probably recycled \citep{tauris2017}. 
The primary-born pulsar population in our models do not show striking differences compared to the second-born pulsars.

\subsection{Selection effects} \label{sec:RadioSel}

\begin{figure*}
	\centering
	\includegraphics[width=1.0\textwidth]{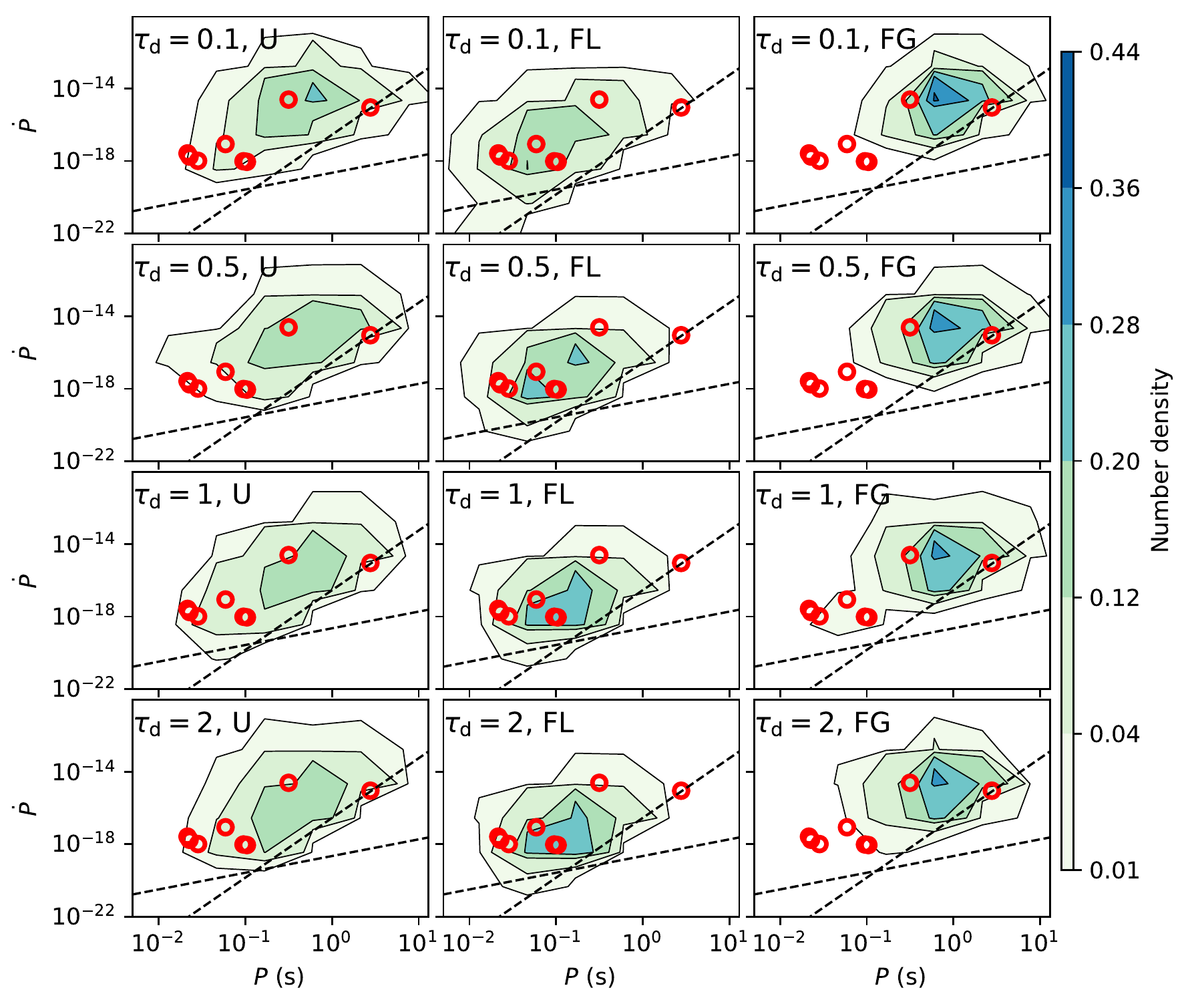}
	\caption{Isocontours showing the number density of detectable pulsars (after applying radio selection effects) in the $P-\dot{P}$ plane. Each column shows the results for a different spin and magnetic field model, from left to right: U, FL, and FG. The rows  assume different values of $\tau_{\rm d}$, from left to right: $\tau_{\rm d}=0.1$, $0.5$, $1$, and $2$ Gyr. All the runs assume $\alpha=3$ and the Emp  model for the MW. The markers show the observed population of pulsars (Table~\ref{tab:pulsars}). The dashed black lines show the  death lines defined in Eq.~\ref{eq:deathlines}. } 
	\label{fig:ppdotiso}
\end{figure*}

\begin{figure}
    \centering
    \includegraphics[width=.5\textwidth]{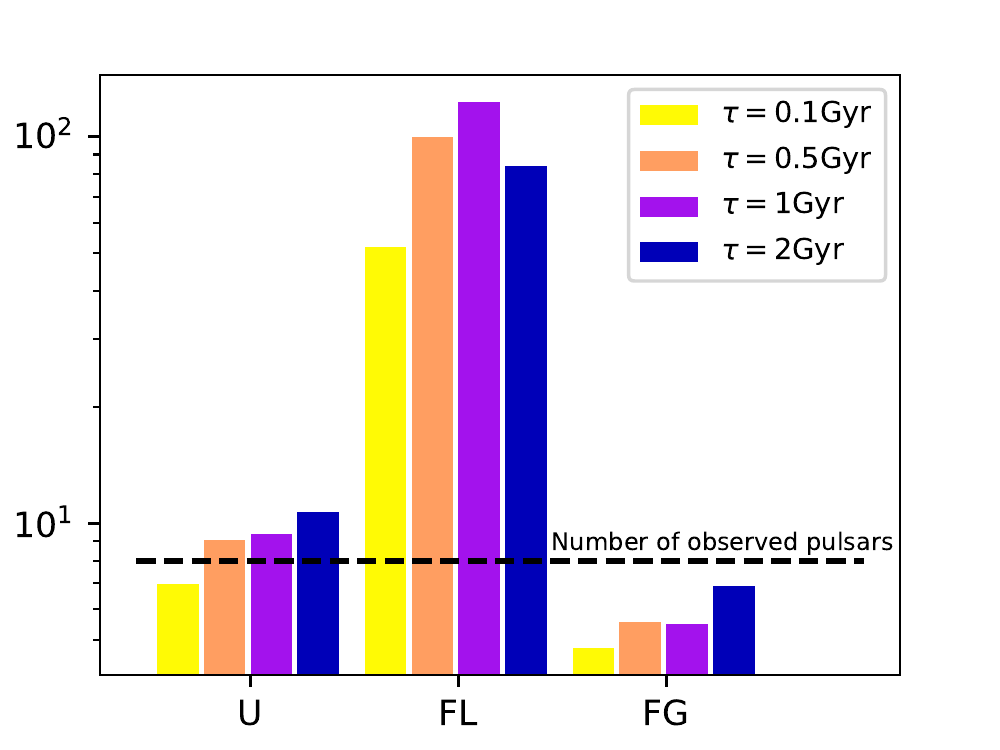}
    \caption{Mean number of radio selected pulsars predicted by our models after averaging over the $N=100$ realizations of the selection effects, assuming the fiducial $\alpha=3$ and the Emp  MW  model.  Each bar is a different model: the $x$-axis  shows the spin and magnetic field initial distributions (U, FL, and FG). Each colour is associated to a different $\tau_{\rm d}$ value: yellow, pink, purple, and blue for $\tau_{\rm d} =0.1$, 0.5, 1, and 2 Gyr, respectively. The black horizontal dashed line shows the number of observed Galactic BNSs in the considered surveys, $N_{\rm obs}=8$. }
    \label{fig:numbers}
\end{figure}

Figure~\ref{fig:ppdot} shows the intrinsic astrophysical population, without accounting for  radio selection effects. In contrast, Figure~\ref{fig:ppdotiso} shows 
the distribution in the $P-\dot{P}$ plane of the detectable pulsars predicted by our models, obtained as  described in Section~\ref{sec:seleffects} and averaging over  $N=100$ realizations of radio-selected pulsar populations. 
Figure \ref{fig:ppdotiso} shows the iso-density contours for models varying $\tau_{\rm d}$ and the initial distributions of spins and magnetic fields (U, FL, and FG). 
The shapes of the contours   depend on the chosen initial spins and magnetic fields: the pulsar density peak shifts in the parameter space because of these two parameters. The FG distribution generally  produces slower spinning pulsars, with larger values of the spin periods $P$. In contrast, the FL distribution models peak at $P \sim 0.1 \text{s}$, where the majority of the observed pulsars lies. The U distribution,  our fiducial model,  shows a broader profile spanning a wider range of spin periods with respect to the other  models.

We also calculate the number of pulsars predicted by our models.
Our approach is  completely self-consistent: we consider the specifics of the chosen surveys to account for radio selection effects, and compare our samples with the pulsars observed by the same surveys. 
Figure~\ref{fig:numbers} shows the mean number of predicted detections  averaged over the $N$ realizations of the radio selection effects. The FL model seems to better fit the observed pulsar  distribution in the $P-\dot{P}$ plane: the predicted pulsar sample peaks at $\sim 0.1$~s, where the majority of pulsars lie. 
However, Figure \ref{fig:numbers} shows that the FL model predicts too many observed BNSs. In fact, it estimates $\sim 10^2$ observable pulsar binary systems, almost $10$ times more than the number of detected Galactic BNSs. On the other hand, the FG model strongly underestimates the number of observed pulsars. 
The U model, instead, not only populates the $P-\dot{P}$ region where the observed pulsars lie, but also  predicts the correct number of BNSs. For this reason, we choose  the U model as the fiducial one. The reason why the FL model predicts a much higher number of detectable BNSs with respect to both FG and U is that it initializes the bulk of the pulsar population with lower values of both  magnetic field and  initial spin. Hence, more pulsars are still above the death line at current time in model FL compared to U and FG \citep{chattopadhyay2020}.

From Figures \ref{fig:ppdotiso} and \ref{fig:numbers},  we can see another feature: the samples of detectable BNSs decrease for lower values of $\tau_{\rm d}$.
This happens because younger pulsars constitute the great majority of the detectable pulsars. Since for lower values of $\tau_{\rm d}$ the pulsars move faster in the $P-\dot{P}$ plane, towards larger spin periods and lower magnetic fields, pulsars cross the death line in shorter intervals of time.  

\cite{chattopadhyay2020} produce, in general, more radio selected pulsars with respect to our models, from 5 to 20 times more pulsars than our predictions (excluding their model CE--Z, which assumes no accretion during CE, as in our work, see Section \ref{sec:discussion}).
Nonetheless, they observe the same trend with $\tau_{\rm d}$. 

\begin{figure}
	\centering
	\includegraphics[width=0.5\textwidth]{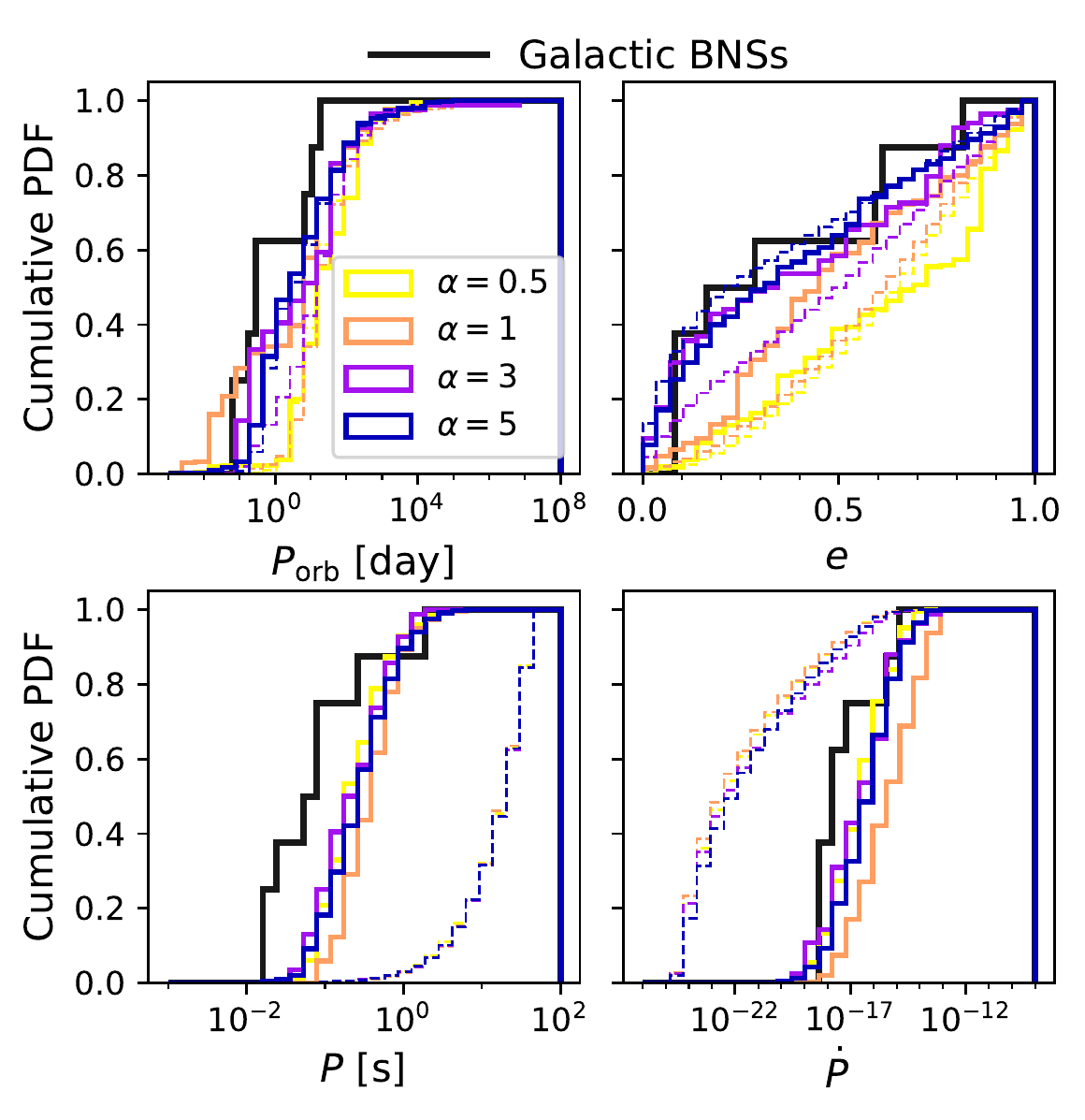}
	\caption{Cumulative distribution of the pulsar parameters $P_{\rm orb}$, $e$, $P$ and $\dot{P}$ (from  top left to  bottom right) for the fiducial model parameters. Yellow, pink, purple, and blue solid lines: $\alpha =0.5$, 1, 3, 5. The dashed thin lines display the underlying astrophysical BNS population, while the solid thick lines show the detectable population, after implementing radio-selection effects. The black solid line marks the cumulative distribution of the observed Galactic BNSs.}  
	\label{fig:cumdistr}
\end{figure}

Figure \ref{fig:cumdistr} shows the cumulative distributions of the orbital period, eccentricity, spin period, and spin period derivative, for our fiducial model. We show the distributions we obtained with and without accounting for the selection effects, for comparison. Radio selection effects select only the non-dead pulsars, characterized in general by smaller spin periods $P$ and larger $\dot{P}$. Furthermore, binary selection effects tend to select circular  with respect to eccentric systems.

In general, our models match the observed distributions, including the eccentricity distribution, when we account for  selection effects. 

Figure \ref{fig:cumdistr} also compares different CE parameters $\alpha=0.5-5$. The choice of $\alpha$   influences the orbital period and especially the eccentricity. The models with $\alpha >1$ better match the data with respect to those with $\alpha\leq{}1$, because the latter produce too many highly eccentric systems. 
This preference for $\alpha{}>1$ when considering the eccentricity distribution points in the same direction as the result of the  BNS merger rate: we found the best match with the MW merger rate for $\alpha=3$.

The distributions of  $P$ and $\dot{P}$   do not change much with $\alpha$, as expected. The Galactic BNSs seem to have a double peaked distribution: the first peak around $P\sim1$~d, the second one at about $10^2$~d \citep{andrews2019}. 
However, given the small sample of detected BNSs we do not have enough statistics to claim that such feature is characteristic of the underlying BNS distribution.

\footnotetext[9]{The analytical methods to compute the GW merging time are implemented in the function \texttt{estimate\textunderscore gw}, in the publicly available \textsc{Python} module \textsc{pyblack}, \url{https://gitlab.com/iogiul/pyblack}.}


\begin{figure}
    \centering
    \begin{subfigure}{0.5\textwidth} 
    \centering
	\includegraphics[width=\textwidth]{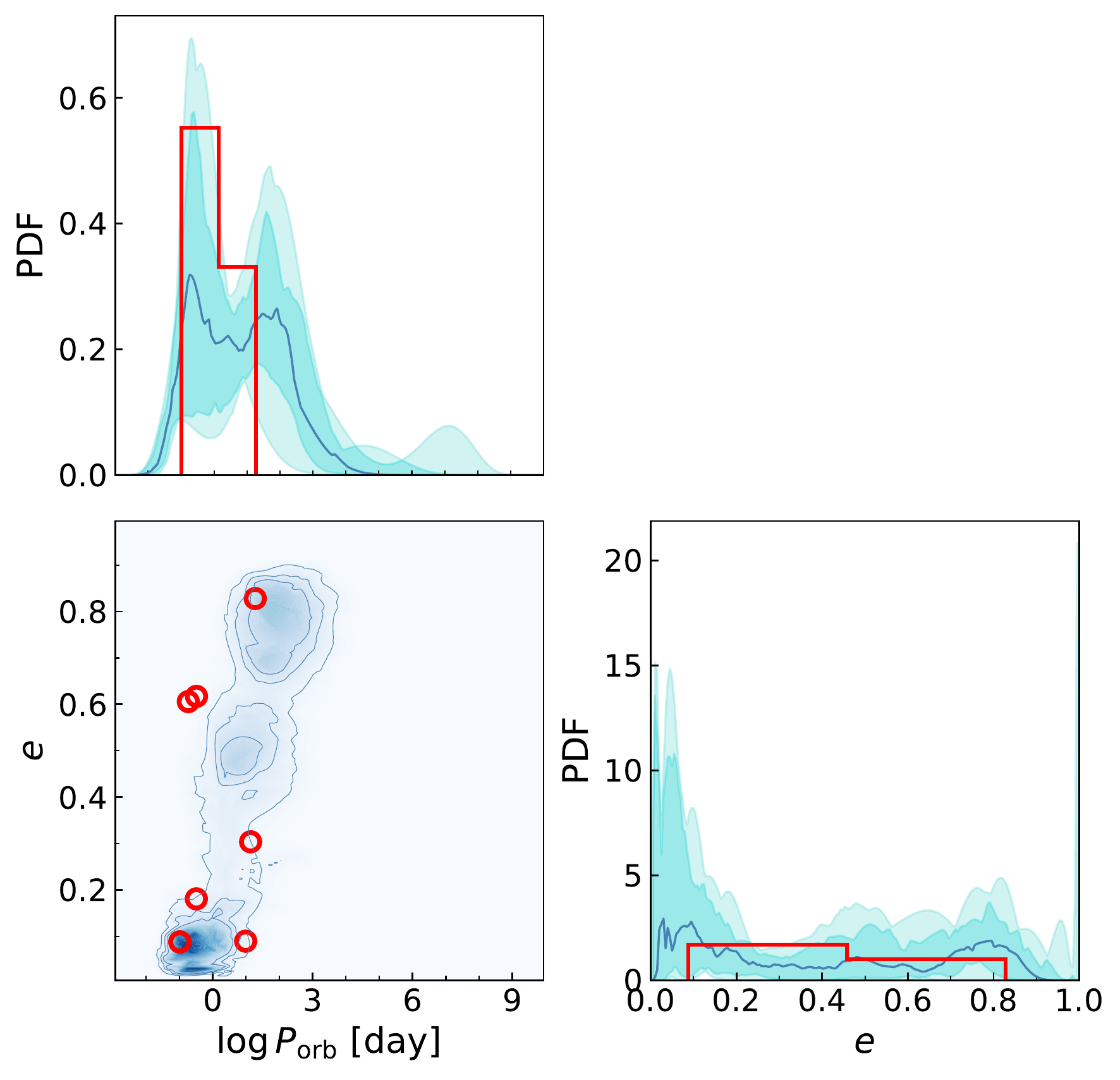}
    \caption{ }\label{fig:GMMporb}
    \end{subfigure}
    
    \begin{subfigure}{0.5\textwidth} 
    \centering
	\includegraphics[width=\textwidth]{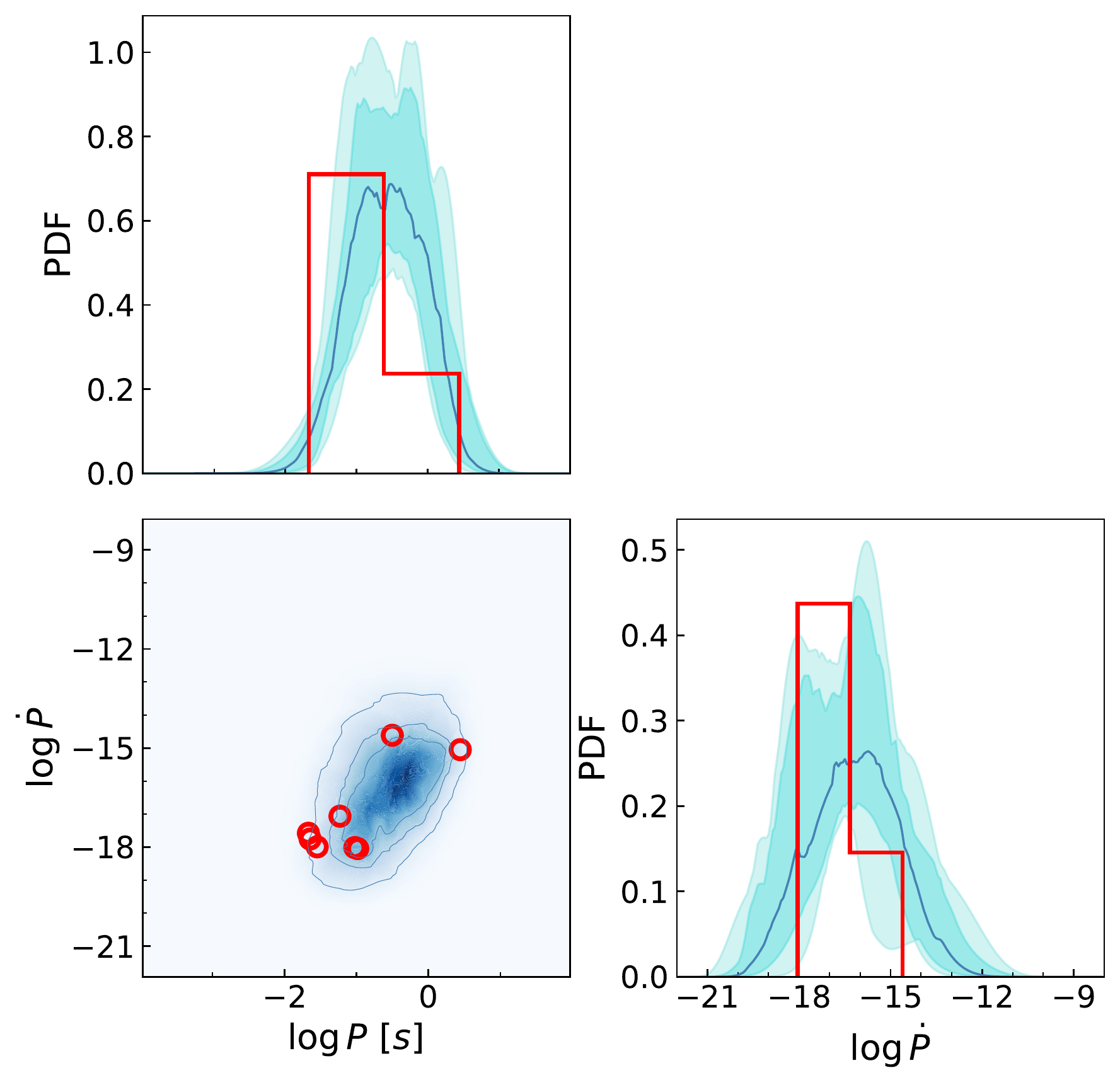}
    \caption{ }\label{fig:GMMppdot}
    \end{subfigure}
    \caption{ Two-dimensional likelihood for $P_{\rm orb}$ and $e$ (panel a) and $P$ and $\dot{P}$ (panel b) averaged over the simulated pulsars catalogues of our fiducial model, as reconstructed by \textsc{figaro}. These likelihoods are obtained by marginalising out $P$ and $\dot P$ or $P_{\rm orb}$ and $e$ from the four-dimensional likelihood, respectively.
    The red circles and histograms show the observed Galactic BNSs (Table~\ref{tab:pulsars}). 
    The blue contour lines in the 2D plot show the 90\%, 68\%, and 50\% credible regions of the median distribution (blue lines in 1D plots). The shaded regions in 1D plots represent the 68\% and 90\% credible intervals for the marginal probability density for each parameter.
    }
\end{figure}

\subsection{Bayes factors}\label{sec:bayes}
 
Table~\ref{tab:bayesfactors} shows the Bayes factors of each model, that we obtained as described in Section~\ref{sec:statisticalanalysis}. For this analysis, we used four BNS parameters: orbital period, eccentricity, spin period and derivative of the spin period. Figures \ref{fig:GMMporb} and \ref{fig:GMMppdot} show two examples of the DPGMMs we applied to these four parameters in our simulations.

The Bayes factors indicate that the FG  distribution is strongly disfavoured in all the models, confirming what we can qualitatively see from Figures \ref{fig:ppdotiso} and \ref{fig:numbers}. 
Also, the U model performs better than the FL model, because the latter produces too many detectable BNSs (Fig.~\ref{fig:numbers}). 

Considering the fiducial model, we find that $\tau_{\rm d}=1$~Gyr produces slightly better results. However, the differences yielded by distinct $\tau_{\rm d}$ values are not as appreciable as those produced by different initial spin and magnetic field distributions. Thus, we cannot confidently discard a particular value of $\tau_{\rm d}$ based only on this result.

\begin{table*}
\centering
\begin{tabular}{c cccc cccc cccc}
\hline
&  \multicolumn{4}{c}{Uniform (U)} & \multicolumn{4}{c}{Flat-in-log (FL)} & \multicolumn{4}{c}{Faucher-Giguere (FG)} \\
& \multicolumn{4}{c}{$\tau_{\rm d} \,{}(\text{Gyr})$} & \multicolumn{4}{c}{$\tau_{\rm d}\,{} (\text{Gyr})$} & \multicolumn{4}{c}{$\tau_{\rm d}\,{} (\text{Gyr})$} \\ \hline

 & 0.1    & 0.5   & 1 & 2   & 0.1    & 0.5   & 1 & 2  & 0.1  &  0.5   & 1 & 2 \\
$\alpha$   &  &  &  &    &  &   &  &  &  &  &  &  \\ \hline

0.5 & -136 (158) & -76 (99) & -129(165) & -57(65)   &   -9 (31) & -25 (26) & -36 (26) & -42 (26)  &  -148 (128) & -173 (129) & -199 (125) & -45 (28) \\

1 & -103 (148) & -69 (85) & -71 (100) & -53(88)   &   -10 (28) & -23 (26) & -37 (27) & -47 (28)  &  -119 (128) & -96 (129) & -97 (125)& -50 (26) \\

3 & -10 (82) & 0.5 (27) & \textbf{0 (29)} & -0.1 (27)   &   -20 (23) & -65 (25) & -86 (25) & -53 (24)  &  -42 (62) & -27 (39) & -25 (45) & -21 (41)\\

5 & -0.5 (27) & -55 (82) & 4 (25) & -25 (27)  &   -27 (23) & -85 (24) & -121 (25) & -172 (26)  &  -37 (87) & -36 (80) & -20 (43)& -32 (46)  \\ 
\hline
\end{tabular}
\caption{Logarithmic Bayes factors ($\ln{B}$) of the simulated models compared to the fiducial model 'Ua3t1Emp', adopting the Emp MW model. The fiducial model is highlighted in bold face in the Table.}
\label{tab:bayesfactors}
\end{table*}

\subsection{Predictions for the SKA}

Within our framework, we can make predictions for the Square Kilometre Array (SKA). We applied the radio selection effects setting the parameters of the major SKA surveys (MID and LOW, see Appendix~\ref{sec:surveyparams}), and studied how the predictions vary for our different models.
Table~\ref{tab:SKApredictions} shows the resulting number of SKA-detectable BNSs  for different models. 
We see a  spread of about two orders of magnitude among the results. Consistently with Figure~\ref{fig:numbers},  the FL  initial distribution predicts $\sim 10$ times more observable pulsars than the other models. Our fiducial model estimates that SKA will be able to observe $30 \pm 6$ BNSs, among which $21 \pm 4$ new detections .

\begin{table*}
\centering
\begin{tabular}{c cccc cccc cccc}
\hline
&  \multicolumn{4}{c}{Uniform (U)} & \multicolumn{4}{c}{Flat-in-log (FL)} & \multicolumn{4}{c}{Faucher-Giguere (FG)} \\
&  \multicolumn{4}{c}{$\tau_{\rm d} \,{}(\text{Gyr})$} & \multicolumn{4}{c}{$\tau_{\rm d} \,{}(\text{Gyr})$} & \multicolumn{4}{c}{$\tau_{\rm d}\,{} (\text{Gyr})$} \\ \hline

 & 0.1    & 0.5   & 1 & 2   & 0.1    & 0.5   & 1 & 2  & 0.1  &  0.5   & 1 & 2 \\
$\alpha$   &  &  &  &    &  &   &  &  &  &  &  &  \\ \hline


0.5 & 6 (2) & 10 (3) & 9 (3) & 11 (3)   &   54 (8) & 121 (10) & 173 (12) & 211 (14) &  3 (1) & 5 (2) & 4 (2) & 210 (14) \\

1 & 9 (3) & 13 (3) & 10 (3) & 14 (3)   &   84 (8) & 145 (13) & 196 (12) & 241 (13)  &  6 (2) & 6 (2) & 7 (3)& 240 (14) \\

3 & 23 (5) & 29 (5) &  \textbf{30(6)} & 36 (5)   &   155 (12) & 316 (19) & 411 (18) & 288 (17)  &  17 (4) & 21 (4) & 20 (4) & 22 (5) \\

5 & 26 (5) & 16 (4) & 39 (6) & 21 (5)  &   178 (14) & 395 (20) & 532 (24) & 721 (29)  &  17 (4) & 19 (4) & 23 (5)& 21 (5)  \\ 

\hline
\end{tabular}
\caption{Predicted number of pulsars in BNSs detectable by the SKA, after averaging over $N=100$ realisations of the radio selection effects. We report  the standard deviation on the mean within brackets. The fiducial model is highlighted in bold face in the Table.}
\label{tab:SKApredictions}
\end{table*}


\section{Discussion} \label{sec:discussion}

We have explored the BNS population properties within four different Galaxy models: a model with constant (Const) SFR, an empirical (Emp) MW model with exponentially decaying SFR, and two models taken from the {\sc eagle} and \illustris{} cosmological simulations, respectively (Table~\ref{tab:mwmodels}). 
In all the four models, the Galactic SFR at present day matches the one of the MW. In addition, the Emp, {\sc eagle}, and \illustris{} models  also match the current total stellar mass of the MW, whereas the Const model results in a factor of ${\sim 2.5}$ lower mass Galaxy. Furthermore, the SFR history of the {\sc eagle} and \illustris{} galaxies deviate  from the simple exponential decay assumed in the Emp  model. 
The adopted models differ from each other in terms of metallicity as well: adopting the FMR, the Emp model is dominated by Solar metallicity stars, whereas the \textsc{eagle} and \illustris{} models contain a large population of metal-poor stars.

Despite these differences, the $P_{\rm orb}-e$ distribution of BNSs in the Emp, {\sc eagle}, and \illustris{} models show similar features. In contrast,  the Const model  
predicts a much lower number of BNSs than the other models. This is a consequence of the lower total stellar mass of the Const model.

The Galactic BNS merger rate is primarily affected by our choice of the common-envelope parameter $\alpha$. If we assume $\alpha=3$, the resulting present-day Galactic merger rate is a factor of ${\sim 10}$ higher than for $\alpha=0.5$, and is more consistent with the value inferred from observations \citep{pol2019,pol2020}. The merger rate is 
also affected by the SFR and metallicity of the  adopted Galaxy model 
(Fig.~\ref{fig:mergerrates}).

Our models still represent a rather simplified description of the MW. \cite{matteucci2019} and  \cite{kobayashi2022}  explore detailed models of the MW, accounting also for its chemical enrichment history. They account for BNS mergers, and study if the latter can explain the r-process element abundance. \cite{kobayashi2022} explain the relative abundances of elements through BNS mergers only with a dependence of the delay time distribution on metallicity,  
with lower metallicities yielding shorter delay times. 
Their  comprehensive treatment goes beyond the purpose of this work. 

Our models 
reproduce the orbital period and eccentricity distribution of Galactic BNSs, 
and match the observed BNS merger rate, in a self-consistent fashion. 
In particular, our fiducial model  
reproduces the double peak in the eccentricity distribution \citep{andrews2019}: observed pulsars cluster around around $e\sim0.1$ and $e\sim 0.6$, showing instead a gap for $0.3 \lesssim e \lesssim 0.6$. In Figure \ref{fig:porbecc} ($\alpha=3$), we can distinguish two more concentrated regions at  eccentricity $e\sim 0.6$ and $0.1$. This feature becomes even more evident after the application of radio selection effects (Figure \ref{fig:GMMporb}).  This suggests the importance of observational biases not  only for the spins and magnetic fields of pulsars, but also for the orbital properties. Moreover, the distribution of pulsars in the $P_{\rm orb}-e$ plane is almost independent of  the chosen MW model, enforcing the robustness of this result.

We can compare our 
cumulative distributions (solid lines in Fig. \ref{fig:cumdistr}) with the green dashed lines of Figure 6 by \cite{chattopadhyay2020}. Our cumulative distribution of eccentricity is very different from  the one reported by \cite{chattopadhyay2020} because they do not account for binary selection effects. 
Our Figure \ref{fig:porbecc} can be compared with those of \cite{vignagomez2018} and \cite{kruckow2018}. The distribution of BNSs in Figure~11 of \cite{kruckow2018} is less populated at high eccentricity with respect to our results. 
This difference is a consequence of the different kicks and binding energy prescriptions, as discussed in \cite{iorio2022}.
Compared to \cite{kruckow2018} and our results, \cite{vignagomez2018} produce much tighter binaries (coloured dots of their Figure~2). In fact, most of the BNSs show orbital periods shorter than $\sim 10$ d. Moreover, the wider systems (P$_{\rm orb} \gtrsim 10^3$~d) are completely missing. This difference is a consequence of the different natal kick models adopted in \cite{vignagomez2018}. 

In our models, we did not include any spin-up prescription during the CE phase. In contrast, \cite{chattopadhyay2020} assume  that a NS accretes between $0.04$ and  $0.1$~M$_{\odot}$ during CE, according to the rates presented in \cite{macleod2015}. As a result, pulsars spin up much more efficiently during a CE event than during RLO, and their distribution peaks toward smaller spin periods. 
For this reason,  \cite{chattopadhyay2020} predict many more detectable pulsars compared to our models when they enable accretion during CE. Nevertheless, the role played by the CE phase during spin-up is controversial \citep{chamandy2018}. \cite{oslowski2011} assume that the accreted matter during CE contributes only to the magnetic field decay, considering accretion during CE a chaotic process, not able to produce spin-up \citep{benensohn1997}. Our spin-up treatment does not include the effect of winds as well. Indeed, we might expect the infalling matter produced by winds to be chaotic and thus not efficiently cause the spin-up of the pulsar \citep{kiel2008}. We can compare Figure~\ref{fig:numbers} with Table~4 from \cite{chattopadhyay2020}. The number of detectable pulsars predicted by their models is generally one order of magnitude higher than our predictions. Our predicted numbers of observable pulsars are consistent only with their model CE--Z, which does not include spin-up during CE, and predicts $13$ observations.

There are large uncertainties about the surface magnetic field: the burial of magnetic field during mass accretion is poorly understood as well as its decay during the isolated pulsar evolution. The $\tau_{\rm d}$ parameter has been varied in the literature from a few Myr \citep[e.g.][]{oslowski2011} up to a few Gyr \citep[e.g.][]{kiel2008}. 
Another important contribution to the overall uncertainty in the pulsar population is given by the initial distribution of spins and magnetic fields. These are even more important than the magnetic-field decay time $\tau_{\rm d}$.  Several distributions have been proposed for the initial spin and magnetic field  \citep[e.g.][]{igoshev2022, fauchergiguere2006}. However, the lack of statistics does not allow  us to draw strong conclusions.

We evaluated  a four-dimensional likelihood  in the $(P_{\rm orb}, e, P, \dot{P})$ space.  
Our approach consistently takes into account  correlations among parameters, which are instead neglected by other tests (e.g., the Kolmogorov-Smirnov test), frequently used in the pulsar literature. The large errors on the  Bayes factor errors reported in Table \ref{tab:bayesfactors}  are a consequence of the limited number of observed pulsars, making it difficult to differentiate among the models. However, despite this limitation, the Bayes factor indicate that the FG and FL models are  disfavoured with respect to the U  prescription.

We can compare our SKA predictions (Table \ref{tab:SKApredictions}) to Table 4 of  \cite{chattopadhyay2021}. Their fiducial model estimates $78$ BNSs containing a radio-detectable pulsar by the SKA telescope. Their result is thus roughly twice our projection. Nevertheless this discrepancy is consistent with the different approach we used. As already stressed, the different CE formalism, accentuates the spin up  of their pulsars, resulting in shorter spin periods. In turn, this translates into a higher number of detectable pulsars. 

\section{Summary} \label{sec:summary}

The BNS population of the MW is the perfect laboratory to test binary-star evolution models. Here, we used the new population synthesis code \sevn{} \citep{iorio2022} to model 
the population of BNSs in the MW. 
We implemented a new model for both the spin down and spin up of pulsars in \sevn{}, while also probing the relevant parameter space (e.g., CE parameter $\alpha{}$, initial spin distribution, initial magnetic field distribution, decay time of the magnetic field $\tau_{\rm d}$). We injected our simulated binaries into four MW models: an empirical model with an exponentially decaying SFR, a model with constant SFR, and two galaxies from the {\sc eagle} and \illustris{} cosmological simulations.

We compared our simulated BNS catalogues with the observed Galactic 
BNSs after applying radio selection effects with \psrpoppy{} \citep{bates2014}. 
In our analysis, we considered  four observable parameters:   orbital period,  eccentricity,  pulsar spin period, and spin period derivative. After modelling our radio-selected pulsars with a Dirichlet process Gaussian mixture model \citep{rinaldi2022figaro}, we evaluated the four-dimensional likelihood associated with each model in the aforementioned  parameter space.  We then   compared our models by computing the Bayes factor with respect to the fiducial model.

We also derived the Galactic BNS merger rate from our models and compared it with the one inferred from the observations \citep{pol2019,pol2020}. 
The CE parameter $\alpha$ has a large impact on both the merger rate and the orbital properties of the BNS population. The present-day BNS merger rate  varies up to one order of magnitude depending on the choice of $\alpha$. Values of $\alpha{}<1$ are disfavoured as they under-predict the merger rates and produce more eccentric systems compared to observations (Fig. \ref{fig:cumdistr}). Assuming the empirical MW prescription, the model with $\alpha=3$ produces a rate   $\mathcal{R}_{\rm MW} = 31.3$ Myr$^{-1}$ (Fig. \ref{fig:mergerrates}), consistent with the rate inferred from the Galactic pulsar binary systems  ($\mathcal{R}_{\rm MW}=37^{+24}_{-11}\,{}\text{Myr}^{-1}$,  \citealt{pol2020}).

The distribution of magnetic field and spin period at pulsar formation play a critical role on the final population of detectable pulsars (Fig. \ref{fig:ppdotiso}). The Bayes factors favour the uniform (U) distribution of spin periods and magnetic fields, which predicts 5--10 detectable pulsars. 
In contrast, the flat-in-log (FL) model predicts $\sim 100$ detectable pulsars ( Fig.~\ref{fig:numbers}) against the 8 observed Galactic pulsars in the considered surveys (the Parkes Multibeam Pulsar Survey, the Swinburne Multibeam Pulsar Survey, and the High Time Resolution Universe Pulsar Survey). 

The magnetic field decay timescale $\tau_{\rm d}$ is another  
free parameter of our model. There are large uncertainties on the physical process leading to the burial of the magnetic field and consequently to the typical timescales associated with it. $\tau_{\rm d}$ dictates the speed at which a pulsar traverses the $P-\dot{P}$ plane and stops emitting radio beams. Given the small sample of Galactic BNSs, we cannot draw strong conclusions on the expected $\tau_{\rm d}$ value.

We have shown that we need to account for radio selection effects 
in order to reproduce the observed spin period and magnetic field distributions of observed pulsars. Moreover, binary selection effects are critical to correctly match the orbital period and especially the eccentricity distribution of the observed Galactic BNSs. 

According to our fiducial model, which matches both the Galactic merger rate and the orbital properties of Galactic BNSs, the SKA will observe $\sim{30}$ BNSs in the Milky Way, among which $\sim{20}$ new detections.

\section*{Acknowledgements}
We thank \cite{hurley2002} for making the BSE code publicly available.
MM and GI acknowledge financial support from the European Research 
Council for the ERC Consolidator grant DEMOBLACK, under contract no. 
770017.  
MM acknowledges support from PRIN-MIUR~2020 METE, under contract no. 2020KB33TP. 
This work has been funded using resources from the INAF Large Grant 2022 “GCjewels” (P.I. Andrea Possenti) approved with the Presidential Decree 30/2022.
MCA acknowledges financial support from the Seal of Excellence @UNIPD 2020 program under the ACROGAL project. 
MS acknowledges financial support from the National Research Centre for High Performance Computing, Big Data and Quantum Computing (ICSC), and from the program "Data Science methods for MultiMessenger Astrophysics \& Multi-Survey Cosmology” funded by the Italian Ministry of University and Research, Programmazione triennale 2021/2023 (DM n.2503 dd. 09/12/2019), Programma Congiunto Scuole.
This research made use of \textsc{NumPy} \citep{harris2020}, \textsc{SciPy} \citep{virtanen2020}. For the plots we used \textsc{Matplotlib} \citep{hunter2007}. To model the radio selection effects we used \psrpoppy{} \citep{bates2014}. 
\section*{Data Availability}

This work made use of the codes \sevn{}  \citep{iorio2022} available at the gitlab repository \url{https://gitlab.com/sevncodes/sevn} (release {\it Sgalletta23}, \url{https://gitlab.com/sevncodes/sevn/-/releases/sgalletta23}), and  \textsc{figaro}, presented in \citet{rinaldi2022figaro}. 
The list of \sevn{} parameters used in this work can be found at the following link: \url{https://doi.org/10.5281/zenodo.7887279} \citep{zenodo_repo}.
Further data will be shared based on reasonable request to the corresponding authors.



\bibliographystyle{mnras}
\bibliography{ref} 

\begin{thebibliography}{}
\makeatletter
\relax
\def\mn@urlcharsother{\let\do\@makeother \do\$\do\&\do\#\do\^\do\_\do\%\do\~}
\def\mn@doi{\begingroup\mn@urlcharsother \@ifnextchar [ {\mn@doi@}
  {\mn@doi@[]}}
\def\mn@doi@[#1]#2{\def\@tempa{#1}\ifx\@tempa\@empty \href
  {http://dx.doi.org/#2} {doi:#2}\else \href {http://dx.doi.org/#2} {#1}\fi
  \endgroup}
\def\mn@eprint#1#2{\mn@eprint@#1:#2::\@nil}
\def\mn@eprint@arXiv#1{\href {http://arxiv.org/abs/#1} {{\tt arXiv:#1}}}
\def\mn@eprint@dblp#1{\href {http://dblp.uni-trier.de/rec/bibtex/#1.xml}
  {dblp:#1}}
\def\mn@eprint@#1:#2:#3:#4\@nil{\def\@tempa {#1}\def\@tempb {#2}\def\@tempc
  {#3}\ifx \@tempc \@empty \let \@tempc \@tempb \let \@tempb \@tempa \fi \ifx
  \@tempb \@empty \def\@tempb {arXiv}\fi \@ifundefined
  {mn@eprint@\@tempb}{\@tempb:\@tempc}{\expandafter \expandafter \csname
  mn@eprint@\@tempb\endcsname \expandafter{\@tempc}}}

\bibitem[\protect\citeauthoryear{{Abbott} et~al.,}{{Abbott}
  et~al.}{2017a}]{abbott2017a}
{Abbott} B.~P.,  et~al., 2017a, \mn@doi [\prl]
  {10.1103/PhysRevLett.119.161101}, \href
  {https://ui.adsabs.harvard.edu/abs/2017PhRvL.119p1101A} {119, 161101}

\bibitem[\protect\citeauthoryear{{Abbott} et~al.,}{{Abbott}
  et~al.}{2017b}]{abbott2017b}
{Abbott} B.~P.,  et~al., 2017b, \mn@doi [\apjl] {10.3847/2041-8213/aa91c9},
  \href {https://ui.adsabs.harvard.edu/abs/2017ApJ...848L..12A} {848, L12}

\bibitem[\protect\citeauthoryear{{Abbott} et~al.,}{{Abbott}
  et~al.}{2018}]{abbott2018}
{Abbott} B.~P.,  et~al., 2018, \mn@doi [\prl] {10.1103/PhysRevLett.121.161101},
  \href {https://ui.adsabs.harvard.edu/abs/2018PhRvL.121p1101A} {121, 161101}

\bibitem[\protect\citeauthoryear{{Abbott} et~al.,}{{Abbott}
  et~al.}{2020}]{abbottGW190425}
{Abbott} B.~P.,  et~al., 2020, \mn@doi [\apjl] {10.3847/2041-8213/ab75f5},
  \href {https://ui.adsabs.harvard.edu/abs/2020ApJ...892L...3A} {892, L3}

\bibitem[\protect\citeauthoryear{{Ade} et~al.,}{{Ade}
  et~al.}{2014}]{planck2014}
{Ade} P.~A.~R.,  et~al., 2014, \mn@doi [\aap] {10.1051/0004-6361/201321591},
  \href {https://ui.adsabs.harvard.edu/abs/2014A&A...571A..16P} {571, A16}

\bibitem[\protect\citeauthoryear{{Ade} et~al.,}{{Ade} et~al.}{2016}]{ade2016}
{Ade} P.~A.~R.,  et~al., 2016, \mn@doi [\aap] {10.1051/0004-6361/201525830},
  \href {https://ui.adsabs.harvard.edu/abs/2016A&A...594A..13P} {594, A13}

\bibitem[\protect\citeauthoryear{{Andersen} \& {Ransom}}{{Andersen} \&
  {Ransom}}{2018}]{andersen2018}
{Andersen} B.~C.,  {Ransom} S.~M.,  2018, \mn@doi [\apjl]
  {10.3847/2041-8213/aad59f}, \href
  {https://ui.adsabs.harvard.edu/abs/2018ApJ...863L..13A} {863, L13}

\bibitem[\protect\citeauthoryear{{Andrews} \& {Mandel}}{{Andrews} \&
  {Mandel}}{2019}]{andrews2019}
{Andrews} J.~J.,  {Mandel} I.,  2019, \mn@doi [\apjl]
  {10.3847/2041-8213/ab2ed1}, \href
  {https://ui.adsabs.harvard.edu/abs/2019ApJ...880L...8A} {880, L8}

\bibitem[\protect\citeauthoryear{Artale, Mapelli, Giacobbo, Sabha, Spera,
  Santoliquido  \& Bressan}{Artale et~al.}{2019}]{artale2019}
Artale M.~C.,  Mapelli M.,  Giacobbo N.,  Sabha N.~B.,  Spera M.,  Santoliquido
  F.,   Bressan A.,  2019, \mn@doi [MNRAS] {10.1093/mnras/stz1382}, 487, 1675

\bibitem[\protect\citeauthoryear{{Artale}, {Mapelli}, {Bouffanais}, {Giacobbo},
  {Pasquato}  \& {Spera}}{{Artale} et~al.}{2020a}]{artale2020a}
{Artale} M.~C.,  {Mapelli} M.,  {Bouffanais} Y.,  {Giacobbo} N.,  {Pasquato}
  M.,   {Spera} M.,  2020a, \mn@doi [\mnras] {10.1093/mnras/stz3190}, \href
  {https://ui.adsabs.harvard.edu/abs/2020MNRAS.491.3419A} {491, 3419}

\bibitem[\protect\citeauthoryear{{Artale}, {Bouffanais}, {Mapelli}, {Giacobbo},
  {Sabha}, {Santoliquido}, {Pasquato}  \& {Spera}}{{Artale}
  et~al.}{2020b}]{artale2020b}
{Artale} M.~C.,  {Bouffanais} Y.,  {Mapelli} M.,  {Giacobbo} N.,  {Sabha}
  N.~B.,  {Santoliquido} F.,  {Pasquato} M.,   {Spera} M.,  2020b, \mn@doi
  [\mnras] {10.1093/mnras/staa1252}, \href
  {https://ui.adsabs.harvard.edu/abs/2020MNRAS.495.1841A} {495, 1841}

\bibitem[\protect\citeauthoryear{{Arzoumanian} et~al.,}{{Arzoumanian}
  et~al.}{2018}]{arzoumanian2018}
{Arzoumanian} Z.,  et~al., 2018, \mn@doi [\apjs] {10.3847/1538-4365/aab5b0},
  \href {https://ui.adsabs.harvard.edu/abs/2018ApJS..235...37A} {235, 37}

\bibitem[\protect\citeauthoryear{{Arzoumanian} et~al.,}{{Arzoumanian}
  et~al.}{2020}]{Arzoumanian2020}
{Arzoumanian} Z.,  et~al., 2020, \mn@doi [\apjl] {10.3847/2041-8213/abd401},
  \href {https://ui.adsabs.harvard.edu/abs/2020ApJ...905L..34A} {905, L34}

\bibitem[\protect\citeauthoryear{{Asplund}, {Grevesse}, {Sauval}  \&
  {Scott}}{{Asplund} et~al.}{2009}]{asplund2009}
{Asplund} M.,  {Grevesse} N.,  {Sauval} A.~J.,   {Scott} P.,  2009, \mn@doi
  [\araa] {10.1146/annurev.astro.46.060407.145222}, \href
  {https://ui.adsabs.harvard.edu/abs/2009ARA&A..47..481A} {47, 481}

\bibitem[\protect\citeauthoryear{Bagchi, Lorimer  \& Wolfe}{Bagchi
  et~al.}{2013}]{bagchi2013}
Bagchi M.,  Lorimer D.,   Wolfe S.,  2013, \mn@doi [Monthly Notices of the
  Royal Astronomical Society] {10.1093/mnras/stt559}, 432

\bibitem[\protect\citeauthoryear{{Balakrishnan}, {Champion}, {Barr}, {Kramer},
  {Venkatraman Krishnan}, {Eatough}, {Sengar}  \& {Bailes}}{{Balakrishnan}
  et~al.}{2022}]{balakrishnan2022}
{Balakrishnan} V.,  {Champion} D.,  {Barr} E.,  {Kramer} M.,  {Venkatraman
  Krishnan} V.,  {Eatough} R.~P.,  {Sengar} R.,   {Bailes} M.,  2022, \mn@doi
  [\mnras] {10.1093/mnras/stab3746}, \href
  {https://ui.adsabs.harvard.edu/abs/2022MNRAS.511.1265B} {511, 1265}

\bibitem[\protect\citeauthoryear{{Bates}, {Lorimer}, {Rane}  \&
  {Swiggum}}{{Bates} et~al.}{2014}]{bates2014}
{Bates} S.~D.,  {Lorimer} D.~R.,  {Rane} A.,   {Swiggum} J.,  2014, \mn@doi
  [\mnras] {10.1093/mnras/stu157}, \href
  {https://ui.adsabs.harvard.edu/abs/2014MNRAS.439.2893B} {439, 2893}

\bibitem[\protect\citeauthoryear{{Bauswein}, {Just}, {Janka}  \&
  {Stergioulas}}{{Bauswein} et~al.}{2017}]{bauswein2017}
{Bauswein} A.,  {Just} O.,  {Janka} H.-T.,   {Stergioulas} N.,  2017, \mn@doi
  [\apjl] {10.3847/2041-8213/aa9994}, \href
  {https://ui.adsabs.harvard.edu/abs/2017ApJ...850L..34B} {850, L34}

\bibitem[\protect\citeauthoryear{{Belczynski} et~al.,}{{Belczynski}
  et~al.}{2018}]{belczynski2018}
{Belczynski} K.,  et~al., 2018, \mn@doi [\aap] {10.1051/0004-6361/201732428},
  \href {https://ui.adsabs.harvard.edu/abs/2018A&A...615A..91B} {615, A91}

\bibitem[\protect\citeauthoryear{{Belczynski} et~al.,}{{Belczynski}
  et~al.}{2020}]{belczynski2020}
{Belczynski} K.,  et~al., 2020, \mn@doi [\aap] {10.1051/0004-6361/201936528},
  \href {https://ui.adsabs.harvard.edu/abs/2020A&A...636A.104B} {636, A104}

\bibitem[\protect\citeauthoryear{{Benensohn}, {Lamb}  \& {Taam}}{{Benensohn}
  et~al.}{1997}]{benensohn1997}
{Benensohn} J.~S.,  {Lamb} D.~Q.,   {Taam} R.~E.,  1997, \mn@doi [\apj]
  {10.1086/303835}, \href
  {https://ui.adsabs.harvard.edu/abs/1997ApJ...478..723B} {478, 723}

\bibitem[\protect\citeauthoryear{{Bhat}, {Cordes}, {Camilo}, {Nice}  \&
  {Lorimer}}{{Bhat} et~al.}{2004}]{bhat2004}
{Bhat} N.~D.~R.,  {Cordes} J.~M.,  {Camilo} F.,  {Nice} D.~J.,   {Lorimer}
  D.~R.,  2004, \mn@doi [\apj] {10.1086/382680}, \href
  {https://ui.adsabs.harvard.edu/abs/2004ApJ...605..759B} {605, 759}

\bibitem[\protect\citeauthoryear{{Bhattacharya}, {Wijers}, {Hartman}  \&
  {Verbunt}}{{Bhattacharya} et~al.}{1992}]{bhattacharya1992}
{Bhattacharya} D.,  {Wijers} R. A.~M.~J.,  {Hartman} J.~W.,   {Verbunt} F.,
  1992, \aap, \href {https://ui.adsabs.harvard.edu/abs/1992A&A...254..198B}
  {254, 198}

\bibitem[\protect\citeauthoryear{{Blaauw}}{{Blaauw}}{1961}]{blaauw1961}
{Blaauw} A.,  1961, \bain, \href
  {https://ui.adsabs.harvard.edu/abs/1961BAN....15..265B} {15, 265}

\bibitem[\protect\citeauthoryear{{Boco}, {Lapi}, {Goswami}, {Perrotta},
  {Baccigalupi}  \& {Danese}}{{Boco} et~al.}{2019}]{boco2019}
{Boco} L.,  {Lapi} A.,  {Goswami} S.,  {Perrotta} F.,  {Baccigalupi} C.,
  {Danese} L.,  2019, \mn@doi [\apj] {10.3847/1538-4357/ab328e}, \href
  {https://ui.adsabs.harvard.edu/abs/2019ApJ...881..157B} {881, 157}

\bibitem[\protect\citeauthoryear{{Boco}, {Lapi}, {Chruslinska}, {Donevski},
  {Sicilia}  \& {Danese}}{{Boco} et~al.}{2021}]{boco2021}
{Boco} L.,  {Lapi} A.,  {Chruslinska} M.,  {Donevski} D.,  {Sicilia} A.,
  {Danese} L.,  2021, \mn@doi [\apj] {10.3847/1538-4357/abd3a0}, \href
  {https://ui.adsabs.harvard.edu/abs/2021ApJ...907..110B} {907, 110}

\bibitem[\protect\citeauthoryear{{Bogdanov}, {Heinke}, {{\"O}zel}  \&
  {G{\"u}ver}}{{Bogdanov} et~al.}{2016}]{Bogdanov2016}
{Bogdanov} S.,  {Heinke} C.~O.,  {{\"O}zel} F.,   {G{\"u}ver} T.,  2016,
  \mn@doi [\apj] {10.3847/0004-637X/831/2/184}, \href
  {https://ui.adsabs.harvard.edu/abs/2016ApJ...831..184B} {831, 184}

\bibitem[\protect\citeauthoryear{Bouffanais, Mapelli, Santoliquido, Giacobbo,
  Iorio  \& Costa}{Bouffanais et~al.}{2021}]{Bouffanais2021}
Bouffanais Y.,  Mapelli M.,  Santoliquido F.,  Giacobbo N.,  Iorio G.,   Costa
  G.,  2021, \mn@doi [Monthly Notices of the Royal Astronomical Society]
  {10.1093/mnras/stab1589}, 505, 3873

\bibitem[\protect\citeauthoryear{{Bovy}}{{Bovy}}{2017}]{bovy2017}
{Bovy} J.,  2017, \mn@doi [\mnras] {10.1093/mnras/stx1277}, \href
  {https://ui.adsabs.harvard.edu/abs/2017MNRAS.470.1360B} {470, 1360}

\bibitem[\protect\citeauthoryear{{Bray} \& {Eldridge}}{{Bray} \&
  {Eldridge}}{2016}]{bray2016}
{Bray} J.~C.,  {Eldridge} J.~J.,  2016, \mn@doi [\mnras]
  {10.1093/mnras/stw1275}, \href
  {https://ui.adsabs.harvard.edu/abs/2016MNRAS.461.3747B} {461, 3747}

\bibitem[\protect\citeauthoryear{{Bray} \& {Eldridge}}{{Bray} \&
  {Eldridge}}{2018}]{bray2018}
{Bray} J.~C.,  {Eldridge} J.~J.,  2018, \mn@doi [\mnras]
  {10.1093/mnras/sty2230}, \href
  {https://ui.adsabs.harvard.edu/abs/2018MNRAS.480.5657B} {480, 5657}

\bibitem[\protect\citeauthoryear{{Bressan}, {Marigo}, {Girardi}, {Salasnich},
  {Dal Cero}, {Rubele}  \& {Nanni}}{{Bressan} et~al.}{2012}]{bressan2012}
{Bressan} A.,  {Marigo} P.,  {Girardi} L.,  {Salasnich} B.,  {Dal Cero} C.,
  {Rubele} S.,   {Nanni} A.,  2012, \mn@doi [\mnras]
  {10.1111/j.1365-2966.2012.21948.x}, \href
  {https://ui.adsabs.harvard.edu/abs/2012MNRAS.427..127B} {427, 127}

\bibitem[\protect\citeauthoryear{{Breton} et~al.,}{{Breton}
  et~al.}{2008}]{breton2008}
{Breton} R.~P.,  et~al., 2008, \mn@doi [Science] {10.1126/science.1159295},
  \href {https://ui.adsabs.harvard.edu/abs/2008Sci...321..104B} {321, 104}

\bibitem[\protect\citeauthoryear{{Broekgaarden} et~al.,}{{Broekgaarden}
  et~al.}{2021}]{broekgaarden2021a}
{Broekgaarden} F.~S.,  et~al., 2021, \mn@doi [\mnras] {10.1093/mnras/stab2716},
  \href {https://ui.adsabs.harvard.edu/abs/2021MNRAS.508.5028B} {508, 5028}

\bibitem[\protect\citeauthoryear{{Broekgaarden} et~al.,}{{Broekgaarden}
  et~al.}{2022}]{broekgaarden2022}
{Broekgaarden} F.~S.,  et~al., 2022, \mn@doi [\mnras] {10.1093/mnras/stac1677},
  \href {https://ui.adsabs.harvard.edu/abs/2022MNRAS.516.5737B} {516, 5737}

\bibitem[\protect\citeauthoryear{{Brown}}{{Brown}}{1995}]{brown1995}
{Brown} G.~E.,  1995, \mn@doi [\apj] {10.1086/175268}, \href
  {https://ui.adsabs.harvard.edu/abs/1995ApJ...440..270B} {440, 270}

\bibitem[\protect\citeauthoryear{{Burgay} et~al.,}{{Burgay}
  et~al.}{2003}]{burgay2003}
{Burgay} M.,  et~al., 2003, \mn@doi [\nat] {10.1038/nature02124}, \href
  {https://ui.adsabs.harvard.edu/abs/2003Natur.426..531B} {426, 531}

\bibitem[\protect\citeauthoryear{{Cameron} et~al.,}{{Cameron}
  et~al.}{2018}]{cameron2018}
{Cameron} A.~D.,  et~al., 2018, \mn@doi [\mnras] {10.1093/mnrasl/sly003}, \href
  {https://ui.adsabs.harvard.edu/abs/2018MNRAS.475L..57C} {475, L57}

\bibitem[\protect\citeauthoryear{Chamandy et~al.,}{Chamandy
  et~al.}{2018}]{chamandy2018}
Chamandy L.,  et~al., 2018, \mn@doi [Monthly Notices of the Royal Astronomical
  Society] {10.1093/mnras/sty1950}, 480, 1898

\bibitem[\protect\citeauthoryear{{Chattopadhyay}, {Stevenson}, {Hurley},
  {Rossi}  \& {Flynn}}{{Chattopadhyay} et~al.}{2020}]{chattopadhyay2020}
{Chattopadhyay} D.,  {Stevenson} S.,  {Hurley} J.~R.,  {Rossi} L.~J.,   {Flynn}
  C.,  2020, \mn@doi [\mnras] {10.1093/mnras/staa756}, \href
  {https://ui.adsabs.harvard.edu/abs/2020MNRAS.494.1587C} {494, 1587}

\bibitem[\protect\citeauthoryear{{Chattopadhyay}, {Stevenson}, {Hurley},
  {Bailes}  \& {Broekgaarden}}{{Chattopadhyay}
  et~al.}{2021}]{chattopadhyay2021}
{Chattopadhyay} D.,  {Stevenson} S.,  {Hurley} J.~R.,  {Bailes} M.,
  {Broekgaarden} F.,  2021, \mn@doi [\mnras] {10.1093/mnras/stab973}, \href
  {https://ui.adsabs.harvard.edu/abs/2021MNRAS.504.3682C} {504, 3682}

\bibitem[\protect\citeauthoryear{{Chiappini}, {Matteucci}  \&
  {Gratton}}{{Chiappini} et~al.}{1997}]{chiappini1997}
{Chiappini} C.,  {Matteucci} F.,   {Gratton} R.,  1997, \mn@doi [\apj]
  {10.1086/303726}, \href
  {https://ui.adsabs.harvard.edu/abs/1997ApJ...477..765C} {477, 765}

\bibitem[\protect\citeauthoryear{{Chruslinska}, {Belczynski}, {Klencki}  \&
  {Benacquista}}{{Chruslinska} et~al.}{2018}]{chruslinska2018}
{Chruslinska} M.,  {Belczynski} K.,  {Klencki} J.,   {Benacquista} M.,  2018,
  \mn@doi [\mnras] {10.1093/mnras/stx2923}, \href
  {https://ui.adsabs.harvard.edu/abs/2018MNRAS.474.2937C} {474, 2937}

\bibitem[\protect\citeauthoryear{{Chru{\'s}li{\'n}ska}, {Nelemans}, {Boco}  \&
  {Lapi}}{{Chru{\'s}li{\'n}ska} et~al.}{2021}]{chruslinska2021}
{Chru{\'s}li{\'n}ska} M.,  {Nelemans} G.,  {Boco} L.,   {Lapi} A.,  2021,
  \mn@doi [\mnras] {10.1093/mnras/stab2690}, \href
  {https://ui.adsabs.harvard.edu/abs/2021MNRAS.508.4994C} {508, 4994}

\bibitem[\protect\citeauthoryear{{Claeys}, {Pols}, {Izzard}, {Vink}  \&
  {Verbunt}}{{Claeys} et~al.}{2014}]{claeys2014}
{Claeys} J.~S.~W.,  {Pols} O.~R.,  {Izzard} R.~G.,  {Vink} J.,   {Verbunt}
  F.~W.~M.,  2014, \mn@doi [\aap] {10.1051/0004-6361/201322714}, \href
  {https://ui.adsabs.harvard.edu/abs/2014A&A...563A..83C} {563, A83}

\bibitem[\protect\citeauthoryear{{Colombo}, {Salafia}, {Gabrielli},
  {Ghirlanda}, {Giacomazzo}, {Perego}  \& {Colpi}}{{Colombo}
  et~al.}{2022}]{colombo2022}
{Colombo} A.,  {Salafia} O.~S.,  {Gabrielli} F.,  {Ghirlanda} G.,  {Giacomazzo}
  B.,  {Perego} A.,   {Colpi} M.,  2022, \mn@doi [\apj]
  {10.3847/1538-4357/ac8d00}, \href
  {https://ui.adsabs.harvard.edu/abs/2022ApJ...937...79C} {937, 79}

\bibitem[\protect\citeauthoryear{{Combi} \& {Siegel}}{{Combi} \&
  {Siegel}}{2023}]{combi2023}
{Combi} L.,  {Siegel} D.~M.,  2023, \mn@doi [\apj] {10.3847/1538-4357/acac29},
  \href {https://ui.adsabs.harvard.edu/abs/2023ApJ...944...28C} {944, 28}

\bibitem[\protect\citeauthoryear{{Cordes}}{{Cordes}}{2004}]{cordes2004}
{Cordes} J.~M.,  2004, in {Clemens} D.,  {Shah} R.,   {Brainerd} T.,  eds,
  Astronomical Society of the Pacific Conference Series Vol. 317, Milky Way
  Surveys: The Structure and Evolution of our Galaxy. p.~211

\bibitem[\protect\citeauthoryear{{Cordes} \& {Lazio}}{{Cordes} \&
  {Lazio}}{2002}]{cordes2002}
{Cordes} J.~M.,  {Lazio} T.~J.~W.,  2002, arXiv e-prints, \href
  {https://ui.adsabs.harvard.edu/abs/2002astro.ph..7156C} {pp
  astro--ph/0207156}

\bibitem[\protect\citeauthoryear{{Corongiu}, {Kramer}, {Stappers}, {Lyne},
  {Jessner}, {Possenti}, {D'Amico}  \& {L{\"o}hmer}}{{Corongiu}
  et~al.}{2007}]{corongiu2007}
{Corongiu} A.,  {Kramer} M.,  {Stappers} B.~W.,  {Lyne} A.~G.,  {Jessner} A.,
  {Possenti} A.,  {D'Amico} N.,   {L{\"o}hmer} O.,  2007, \mn@doi [\aap]
  {10.1051/0004-6361:20054385}, \href
  {https://ui.adsabs.harvard.edu/abs/2007A&A...462..703C} {462, 703}

\bibitem[\protect\citeauthoryear{Costa, Girardi, Bressan, Marigo, Rodrigues,
  Chen, Lanza  \& Goudfrooij}{Costa et~al.}{2019}]{costa2019}
Costa G.,  Girardi L.,  Bressan A.,  Marigo P.,  Rodrigues T.~S.,  Chen Y.,
  Lanza A.,   Goudfrooij P.,  2019, \mn@doi [Monthly Notices of the Royal
  Astronomical Society] {10.1093/mnras/stz728}, 485, 4641

\bibitem[\protect\citeauthoryear{{Costa}, {Bressan}, {Mapelli}, {Marigo},
  {Iorio}  \& {Spera}}{{Costa} et~al.}{2021}]{costa2021}
{Costa} G.,  {Bressan} A.,  {Mapelli} M.,  {Marigo} P.,  {Iorio} G.,   {Spera}
  M.,  2021, \mn@doi [\mnras] {10.1093/mnras/staa3916}, \href
  {https://ui.adsabs.harvard.edu/abs/2021MNRAS.501.4514C} {501, 4514}

\bibitem[\protect\citeauthoryear{{C{\^o}t{\'e}} et~al.,}{{C{\^o}t{\'e}}
  et~al.}{2019}]{cote2019}
{C{\^o}t{\'e}} B.,  et~al., 2019, \mn@doi [\apj] {10.3847/1538-4357/ab10db},
  \href {https://ui.adsabs.harvard.edu/abs/2019ApJ...875..106C} {875, 106}

\bibitem[\protect\citeauthoryear{{Courteau} et~al.,}{{Courteau}
  et~al.}{2014}]{courteau2014}
{Courteau} S.,  et~al., 2014, \mn@doi [Reviews of Modern Physics]
  {10.1103/RevModPhys.86.47}, \href
  {https://ui.adsabs.harvard.edu/abs/2014RvMP...86...47C} {86, 47}

\bibitem[\protect\citeauthoryear{{Dewey}, {Taylor}, {Weisberg}  \&
  {Stokes}}{{Dewey} et~al.}{1985}]{dewey1985}
{Dewey} R.~J.,  {Taylor} J.~H.,  {Weisberg} J.~M.,   {Stokes} G.~H.,  1985,
  \mn@doi [\apjl] {10.1086/184502}, \href
  {https://ui.adsabs.harvard.edu/abs/1985ApJ...294L..25D} {294, L25}

\bibitem[\protect\citeauthoryear{{Dewi}, {Podsiadlowski}  \& {Sena}}{{Dewi}
  et~al.}{2006}]{dewi2006}
{Dewi} J.~D.~M.,  {Podsiadlowski} P.,   {Sena} A.,  2006, \mn@doi [\mnras]
  {10.1111/j.1365-2966.2006.10233.x}, \href
  {https://ui.adsabs.harvard.edu/abs/2006MNRAS.368.1742D} {368, 1742}

\bibitem[\protect\citeauthoryear{{Di Stefano}, {Kruckow}, {Gao}, {Neunteufel}
  \& {Kobayashi}}{{Di Stefano} et~al.}{2023}]{distefano2023}
{Di Stefano} R.,  {Kruckow} M.~U.,  {Gao} Y.,  {Neunteufel} P.~G.,
  {Kobayashi} C.,  2023, \mn@doi [\apj] {10.3847/1538-4357/acae9b}, \href
  {https://ui.adsabs.harvard.edu/abs/2023ApJ...944...87D} {944, 87}

\bibitem[\protect\citeauthoryear{{Edwards}, {Bailes}, {van Straten}  \&
  {Britton}}{{Edwards} et~al.}{2001}]{edwards2001}
{Edwards} R.~T.,  {Bailes} M.,  {van Straten} W.,   {Britton} M.~C.,  2001,
  \mn@doi [\mnras] {10.1046/j.1365-8711.2001.04637.x}, \href
  {https://ui.adsabs.harvard.edu/abs/2001MNRAS.326..358E} {326, 358}

\bibitem[\protect\citeauthoryear{{Eggleton}}{{Eggleton}}{1983}]{eggleton1983}
{Eggleton} P.~P.,  1983, \mn@doi [\apj] {10.1086/160960}, \href
  {https://ui.adsabs.harvard.edu/abs/1983ApJ...268..368E} {268, 368}

\bibitem[\protect\citeauthoryear{{Eichler}, {Livio}, {Piran}  \&
  {Schramm}}{{Eichler} et~al.}{1989}]{eichler1989}
{Eichler} D.,  {Livio} M.,  {Piran} T.,   {Schramm} D.~N.,  1989, \mn@doi
  [\nat] {10.1038/340126a0}, \href
  {https://ui.adsabs.harvard.edu/abs/1989Natur.340..126E} {340, 126}

\bibitem[\protect\citeauthoryear{{Faucher-Gigu{\`e}re} \&
  {Kaspi}}{{Faucher-Gigu{\`e}re} \& {Kaspi}}{2006}]{fauchergiguere2006}
{Faucher-Gigu{\`e}re} C.-A.,  {Kaspi} V.~M.,  2006, \mn@doi [\apj]
  {10.1086/501516}, \href
  {https://ui.adsabs.harvard.edu/abs/2006ApJ...643..332F} {643, 332}

\bibitem[\protect\citeauthoryear{{Faulkner} et~al.,}{{Faulkner}
  et~al.}{2004}]{faulkner2004}
{Faulkner} A.~J.,  et~al., 2004, \mn@doi [\mnras]
  {10.1111/j.1365-2966.2004.08310.x}, \href
  {https://ui.adsabs.harvard.edu/abs/2004MNRAS.355..147F} {355, 147}

\bibitem[\protect\citeauthoryear{{Fragos}, {Andrews}, {Ramirez-Ruiz}, {Meynet},
  {Kalogera}, {Taam}  \& {Zezas}}{{Fragos} et~al.}{2019}]{fragos2019}
{Fragos} T.,  {Andrews} J.~J.,  {Ramirez-Ruiz} E.,  {Meynet} G.,  {Kalogera}
  V.,  {Taam} R.~E.,   {Zezas} A.,  2019, \mn@doi [\apjl]
  {10.3847/2041-8213/ab40d1}, \href
  {https://ui.adsabs.harvard.edu/abs/2019ApJ...883L..45F} {883, L45}

\bibitem[\protect\citeauthoryear{{Fryer}, {Belczynski}, {Wiktorowicz},
  {Dominik}, {Kalogera}  \& {Holz}}{{Fryer} et~al.}{2012}]{fryer2012}
{Fryer} C.~L.,  {Belczynski} K.,  {Wiktorowicz} G.,  {Dominik} M.,  {Kalogera}
  V.,   {Holz} D.~E.,  2012, \mn@doi [\apj] {10.1088/0004-637X/749/1/91}, \href
  {https://ui.adsabs.harvard.edu/abs/2012ApJ...749...91F} {749, 91}

\bibitem[\protect\citeauthoryear{{Fujibayashi}, {Kiuchi}, {Wanajo}, {Kyutoku},
  {Sekiguchi}  \& {Shibata}}{{Fujibayashi} et~al.}{2023}]{fujibayashi2023}
{Fujibayashi} S.,  {Kiuchi} K.,  {Wanajo} S.,  {Kyutoku} K.,  {Sekiguchi} Y.,
  {Shibata} M.,  2023, \mn@doi [\apj] {10.3847/1538-4357/ac9ce0}, \href
  {https://ui.adsabs.harvard.edu/abs/2023ApJ...942...39F} {942, 39}

\bibitem[\protect\citeauthoryear{{Gessner} \& {Janka}}{{Gessner} \&
  {Janka}}{2018}]{gessner2018}
{Gessner} A.,  {Janka} H.-T.,  2018, \mn@doi [\apj] {10.3847/1538-4357/aadbae},
  \href {https://ui.adsabs.harvard.edu/abs/2018ApJ...865...61G} {865, 61}

\bibitem[\protect\citeauthoryear{{Giacobbo} \& {Mapelli}}{{Giacobbo} \&
  {Mapelli}}{2018}]{giacobbo2018b}
{Giacobbo} N.,  {Mapelli} M.,  2018, \mn@doi [\mnras] {10.1093/mnras/sty1999},
  \href {https://ui.adsabs.harvard.edu/abs/2018MNRAS.480.2011G} {480, 2011}

\bibitem[\protect\citeauthoryear{{Giacobbo} \& {Mapelli}}{{Giacobbo} \&
  {Mapelli}}{2019}]{giacobbo2019}
{Giacobbo} N.,  {Mapelli} M.,  2019, \mn@doi [\mnras] {10.1093/mnras/sty2848},
  \href {https://ui.adsabs.harvard.edu/abs/2019MNRAS.482.2234G} {482, 2234}

\bibitem[\protect\citeauthoryear{{Giacobbo} \& {Mapelli}}{{Giacobbo} \&
  {Mapelli}}{2020}]{giacobbo2020}
{Giacobbo} N.,  {Mapelli} M.,  2020, \mn@doi [\apj] {10.3847/1538-4357/ab7335},
  \href {https://ui.adsabs.harvard.edu/abs/2020ApJ...891..141G} {891, 141}

\bibitem[\protect\citeauthoryear{{Giacobbo}, {Mapelli}  \& {Spera}}{{Giacobbo}
  et~al.}{2018}]{giacobbo2018}
{Giacobbo} N.,  {Mapelli} M.,   {Spera} M.,  2018, \mn@doi [\mnras]
  {10.1093/mnras/stx2933}, \href
  {https://ui.adsabs.harvard.edu/abs/2018MNRAS.474.2959G} {474, 2959}

\bibitem[\protect\citeauthoryear{{Goldreich} \& {Julian}}{{Goldreich} \&
  {Julian}}{1969}]{goldreich1969}
{Goldreich} P.,  {Julian} W.~H.,  1969, \mn@doi [\apj] {10.1086/150119}, \href
  {https://ui.adsabs.harvard.edu/abs/1969ApJ...157..869G} {157, 869}

\bibitem[\protect\citeauthoryear{{Goldreich} \& {Reisenegger}}{{Goldreich} \&
  {Reisenegger}}{1992}]{goldreich1992}
{Goldreich} P.,  {Reisenegger} A.,  1992, \mn@doi [\apj] {10.1086/171646},
  \href {https://ui.adsabs.harvard.edu/abs/1992ApJ...395..250G} {395, 250}

\bibitem[\protect\citeauthoryear{{Goldstein} et~al.,}{{Goldstein}
  et~al.}{2017}]{goldstein2017}
{Goldstein} A.,  et~al., 2017, \mn@doi [\apjl] {10.3847/2041-8213/aa8f41},
  \href {https://ui.adsabs.harvard.edu/abs/2017ApJ...848L..14G} {848, L14}

\bibitem[\protect\citeauthoryear{{Gonthier}, {Van Guilder}  \&
  {Harding}}{{Gonthier} et~al.}{2004}]{gonthier2004}
{Gonthier} P.~L.,  {Van Guilder} R.,   {Harding} A.~K.,  2004, \mn@doi [\apj]
  {10.1086/382070}, \href
  {https://ui.adsabs.harvard.edu/abs/2004ApJ...604..775G} {604, 775}

\bibitem[\protect\citeauthoryear{{Grisoni}, {Spitoni}, {Matteucci},
  {Recio-Blanco}, {de Laverny}, {Hayden}, {Mikolaitis}  \& {Worley}}{{Grisoni}
  et~al.}{2017}]{grisoni2017}
{Grisoni} V.,  {Spitoni} E.,  {Matteucci} F.,  {Recio-Blanco} A.,  {de Laverny}
  P.,  {Hayden} M.,  {Mikolaitis} {\^{S}}.,   {Worley} C.~C.,  2017, \mn@doi
  [\mnras] {10.1093/mnras/stx2201}, \href
  {https://ui.adsabs.harvard.edu/abs/2017MNRAS.472.3637G} {472, 3637}

\bibitem[\protect\citeauthoryear{{Harris} et~al.,}{{Harris}
  et~al.}{2020}]{harris2020}
{Harris} C.~R.,  et~al., 2020, \mn@doi [\nat] {10.1038/s41586-020-2649-2},
  \href {https://ui.adsabs.harvard.edu/abs/2020Natur.585..357H} {585, 357}

\bibitem[\protect\citeauthoryear{{Haslam}, {Klein}, {Salter}, {Stoffel},
  {Wilson}, {Cleary}, {Cooke}  \& {Thomasson}}{{Haslam}
  et~al.}{1981}]{haslam1981}
{Haslam} C.~G.~T.,  {Klein} U.,  {Salter} C.~J.,  {Stoffel} H.,  {Wilson}
  W.~E.,  {Cleary} M.~N.,  {Cooke} D.~J.,   {Thomasson} P.,  1981, \aap, \href
  {https://ui.adsabs.harvard.edu/abs/1981A&A...100..209H} {100, 209}

\bibitem[\protect\citeauthoryear{{Hewish}, {Bell}, {Pilkington}, {Scott}  \&
  {Collins}}{{Hewish} et~al.}{1968}]{hewish1968}
{Hewish} A.,  {Bell} S.~J.,  {Pilkington} J.~D.~H.,  {Scott} P.~F.,   {Collins}
  R.~A.,  1968, \mn@doi [\nat] {10.1038/217709a0}, \href
  {https://ui.adsabs.harvard.edu/abs/1968Natur.217..709H} {217, 709}

\bibitem[\protect\citeauthoryear{{Hirai} \& {Mandel}}{{Hirai} \&
  {Mandel}}{2022}]{hirai2022}
{Hirai} R.,  {Mandel} I.,  2022, \mn@doi [\apjl] {10.3847/2041-8213/ac9519},
  \href {https://ui.adsabs.harvard.edu/abs/2022ApJ...937L..42H} {937, L42}

\bibitem[\protect\citeauthoryear{{Hobbs}}{{Hobbs}}{2013}]{hobbs2013}
{Hobbs} G.,  2013, \mn@doi [Classical and Quantum Gravity]
  {10.1088/0264-9381/30/22/224007}, \href
  {https://ui.adsabs.harvard.edu/abs/2013CQGra..30v4007H} {30, 224007}

\bibitem[\protect\citeauthoryear{Hobbs, Lorimer, Lyne  \& Kramer}{Hobbs
  et~al.}{2005}]{hobbs2005}
Hobbs G.,  Lorimer D.~R.,  Lyne A.~G.,   Kramer M.,  2005, \mn@doi [Monthly
  Notices of the Royal Astronomical Society]
  {10.1111/j.1365-2966.2005.09087.x}, 360, 974

\bibitem[\protect\citeauthoryear{{Hobbs} et~al.,}{{Hobbs}
  et~al.}{2010}]{hobbs2010}
{Hobbs} G.,  et~al., 2010, \mn@doi [Classical and Quantum Gravity]
  {10.1088/0264-9381/27/8/084013}, \href
  {https://ui.adsabs.harvard.edu/abs/2010CQGra..27h4013H} {27, 084013}

\bibitem[\protect\citeauthoryear{{Hotokezaka}, {Beniamini}  \&
  {Piran}}{{Hotokezaka} et~al.}{2018}]{hotokezaka2018}
{Hotokezaka} K.,  {Beniamini} P.,   {Piran} T.,  2018, \mn@doi [International
  Journal of Modern Physics D] {10.1142/S0218271818420051}, \href
  {https://ui.adsabs.harvard.edu/abs/2018IJMPD..2742005H} {27, 1842005}

\bibitem[\protect\citeauthoryear{{Hulse} \& {Taylor}}{{Hulse} \&
  {Taylor}}{1975}]{hulse1975a}
{Hulse} R.~A.,  {Taylor} J.~H.,  1975, \mn@doi [\apjl] {10.1086/181708}, \href
  {https://ui.adsabs.harvard.edu/abs/1975ApJ...195L..51H} {195, L51}

\bibitem[\protect\citeauthoryear{{Hunt}, {Dayal}, {Magrini}  \&
  {Ferrara}}{{Hunt} et~al.}{2016}]{hunt2016}
{Hunt} L.,  {Dayal} P.,  {Magrini} L.,   {Ferrara} A.,  2016, \mn@doi [\mnras]
  {10.1093/mnras/stw2091}, \href
  {https://ui.adsabs.harvard.edu/abs/2016MNRAS.463.2020H} {463, 2020}

\bibitem[\protect\citeauthoryear{{Hunter}}{{Hunter}}{2007}]{hunter2007}
{Hunter} J.~D.,  2007, \mn@doi [Computing in Science and Engineering]
  {10.1109/MCSE.2007.55}, \href
  {https://ui.adsabs.harvard.edu/abs/2007CSE.....9...90H} {9, 90}

\bibitem[\protect\citeauthoryear{{Hurley}, {Tout}  \& {Pols}}{{Hurley}
  et~al.}{2002}]{hurley2002}
{Hurley} J.~R.,  {Tout} C.~A.,   {Pols} O.~R.,  2002, \mn@doi [\mnras]
  {10.1046/j.1365-8711.2002.05038.x}, \href
  {https://ui.adsabs.harvard.edu/abs/2002MNRAS.329..897H} {329, 897}

\bibitem[\protect\citeauthoryear{{Igoshev}, {Frantsuzova}, {Gourgouliatos},
  {Tsichli}, {Konstantinou}  \& {Popov}}{{Igoshev} et~al.}{2022}]{igoshev2022}
{Igoshev} A.~P.,  {Frantsuzova} A.,  {Gourgouliatos} K.~N.,  {Tsichli} S.,
  {Konstantinou} L.,   {Popov} S.~B.,  2022, \mn@doi [\mnras]
  {10.1093/mnras/stac1648}, \href
  {https://ui.adsabs.harvard.edu/abs/2022MNRAS.514.4606I} {514, 4606}

\bibitem[\protect\citeauthoryear{{Illarionov} \& {Sunyaev}}{{Illarionov} \&
  {Sunyaev}}{1975}]{illarionov1975}
{Illarionov} A.~F.,  {Sunyaev} R.~A.,  1975, \aap, \href
  {https://ui.adsabs.harvard.edu/abs/1975A&A....39..185I} {39, 185}

\bibitem[\protect\citeauthoryear{{Iorio} et~al.,}{{Iorio}
  et~al.}{2022}]{iorio2022}
{Iorio} G.,  et~al., 2022, arXiv e-prints, \href
  {https://ui.adsabs.harvard.edu/abs/2022arXiv221111774I} {p. arXiv:2211.11774}

\bibitem[\protect\citeauthoryear{{Ivanova} et~al.,}{{Ivanova}
  et~al.}{2013}]{ivanova2013}
{Ivanova} N.,  et~al., 2013, \mn@doi [\aapr] {10.1007/s00159-013-0059-2}, \href
  {https://ui.adsabs.harvard.edu/abs/2013A&ARv..21...59I} {21, 59}

\bibitem[\protect\citeauthoryear{{Janka} \& {Mueller}}{{Janka} \&
  {Mueller}}{1994}]{janka1994}
{Janka} H.~T.,  {Mueller} E.,  1994, \aap, \href
  {https://ui.adsabs.harvard.edu/abs/1994A&A...290..496J} {290, 496}

\bibitem[\protect\citeauthoryear{{Justham}, {Podsiadlowski}  \&
  {Han}}{{Justham} et~al.}{2011}]{justham2011}
{Justham} S.,  {Podsiadlowski} P.,   {Han} Z.,  2011, \mn@doi [\mnras]
  {10.1111/j.1365-2966.2010.17497.x}, \href
  {https://ui.adsabs.harvard.edu/abs/2011MNRAS.410..984J} {410, 984}

\bibitem[\protect\citeauthoryear{{Kapil}, {Mandel}, {Berti}  \&
  {M{\"u}ller}}{{Kapil} et~al.}{2023}]{kapil2023}
{Kapil} V.,  {Mandel} I.,  {Berti} E.,   {M{\"u}ller} B.,  2023, \mn@doi
  [\mnras] {10.1093/mnras/stad019}, \href
  {https://ui.adsabs.harvard.edu/abs/2023MNRAS.519.5893K} {519, 5893}

\bibitem[\protect\citeauthoryear{{Kasen}, {Metzger}, {Barnes}, {Quataert}  \&
  {Ramirez-Ruiz}}{{Kasen} et~al.}{2017}]{kasen2017}
{Kasen} D.,  {Metzger} B.,  {Barnes} J.,  {Quataert} E.,   {Ramirez-Ruiz} E.,
  2017, \mn@doi [\nat] {10.1038/nature24453}, \href
  {https://ui.adsabs.harvard.edu/abs/2017Natur.551...80K} {551, 80}

\bibitem[\protect\citeauthoryear{{Kaspi}, {Taylor}  \& {Ryba}}{{Kaspi}
  et~al.}{1994}]{kaspi1994}
{Kaspi} V.~M.,  {Taylor} J.~H.,   {Ryba} M.~F.,  1994, \mn@doi [\apj]
  {10.1086/174280}, \href
  {https://ui.adsabs.harvard.edu/abs/1994ApJ...428..713K} {428, 713}

\bibitem[\protect\citeauthoryear{{Keith}, {Eatough}, {Lyne}, {Kramer},
  {Possenti}, {Camilo}  \& {Manchester}}{{Keith} et~al.}{2009}]{keith2009}
{Keith} M.~J.,  {Eatough} R.~P.,  {Lyne} A.~G.,  {Kramer} M.,  {Possenti} A.,
  {Camilo} F.,   {Manchester} R.~N.,  2009, \mn@doi [\mnras]
  {10.1111/j.1365-2966.2009.14543.x}, \href
  {https://ui.adsabs.harvard.edu/abs/2009MNRAS.395..837K} {395, 837}

\bibitem[\protect\citeauthoryear{{Keith} et~al.,}{{Keith}
  et~al.}{2010}]{keith2010}
{Keith} M.~J.,  et~al., 2010, \mn@doi [\mnras]
  {10.1111/j.1365-2966.2010.17325.x}, \href
  {https://ui.adsabs.harvard.edu/abs/2010MNRAS.409..619K} {409, 619}

\bibitem[\protect\citeauthoryear{Kiel, Hurley, Bailes  \& Murray}{Kiel
  et~al.}{2008}]{kiel2008}
Kiel P.~D.,  Hurley J.~R.,  Bailes M.,   Murray J.~R.,  2008, \mn@doi [Monthly
  Notices of the Royal Astronomical Society]
  {10.1111/j.1365-2966.2008.13402.x}, 388, 393

\bibitem[\protect\citeauthoryear{{Klencki}, {Nelemans}, {Istrate}  \&
  {Chruslinska}}{{Klencki} et~al.}{2021}]{klencki2021}
{Klencki} J.,  {Nelemans} G.,  {Istrate} A.~G.,   {Chruslinska} M.,  2021,
  \mn@doi [\aap] {10.1051/0004-6361/202038707}, \href
  {https://ui.adsabs.harvard.edu/abs/2021A&A...645A..54K} {645, A54}

\bibitem[\protect\citeauthoryear{{Kobayashi} et~al.,}{{Kobayashi}
  et~al.}{2023}]{kobayashi2022}
{Kobayashi} C.,  et~al., 2023, \mn@doi [\apjl] {10.3847/2041-8213/acad82},
  \href {https://ui.adsabs.harvard.edu/abs/2023ApJ...943L..12K} {943, L12}

\bibitem[\protect\citeauthoryear{Konar \& Bhattacharya}{Konar \&
  Bhattacharya}{1997}]{konar1997}
Konar S.,  Bhattacharya D.,  1997, \mn@doi [Monthly Notices of the Royal
  Astronomical Society] {10.1093/mnras/284.2.311}, 284, 311

\bibitem[\protect\citeauthoryear{Konar \& Bhattacharya}{Konar \&
  Bhattacharya}{1999}]{konar1999}
Konar S.,  Bhattacharya D.,  1999, \mn@doi [Monthly Notices of the Royal
  Astronomical Society] {10.1046/j.1365-8711.1999.02287.x}, 303, 588

\bibitem[\protect\citeauthoryear{{Kramer} et~al.,}{{Kramer}
  et~al.}{2006}]{kramer2006}
{Kramer} M.,  et~al., 2006, \mn@doi [Science] {10.1126/science.1132305}, \href
  {https://ui.adsabs.harvard.edu/abs/2006Sci...314...97K} {314, 97}

\bibitem[\protect\citeauthoryear{{Kramer} et~al.,}{{Kramer}
  et~al.}{2021}]{kramer2021}
{Kramer} M.,  et~al., 2021, \mn@doi [Physical Review X]
  {10.1103/PhysRevX.11.041050}, \href
  {https://ui.adsabs.harvard.edu/abs/2021PhRvX..11d1050K} {11, 041050}

\bibitem[\protect\citeauthoryear{{Kroupa}}{{Kroupa}}{2001}]{kroupa2001}
{Kroupa} P.,  2001, \mn@doi [\mnras] {10.1046/j.1365-8711.2001.04022.x}, \href
  {https://ui.adsabs.harvard.edu/abs/2001MNRAS.322..231K} {322, 231}

\bibitem[\protect\citeauthoryear{{Kruckow}, {Tauris}, {Langer}, {Kramer}  \&
  {Izzard}}{{Kruckow} et~al.}{2018}]{kruckow2018}
{Kruckow} M.~U.,  {Tauris} T.~M.,  {Langer} N.,  {Kramer} M.,   {Izzard} R.~G.,
   2018, \mn@doi [\mnras] {10.1093/mnras/sty2190}, \href
  {https://ui.adsabs.harvard.edu/abs/2018MNRAS.481.1908K} {481, 1908}

\bibitem[\protect\citeauthoryear{{Lai}, {Chernoff}  \& {Cordes}}{{Lai}
  et~al.}{2001}]{lai2001}
{Lai} D.,  {Chernoff} D.~F.,   {Cordes} J.~M.,  2001, \mn@doi [\apj]
  {10.1086/319455}, \href
  {https://ui.adsabs.harvard.edu/abs/2001ApJ...549.1111L} {549, 1111}

\bibitem[\protect\citeauthoryear{{Lattimer}}{{Lattimer}}{2021}]{lattimer2021}
{Lattimer} J.~M.,  2021, \mn@doi [Annual Review of Nuclear and Particle
  Science] {10.1146/annurev-nucl-102419-124827}, \href
  {https://ui.adsabs.harvard.edu/abs/2021ARNPS..71..433L} {71, 433}

\bibitem[\protect\citeauthoryear{{Lentati} et~al.,}{{Lentati}
  et~al.}{2015}]{lentati2015}
{Lentati} L.,  et~al., 2015, \mn@doi [\mnras] {10.1093/mnras/stv1538}, \href
  {https://ui.adsabs.harvard.edu/abs/2015MNRAS.453.2576L} {453, 2576}

\bibitem[\protect\citeauthoryear{{Licquia} \& {Newman}}{{Licquia} \&
  {Newman}}{2015}]{licquia2015}
{Licquia} T.~C.,  {Newman} J.~A.,  2015, \mn@doi [\apj]
  {10.1088/0004-637X/806/1/96}, \href
  {https://ui.adsabs.harvard.edu/abs/2015ApJ...806...96L} {806, 96}

\bibitem[\protect\citeauthoryear{{Lorimer}}{{Lorimer}}{2008}]{lorimer2008}
{Lorimer} D.~R.,  2008, \mn@doi [Living Reviews in Relativity]
  {10.12942/lrr-2008-8}, \href
  {https://ui.adsabs.harvard.edu/abs/2008LRR....11....8L} {11, 8}

\bibitem[\protect\citeauthoryear{Lorimer}{Lorimer}{2011}]{lorimer2011}
Lorimer D.,  2011, Astrophysics Source Code Library, pp 07019--

\bibitem[\protect\citeauthoryear{{Lorimer} \& {Kramer}}{{Lorimer} \&
  {Kramer}}{2004}]{lorimer2004}
{Lorimer} D.~R.,  {Kramer} M.,  2004, {Handbook of Pulsar Astronomy}.
~ Vol. 4

\bibitem[\protect\citeauthoryear{{Lyne} \& {Manchester}}{{Lyne} \&
  {Manchester}}{1988}]{lyne1988}
{Lyne} A.~G.,  {Manchester} R.~N.,  1988, \mn@doi [\mnras]
  {10.1093/mnras/234.3.477}, \href
  {https://ui.adsabs.harvard.edu/abs/1988MNRAS.234..477L} {234, 477}

\bibitem[\protect\citeauthoryear{{Lyne} et~al.,}{{Lyne}
  et~al.}{2004}]{lyne2004}
{Lyne} A.~G.,  et~al., 2004, \mn@doi [Science] {10.1126/science.1094645}, \href
  {https://ui.adsabs.harvard.edu/abs/2004Sci...303.1153L} {303, 1153}

\bibitem[\protect\citeauthoryear{{Lyne}, {Hobbs}, {Kramer}, {Stairs}  \&
  {Stappers}}{{Lyne} et~al.}{2010}]{lyne2010}
{Lyne} A.,  {Hobbs} G.,  {Kramer} M.,  {Stairs} I.,   {Stappers} B.,  2010,
  \mn@doi [Science] {10.1126/science.1186683}, \href
  {https://ui.adsabs.harvard.edu/abs/2010Sci...329..408L} {329, 408}

\bibitem[\protect\citeauthoryear{{MacLeod} \& {Ramirez-Ruiz}}{{MacLeod} \&
  {Ramirez-Ruiz}}{2015}]{macleod2015}
{MacLeod} M.,  {Ramirez-Ruiz} E.,  2015, \mn@doi [\apj]
  {10.1088/0004-637X/803/1/41}, \href
  {https://ui.adsabs.harvard.edu/abs/2015ApJ...803...41M} {803, 41}

\bibitem[\protect\citeauthoryear{{Manchester} et~al.,}{{Manchester}
  et~al.}{2001}]{manchester2001}
{Manchester} R.~N.,  et~al., 2001, \mn@doi [\mnras]
  {10.1046/j.1365-8711.2001.04751.x}, \href
  {https://ui.adsabs.harvard.edu/abs/2001MNRAS.328...17M} {328, 17}

\bibitem[\protect\citeauthoryear{{Manchester}, {Hobbs}, {Teoh}  \&
  {Hobbs}}{{Manchester} et~al.}{2005}]{manchester2005}
{Manchester} R.~N.,  {Hobbs} G.~B.,  {Teoh} A.,   {Hobbs} M.,  2005, \mn@doi
  [\aj] {10.1086/428488}, \href
  {https://ui.adsabs.harvard.edu/abs/2005AJ....129.1993M} {129, 1993}

\bibitem[\protect\citeauthoryear{{Manchester} et~al.,}{{Manchester}
  et~al.}{2013}]{manchester2013}
{Manchester} R.~N.,  et~al., 2013, \mn@doi [\pasa] {10.1017/pasa.2012.017},
  \href {https://ui.adsabs.harvard.edu/abs/2013PASA...30...17M} {30, e017}

\bibitem[\protect\citeauthoryear{{Mandel} \& {Broekgaarden}}{{Mandel} \&
  {Broekgaarden}}{2022}]{mandel2022}
{Mandel} I.,  {Broekgaarden} F.~S.,  2022, \mn@doi [Living Reviews in
  Relativity] {10.1007/s41114-021-00034-3}, \href
  {https://ui.adsabs.harvard.edu/abs/2022LRR....25....1M} {25, 1}

\bibitem[\protect\citeauthoryear{{Mannucci}, {Cresci}, {Maiolino}, {Marconi}
  \& {Gnerucci}}{{Mannucci} et~al.}{2010}]{mannucci2010}
{Mannucci} F.,  {Cresci} G.,  {Maiolino} R.,  {Marconi} A.,   {Gnerucci} A.,
  2010, \mn@doi [\mnras] {10.1111/j.1365-2966.2010.17291.x}, \href
  {https://ui.adsabs.harvard.edu/abs/2010MNRAS.408.2115M} {408, 2115}

\bibitem[\protect\citeauthoryear{{Mannucci}, {Salvaterra}  \&
  {Campisi}}{{Mannucci} et~al.}{2011}]{mannucci2011}
{Mannucci} F.,  {Salvaterra} R.,   {Campisi} M.~A.,  2011, \mn@doi [\mnras]
  {10.1111/j.1365-2966.2011.18459.x}, \href
  {https://ui.adsabs.harvard.edu/abs/2011MNRAS.414.1263M} {414, 1263}

\bibitem[\protect\citeauthoryear{{Mapelli} \& {Giacobbo}}{{Mapelli} \&
  {Giacobbo}}{2018}]{mapelli2018}
{Mapelli} M.,  {Giacobbo} N.,  2018, \mn@doi [\mnras] {10.1093/mnras/sty1613},
  \href {https://ui.adsabs.harvard.edu/abs/2018MNRAS.479.4391M} {479, 4391}

\bibitem[\protect\citeauthoryear{{Mapelli}, {Giacobbo}, {Ripamonti}  \&
  {Spera}}{{Mapelli} et~al.}{2017}]{mapelli2017}
{Mapelli} M.,  {Giacobbo} N.,  {Ripamonti} E.,   {Spera} M.,  2017, \mn@doi
  [\mnras] {10.1093/mnras/stx2123}, \href
  {https://ui.adsabs.harvard.edu/abs/2017MNRAS.472.2422M} {472, 2422}

\bibitem[\protect\citeauthoryear{{Mapelli}, {Giacobbo}, {Santoliquido}  \&
  {Artale}}{{Mapelli} et~al.}{2019}]{mapelli2019}
{Mapelli} M.,  {Giacobbo} N.,  {Santoliquido} F.,   {Artale} M.~C.,  2019,
  \mn@doi [\mnras] {10.1093/mnras/stz1150}, \href
  {https://ui.adsabs.harvard.edu/abs/2019MNRAS.487....2M} {487, 2}

\bibitem[\protect\citeauthoryear{{Mapelli}, {Spera}, {Montanari}, {Limongi},
  {Chieffi}, {Giacobbo}, {Bressan}  \& {Bouffanais}}{{Mapelli}
  et~al.}{2020}]{mapelli2020}
{Mapelli} M.,  {Spera} M.,  {Montanari} E.,  {Limongi} M.,  {Chieffi} A.,
  {Giacobbo} N.,  {Bressan} A.,   {Bouffanais} Y.,  2020, \mn@doi [\apj]
  {10.3847/1538-4357/ab584d}, \href
  {https://ui.adsabs.harvard.edu/abs/2020ApJ...888...76M} {888, 76}

\bibitem[\protect\citeauthoryear{{Martinez} et~al.,}{{Martinez}
  et~al.}{2015}]{martinez2015}
{Martinez} J.~G.,  et~al., 2015, \mn@doi [\apj] {10.1088/0004-637X/812/2/143},
  \href {https://ui.adsabs.harvard.edu/abs/2015ApJ...812..143M} {812, 143}

\bibitem[\protect\citeauthoryear{{Matteucci}, {Romano}, {Cescutti}  \&
  {Simonetti}}{{Matteucci} et~al.}{2019}]{matteucci2019}
{Matteucci} F.,  {Romano} D.,  {Cescutti} G.,   {Simonetti} P.,  2019, \mn@doi
  [Rendiconti Lincei. Scienze Fisiche e Naturali] {10.1007/s12210-018-0754-z},
  \href {https://ui.adsabs.harvard.edu/abs/2019RLSFN..30S..85M} {30, 85}

\bibitem[\protect\citeauthoryear{{McMillan}}{{McMillan}}{2017}]{mcmillan2017}
{McMillan} P.~J.,  2017, \mn@doi [\mnras] {10.1093/mnras/stw2759}, \href
  {https://ui.adsabs.harvard.edu/abs/2017MNRAS.465...76M} {465, 76}

\bibitem[\protect\citeauthoryear{{Metzger}}{{Metzger}}{2017}]{metzger2017}
{Metzger} B.~D.,  2017, \mn@doi [Living Reviews in Relativity]
  {10.1007/s41114-017-0006-z}, \href
  {https://ui.adsabs.harvard.edu/abs/2017LRR....20....3M} {20, 3}

\bibitem[\protect\citeauthoryear{{Metzger}}{{Metzger}}{2019}]{metzger2019}
{Metzger} B.~D.,  2019, \mn@doi [Living Reviews in Relativity]
  {10.1007/s41114-019-0024-0}, \href
  {https://ui.adsabs.harvard.edu/abs/2019LRR....23....1M} {23, 1}

\bibitem[\protect\citeauthoryear{{Miles} et~al.,}{{Miles}
  et~al.}{2023}]{miles2023}
{Miles} M.~T.,  et~al., 2023, \mn@doi [\mnras] {10.1093/mnras/stac3644}, \href
  {https://ui.adsabs.harvard.edu/abs/2023MNRAS.519.3976M} {519, 3976}

\bibitem[\protect\citeauthoryear{{M{\"u}ller} et~al.,}{{M{\"u}ller}
  et~al.}{2019}]{muller2019}
{M{\"u}ller} B.,  et~al., 2019, \mn@doi [\mnras] {10.1093/mnras/stz216}, \href
  {https://ui.adsabs.harvard.edu/abs/2019MNRAS.484.3307M} {484, 3307}

\bibitem[\protect\citeauthoryear{{Murase} et~al.,}{{Murase}
  et~al.}{2018}]{murase2018}
{Murase} K.,  et~al., 2018, \mn@doi [\apj] {10.3847/1538-4357/aaa48a}, \href
  {https://ui.adsabs.harvard.edu/abs/2018ApJ...854...60M} {854, 60}

\bibitem[\protect\citeauthoryear{{Nedora}, {Dietrich}, {Shibata}, {Pohl}  \&
  {Crosato Menegazzi}}{{Nedora} et~al.}{2023}]{nedora2023}
{Nedora} V.,  {Dietrich} T.,  {Shibata} M.,  {Pohl} M.,   {Crosato Menegazzi}
  L.,  2023, \mn@doi [\mnras] {10.1093/mnras/stad175}, \href
  {https://ui.adsabs.harvard.edu/abs/2023MNRAS.520.2727N} {520, 2727}

\bibitem[\protect\citeauthoryear{{Neijssel} et~al.,}{{Neijssel}
  et~al.}{2019}]{coenrad2019}
{Neijssel} C.~J.,  et~al., 2019, \mn@doi [\mnras] {10.1093/mnras/stz2840},
  \href {https://ui.adsabs.harvard.edu/abs/2019MNRAS.490.3740N} {490, 3740}

\bibitem[\protect\citeauthoryear{{Nelemans}, {Verbunt}, {Yungelson}  \&
  {Portegies Zwart}}{{Nelemans} et~al.}{2000}]{nelemans2000}
{Nelemans} G.,  {Verbunt} F.,  {Yungelson} L.~R.,   {Portegies Zwart} S.~F.,
  2000, \mn@doi [\aap] {10.48550/arXiv.astro-ph/0006216}, \href
  {https://ui.adsabs.harvard.edu/abs/2000A&A...360.1011N} {360, 1011}

\bibitem[\protect\citeauthoryear{{Nelson} et~al.,}{{Nelson}
  et~al.}{2019a}]{nelson2019}
{Nelson} D.,  et~al., 2019a, \mn@doi [Computational Astrophysics and Cosmology]
  {10.1186/s40668-019-0028-x}, \href
  {https://ui.adsabs.harvard.edu/abs/2019ComAC...6....2N} {6, 2}

\bibitem[\protect\citeauthoryear{{Nelson} et~al.,}{{Nelson}
  et~al.}{2019b}]{nelson2019tng50}
{Nelson} D.,  et~al., 2019b, \mn@doi [\mnras] {10.1093/mnras/stz2306}, \href
  {https://ui.adsabs.harvard.edu/abs/2019MNRAS.490.3234N} {490, 3234}

\bibitem[\protect\citeauthoryear{{Ng} et~al.,}{{Ng} et~al.}{2015}]{ng2015}
{Ng} C.,  et~al., 2015, \mn@doi [\mnras] {10.1093/mnras/stv753}, \href
  {https://ui.adsabs.harvard.edu/abs/2015MNRAS.450.2922N} {450, 2922}

\bibitem[\protect\citeauthoryear{{Ng} et~al.,}{{Ng} et~al.}{2018}]{ng2018}
{Ng} C.,  et~al., 2018, \mn@doi [\mnras] {10.1093/mnras/sty482}, \href
  {https://ui.adsabs.harvard.edu/abs/2018MNRAS.476.4315N} {476, 4315}

\bibitem[\protect\citeauthoryear{{Nguyen} et~al.,}{{Nguyen}
  et~al.}{2022}]{nguyen2022}
{Nguyen} C.~T.,  et~al., 2022, \mn@doi [\aap] {10.1051/0004-6361/202244166},
  \href {https://ui.adsabs.harvard.edu/abs/2022A&A...665A.126N} {665, A126}

\bibitem[\protect\citeauthoryear{{O'Doherty}, {Bahramian}, {Miller-Jones},
  {Goodwin}, {Mandel}, {Willcox}, {Atri}  \& {Strader}}{{O'Doherty}
  et~al.}{2023}]{odoherty2023}
{O'Doherty} T.~N.,  {Bahramian} A.,  {Miller-Jones} J. C.~A.,  {Goodwin} A.~J.,
   {Mandel} I.,  {Willcox} R.,  {Atri} P.,   {Strader} J.,  2023, \mn@doi
  [\mnras] {10.1093/mnras/stad680}, \href
  {https://ui.adsabs.harvard.edu/abs/2023MNRAS.521.2504O} {521, 2504}

\bibitem[\protect\citeauthoryear{{Olejak}, {Belczynski}  \& {Ivanova}}{{Olejak}
  et~al.}{2021}]{olejak2021}
{Olejak} A.,  {Belczynski} K.,   {Ivanova} N.,  2021, \mn@doi [\aap]
  {10.1051/0004-6361/202140520}, \href
  {https://ui.adsabs.harvard.edu/abs/2021A&A...651A.100O} {651, A100}

\bibitem[\protect\citeauthoryear{{Olejak}, {Fryer}, {Belczynski}  \&
  {Baibhav}}{{Olejak} et~al.}{2022}]{olejak2022}
{Olejak} A.,  {Fryer} C.~L.,  {Belczynski} K.,   {Baibhav} V.,  2022, \mn@doi
  [\mnras] {10.1093/mnras/stac2359}, \href
  {https://ui.adsabs.harvard.edu/abs/2022MNRAS.516.2252O} {516, 2252}

\bibitem[\protect\citeauthoryear{{Os{\l}owski}, {Bulik}, {Gondek-Rosi{\'n}ska}
  \& {Belczy{\'n}ski}}{{Os{\l}owski} et~al.}{2011}]{oslowski2011}
{Os{\l}owski} S.,  {Bulik} T.,  {Gondek-Rosi{\'n}ska} D.,   {Belczy{\'n}ski}
  K.,  2011, \mn@doi [\mnras] {10.1111/j.1365-2966.2010.18147.x}, \href
  {https://ui.adsabs.harvard.edu/abs/2011MNRAS.413..461O} {413, 461}

\bibitem[\protect\citeauthoryear{{Ostriker} \& {Gunn}}{{Ostriker} \&
  {Gunn}}{1969}]{ostriker1969}
{Ostriker} J.~P.,  {Gunn} J.~E.,  1969, \mn@doi [\apj] {10.1086/150160}, \href
  {https://ui.adsabs.harvard.edu/abs/1969ApJ...157.1395O} {157, 1395}

\bibitem[\protect\citeauthoryear{{{\"O}zel}, {Psaltis}, {Narayan}  \& {Santos
  Villarreal}}{{{\"O}zel} et~al.}{2012}]{ozel2012}
{{\"O}zel} F.,  {Psaltis} D.,  {Narayan} R.,   {Santos Villarreal} A.,  2012,
  \mn@doi [\apj] {10.1088/0004-637X/757/1/55}, \href
  {https://ui.adsabs.harvard.edu/abs/2012ApJ...757...55O} {757, 55}

\bibitem[\protect\citeauthoryear{{{\"O}zel}, {Psaltis}, {G{\"u}ver}, {Baym},
  {Heinke}  \& {Guillot}}{{{\"O}zel} et~al.}{2016}]{ozel2016}
{{\"O}zel} F.,  {Psaltis} D.,  {G{\"u}ver} T.,  {Baym} G.,  {Heinke} C.,
  {Guillot} S.,  2016, \mn@doi [\apj] {10.3847/0004-637X/820/1/28}, \href
  {https://ui.adsabs.harvard.edu/abs/2016ApJ...820...28O} {820, 28}

\bibitem[\protect\citeauthoryear{{Pacini}}{{Pacini}}{1968}]{pacini1968}
{Pacini} F.,  1968, \mn@doi [\nat] {10.1038/219145a0}, \href
  {https://ui.adsabs.harvard.edu/abs/1968Natur.219..145P} {219, 145}

\bibitem[\protect\citeauthoryear{{Paczynski}}{{Paczynski}}{1976}]{Paczynski1976}
{Paczynski} B.,  1976, in {Eggleton} P.,  {Mitton} S.,   {Whelan} J.,  eds, ~
  Vol. 73, Structure and Evolution of Close Binary Systems. p.~75

\bibitem[\protect\citeauthoryear{{Perna}, {Artale}, {Wang}, {Mapelli},
  {Lazzati}, {Sgalletta}  \& {Santoliquido}}{{Perna} et~al.}{2022}]{perna2022}
{Perna} R.,  {Artale} M.~C.,  {Wang} Y.-H.,  {Mapelli} M.,  {Lazzati} D.,
  {Sgalletta} C.,   {Santoliquido} F.,  2022, \mn@doi [\mnras]
  {10.1093/mnras/stac685}, \href
  {https://ui.adsabs.harvard.edu/abs/2022MNRAS.512.2654P} {512, 2654}

\bibitem[\protect\citeauthoryear{Peters}{Peters}{1964}]{peters1964}
Peters P.~C.,  1964, \mn@doi [Phys. Rev.] {10.1103/PhysRev.136.B1224}, 136,
  B1224

\bibitem[\protect\citeauthoryear{Pezzulli \& Fraternali}{Pezzulli \&
  Fraternali}{2015}]{pezzulli2015}
Pezzulli G.,  Fraternali F.,  2015, \mn@doi [Monthly Notices of the Royal
  Astronomical Society] {10.1093/mnras/stv2397}, 455, 2308

\bibitem[\protect\citeauthoryear{{Pillepich} et~al.,}{{Pillepich}
  et~al.}{2019}]{pillepich2019}
{Pillepich} A.,  et~al., 2019, \mn@doi [\mnras] {10.1093/mnras/stz2338}, \href
  {https://ui.adsabs.harvard.edu/abs/2019MNRAS.490.3196P} {490, 3196}

\bibitem[\protect\citeauthoryear{{Pillepich} et~al.,}{{Pillepich}
  et~al.}{2023}]{pillepich2023}
{Pillepich} A.,  et~al., 2023, \mn@doi [arXiv e-prints]
  {10.48550/arXiv.2303.16217}, \href
  {https://ui.adsabs.harvard.edu/abs/2023arXiv230316217P} {p. arXiv:2303.16217}

\bibitem[\protect\citeauthoryear{{Pol}, {McLaughlin}  \& {Lorimer}}{{Pol}
  et~al.}{2019}]{pol2019}
{Pol} N.,  {McLaughlin} M.,   {Lorimer} D.~R.,  2019, \mn@doi [\apj]
  {10.3847/1538-4357/aaf006}, \href
  {https://ui.adsabs.harvard.edu/abs/2019ApJ...870...71P} {870, 71}

\bibitem[\protect\citeauthoryear{{Pol}, {McLaughlin}  \& {Lorimer}}{{Pol}
  et~al.}{2020}]{pol2020}
{Pol} N.,  {McLaughlin} M.,   {Lorimer} D.~R.,  2020, \mn@doi [Research Notes
  of the American Astronomical Society] {10.3847/2515-5172/ab7307}, \href
  {https://ui.adsabs.harvard.edu/abs/2020RNAAS...4...22P} {4, 22}

\bibitem[\protect\citeauthoryear{{Radice}, {Bernuzzi}  \& {Perego}}{{Radice}
  et~al.}{2020}]{radice2020}
{Radice} D.,  {Bernuzzi} S.,   {Perego} A.,  2020, \mn@doi [Annual Review of
  Nuclear and Particle Science] {10.1146/annurev-nucl-013120-114541}, \href
  {https://ui.adsabs.harvard.edu/abs/2020ARNPS..70...95R} {70, 95}

\bibitem[\protect\citeauthoryear{{Riley} et~al.,}{{Riley}
  et~al.}{2022}]{riley2022}
{Riley} J.,  et~al., 2022, \mn@doi [\apjs] {10.3847/1538-4365/ac416c}, \href
  {https://ui.adsabs.harvard.edu/abs/2022ApJS..258...34R} {258, 34}

\bibitem[\protect\citeauthoryear{{Rinaldi} \& {Del Pozzo}}{{Rinaldi} \& {Del
  Pozzo}}{2022a}]{Rinaldi2022}
{Rinaldi} S.,  {Del Pozzo} W.,  2022a, \mn@doi [\mnras]
  {10.1093/mnras/stab3224}, \href
  {https://ui.adsabs.harvard.edu/abs/2022MNRAS.509.5454R} {509, 5454}

\bibitem[\protect\citeauthoryear{{Rinaldi} \& {Del Pozzo}}{{Rinaldi} \& {Del
  Pozzo}}{2022b}]{rinaldi2022figaro}
{Rinaldi} S.,  {Del Pozzo} W.,  2022b, \mn@doi [\mnras]
  {10.1093/mnrasl/slac101}, \href
  {https://ui.adsabs.harvard.edu/abs/2022MNRAS.517L...5R} {517, L5}

\bibitem[\protect\citeauthoryear{{Roepke} \& {De Marco}}{{Roepke} \& {De
  Marco}}{2022}]{roepke2022}
{Roepke} F.~K.,  {De Marco} O.,  2022, \mn@doi [arXiv e-prints]
  {10.48550/arXiv.2212.07308}, \href
  {https://ui.adsabs.harvard.edu/abs/2022arXiv221207308R} {p. arXiv:2212.07308}

\bibitem[\protect\citeauthoryear{Rudak \& Ritter}{Rudak \&
  Ritter}{1994}]{rudak1994}
Rudak B.,  Ritter H.,  1994, \mn@doi [Monthly Notices of the Royal Astronomical
  Society] {10.1093/mnras/267.3.513}, 267, 513

\bibitem[\protect\citeauthoryear{{Sana} et~al.,}{{Sana}
  et~al.}{2012}]{sana2012}
{Sana} H.,  et~al., 2012, \mn@doi [Science] {10.1126/science.1223344}, \href
  {https://ui.adsabs.harvard.edu/abs/2012Sci...337..444S} {337, 444}

\bibitem[\protect\citeauthoryear{Santoliquido, Mapelli, Artale  \&
  Boco}{Santoliquido et~al.}{2022}]{santoliquido2022}
Santoliquido F.,  Mapelli M.,  Artale M.~C.,   Boco L.,  2022, \mn@doi [Monthly
  Notices of the Royal Astronomical Society] {10.1093/mnras/stac2384}, 516,
  3297

\bibitem[\protect\citeauthoryear{Schaye et~al.,}{Schaye
  et~al.}{2014}]{schaye2015}
Schaye J.,  et~al., 2014, \mn@doi [Monthly Notices of the Royal Astronomical
  Society] {10.1093/mnras/stu2058}, 446, 521

\bibitem[\protect\citeauthoryear{Sgalletta et~al.,}{Sgalletta
  et~al.}{2023}]{zenodo_repo}
Sgalletta C.,  et~al., 2023, {SEVN parameter file from the paper "Binary
  neutron star populations in the Milky Way" by Sgalletta et al., 2023},
  \mn@doi{10.5281/zenodo.7887279}, \url
  {https://doi.org/10.5281/zenodo.7887279}

\bibitem[\protect\citeauthoryear{{Shao} \& {Li}}{{Shao} \&
  {Li}}{2018}]{shao2018}
{Shao} Y.,  {Li} X.-D.,  2018, \mn@doi [\apj] {10.3847/1538-4357/aae648}, \href
  {https://ui.adsabs.harvard.edu/abs/2018ApJ...867..124S} {867, 124}

\bibitem[\protect\citeauthoryear{{Smartt} et~al.,}{{Smartt}
  et~al.}{2017}]{smartt2017}
{Smartt} S.~J.,  et~al., 2017, \mn@doi [\nat] {10.1038/nature24303}, \href
  {https://ui.adsabs.harvard.edu/abs/2017Natur.551...75S} {551, 75}

\bibitem[\protect\citeauthoryear{{Spera} \& {Mapelli}}{{Spera} \&
  {Mapelli}}{2017}]{spera2017}
{Spera} M.,  {Mapelli} M.,  2017, \mn@doi [\mnras] {10.1093/mnras/stx1576},
  \href {https://ui.adsabs.harvard.edu/abs/2017MNRAS.470.4739S} {470, 4739}

\bibitem[\protect\citeauthoryear{{Spera}, {Mapelli}, {Giacobbo}, {Trani},
  {Bressan}  \& {Costa}}{{Spera} et~al.}{2019}]{spera2019}
{Spera} M.,  {Mapelli} M.,  {Giacobbo} N.,  {Trani} A.~A.,  {Bressan} A.,
  {Costa} G.,  2019, \mn@doi [\mnras] {10.1093/mnras/stz359}, \href
  {https://ui.adsabs.harvard.edu/abs/2019MNRAS.485..889S} {485, 889}

\bibitem[\protect\citeauthoryear{{Spera}, {Trani}  \& {Mencagli}}{{Spera}
  et~al.}{2022}]{spera2022}
{Spera} M.,  {Trani} A.~A.,   {Mencagli} M.,  2022, \mn@doi [Galaxies]
  {10.3390/galaxies10040076}, \href
  {https://ui.adsabs.harvard.edu/abs/2022Galax..10...76S} {10, 76}

\bibitem[\protect\citeauthoryear{{Springel}}{{Springel}}{2005}]{springel2005}
{Springel} V.,  2005, \mn@doi [\mnras] {10.1111/j.1365-2966.2005.09655.x},
  \href {https://ui.adsabs.harvard.edu/abs/2005MNRAS.364.1105S} {364, 1105}

\bibitem[\protect\citeauthoryear{{Springel}}{{Springel}}{2010}]{springel2010}
{Springel} V.,  2010, \mn@doi [\mnras] {10.1111/j.1365-2966.2009.15715.x},
  \href {https://ui.adsabs.harvard.edu/abs/2010MNRAS.401..791S} {401, 791}

\bibitem[\protect\citeauthoryear{{Stairs}, {Thorsett}, {Taylor}  \&
  {Wolszczan}}{{Stairs} et~al.}{2002}]{stairs2002}
{Stairs} I.~H.,  {Thorsett} S.~E.,  {Taylor} J.~H.,   {Wolszczan} A.,  2002,
  \mn@doi [\apj] {10.1086/344157}, \href
  {https://ui.adsabs.harvard.edu/abs/2002ApJ...581..501S} {581, 501}

\bibitem[\protect\citeauthoryear{{Stappers}, {Keane}, {Kramer}, {Possenti}  \&
  {Stairs}}{{Stappers} et~al.}{2018}]{stappers2018}
{Stappers} B.~W.,  {Keane} E.~F.,  {Kramer} M.,  {Possenti} A.,   {Stairs}
  I.~H.,  2018, \mn@doi [Philosophical Transactions of the Royal Society of
  London Series A] {10.1098/rsta.2017.0293}, \href
  {https://ui.adsabs.harvard.edu/abs/2018RSPTA.37670293S} {376, 20170293}

\bibitem[\protect\citeauthoryear{{Stevenson}, {Willcox}, {Vigna-G{\'o}mez}  \&
  {Broekgaarden}}{{Stevenson} et~al.}{2022}]{stevenson2022}
{Stevenson} S.,  {Willcox} R.,  {Vigna-G{\'o}mez} A.,   {Broekgaarden} F.,
  2022, \mn@doi [\mnras] {10.1093/mnras/stac1322}, \href
  {https://ui.adsabs.harvard.edu/abs/2022MNRAS.513.6105S} {513, 6105}

\bibitem[\protect\citeauthoryear{{Szary}, {Zhang}, {Melikidze}, {Gil}  \&
  {Xu}}{{Szary} et~al.}{2014}]{szary2014}
{Szary} A.,  {Zhang} B.,  {Melikidze} G.~I.,  {Gil} J.,   {Xu} R.-X.,  2014,
  \mn@doi [\apj] {10.1088/0004-637X/784/1/59}, \href
  {https://ui.adsabs.harvard.edu/abs/2014ApJ...784...59S} {784, 59}

\bibitem[\protect\citeauthoryear{{Tanvir}, {Levan}, {Fruchter}, {Hjorth},
  {Hounsell}, {Wiersema}  \& {Tunnicliffe}}{{Tanvir} et~al.}{2013}]{tanvir2013}
{Tanvir} N.~R.,  {Levan} A.~J.,  {Fruchter} A.~S.,  {Hjorth} J.,  {Hounsell}
  R.~A.,  {Wiersema} K.,   {Tunnicliffe} R.~L.,  2013, \mn@doi [\nat]
  {10.1038/nature12505}, \href
  {https://ui.adsabs.harvard.edu/abs/2013Natur.500..547T} {500, 547}

\bibitem[\protect\citeauthoryear{Tauris \& Manchester}{Tauris \&
  Manchester}{1998}]{tauris1998}
Tauris T.~M.,  Manchester R.~N.,  1998, \mn@doi [Monthly Notices of the Royal
  Astronomical Society] {10.1046/j.1365-8711.1998.01369.x}, 298, 625

\bibitem[\protect\citeauthoryear{{Tauris}, {Langer}  \&
  {Podsiadlowski}}{{Tauris} et~al.}{2015}]{tauris2015}
{Tauris} T.~M.,  {Langer} N.,   {Podsiadlowski} P.,  2015, \mn@doi [\mnras]
  {10.1093/mnras/stv990}, \href
  {https://ui.adsabs.harvard.edu/abs/2015MNRAS.451.2123T} {451, 2123}

\bibitem[\protect\citeauthoryear{{Tauris} et~al.,}{{Tauris}
  et~al.}{2017}]{tauris2017}
{Tauris} T.~M.,  et~al., 2017, \mn@doi [\apj] {10.3847/1538-4357/aa7e89}, \href
  {https://ui.adsabs.harvard.edu/abs/2017ApJ...846..170T} {846, 170}

\bibitem[\protect\citeauthoryear{{The EAGLE team}}{{The EAGLE
  team}}{2017}]{eagle2017}
{The EAGLE team} 2017, \mn@doi [arXiv e-prints] {10.48550/arXiv.1706.09899},
  \href {https://ui.adsabs.harvard.edu/abs/2017arXiv170609899T} {p.
  arXiv:1706.09899}

\bibitem[\protect\citeauthoryear{{Thrane}, {Os{\l}owski}  \& {Lasky}}{{Thrane}
  et~al.}{2020}]{thrane2020}
{Thrane} E.,  {Os{\l}owski} S.,   {Lasky} P.~D.,  2020, \mn@doi [\mnras]
  {10.1093/mnras/staa593}, \href
  {https://ui.adsabs.harvard.edu/abs/2020MNRAS.493.5408T} {493, 5408}

\bibitem[\protect\citeauthoryear{{Tout}, {Aarseth}, {Pols}  \&
  {Eggleton}}{{Tout} et~al.}{1997}]{tout1997}
{Tout} C.~A.,  {Aarseth} S.~J.,  {Pols} O.~R.,   {Eggleton} P.~P.,  1997,
  \mn@doi [\mnras] {10.1093/mnras/291.4.732}, \href
  {https://ui.adsabs.harvard.edu/abs/1997MNRAS.291..732T} {291, 732}

\bibitem[\protect\citeauthoryear{{Troja} et~al.,}{{Troja}
  et~al.}{2017}]{troja2017}
{Troja} E.,  et~al., 2017, \mn@doi [\nat] {10.1038/nature24290}, \href
  {https://ui.adsabs.harvard.edu/abs/2017Natur.551...71T} {551, 71}

\bibitem[\protect\citeauthoryear{Urpin \& Konenkov}{Urpin \&
  Konenkov}{1997}]{urpin1997}
Urpin V.,  Konenkov D.,  1997, \mn@doi [Monthly Notices of the Royal
  Astronomical Society] {10.1093/mnras/292.1.167}, 292, 167

\bibitem[\protect\citeauthoryear{{Vigna-G{\'o}mez} et~al.,}{{Vigna-G{\'o}mez}
  et~al.}{2018}]{vignagomez2018}
{Vigna-G{\'o}mez} A.,  et~al., 2018, \mn@doi [\mnras] {10.1093/mnras/sty2463},
  \href {https://ui.adsabs.harvard.edu/abs/2018MNRAS.481.4009V} {481, 4009}

\bibitem[\protect\citeauthoryear{{Vigna-G{\'o}mez} et~al.,}{{Vigna-G{\'o}mez}
  et~al.}{2020}]{vignagomez2020}
{Vigna-G{\'o}mez} A.,  et~al., 2020, \mn@doi [\pasa] {10.1017/pasa.2020.31},
  \href {https://ui.adsabs.harvard.edu/abs/2020PASA...37...38V} {37, e038}

\bibitem[\protect\citeauthoryear{{Virtanen} et~al.,}{{Virtanen}
  et~al.}{2020}]{virtanen2020}
{Virtanen} P.,  et~al., 2020, \mn@doi [Nature Methods]
  {10.1038/s41592-019-0686-2}, \href
  {https://ui.adsabs.harvard.edu/abs/2020NatMe..17..261V} {17, 261}

\bibitem[\protect\citeauthoryear{{Webbink}}{{Webbink}}{1985}]{webbink10985}
{Webbink} R.~F.,  1985, in {Pringle} J.~E.,  {Wade} R.~A.,  eds, , Interacting
  Binary Stars.
p.~39

\bibitem[\protect\citeauthoryear{{Weisberg} \& {Huang}}{{Weisberg} \&
  {Huang}}{2016}]{weisberg2016}
{Weisberg} J.~M.,  {Huang} Y.,  2016, \mn@doi [\apj]
  {10.3847/0004-637X/829/1/55}, \href
  {https://ui.adsabs.harvard.edu/abs/2016ApJ...829...55W} {829, 55}

\bibitem[\protect\citeauthoryear{{Woosley}}{{Woosley}}{1987}]{woosley1987}
{Woosley} S.~E.,  1987, in {Helfand} D.~J.,  {Huang} J.~H.,  eds, ~ Vol. 125,
  The Origin and Evolution of Neutron Stars. p.~255

\bibitem[\protect\citeauthoryear{{Yusifov} \& {K{\"u}{\c{c}}{\"u}k}}{{Yusifov}
  \& {K{\"u}{\c{c}}{\"u}k}}{2004}]{yusifov2004}
{Yusifov} I.,  {K{\"u}{\c{c}}{\"u}k} I.,  2004, \mn@doi [\aap]
  {10.1051/0004-6361:20040152}, \href
  {https://ui.adsabs.harvard.edu/abs/2004A&A...422..545Y} {422, 545}

\bibitem[\protect\citeauthoryear{{Zhang} \& {Kojima}}{{Zhang} \&
  {Kojima}}{2006}]{zhang2006}
{Zhang} C.~M.,  {Kojima} Y.,  2006, \mn@doi [\mnras]
  {10.1111/j.1365-2966.2005.09802.x}, \href
  {https://ui.adsabs.harvard.edu/abs/2006MNRAS.366..137Z} {366, 137}

\makeatother
\end{thebibliography}




\appendix

\section{Masses} \label{sec:masses}

\begin{figure}
    \centering
    \includegraphics[width=0.5\textwidth]{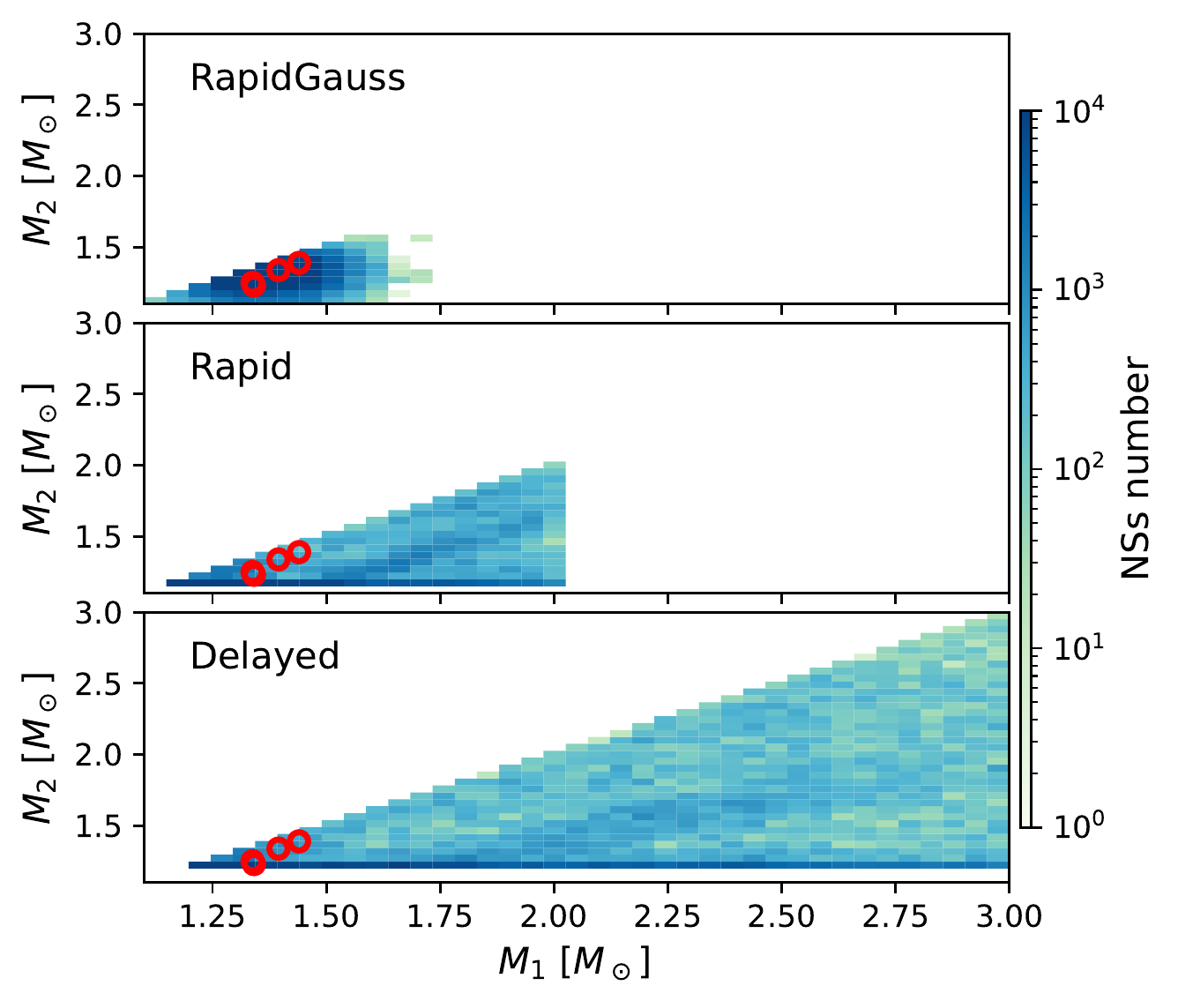}
    \caption{Secondary versus primary mass of the simulated BNSs, for the three different SN prescriptions adopted in this work. From top to bottom: rapid-gauss, rapid, and delayed. Here, the primary (secondary) is the most (least) massive NS of each BNS. The colour shows the number of binaries in each cell. The red circles are the observed Galactic BNS masses (some of the circles overlap in this Figure).}
    \label{fig:MassesDistr}
\end{figure}

Figure~\ref{fig:MassesDistr} shows the BNS masses we obtained for the three core-collapse SN models we assumed (Section~\ref{sec:sevn}). 
 The distributions computed with the rapid and delayed models by \cite{fryer2012} produce a strong peak of the secondary mass at about $1.2\text{M}_{\odot}$, failing to reproduce the observed masses of the Galactic BNSs. In contrast, the rapid-gauss model matches the observed population by construction. We decided to exclude the masses from the statistical analysis in Section~\ref{sec:bayes} because the only models that matches the observation does it by construction.

\section{SKA survey parameters}\label{sec:surveyparams}

Table~\ref{tab:surveysSKA} shows the SKA survey parameters in the MID- and LOW- frequency ranges \citep{stappers2018} we used to obtain~Table~\ref{tab:SKApredictions}. 

\begin{table} \centering
\begin{tabular}{lcccccc}
\hline
Survey & $\Delta \nu$  & $G_{\rm A}$ & $T_{\rm rec}$ & $\tau_{\rm sampl}$ & $t_{\rm int}$ & Sky \\ 
& (MHz) & (K/Jy) & (K) & (ms) & (s) & coverage\\ \hline
LOW  & 100 &  26.85  &  56  &  0.1  &  600  &  -90$^{\circ}$ < $\delta$ < 30$^{\circ}$ \\
MID   &  300 &  3.92  &  20  &  0.1  &  600  & -90$^{\circ}$ < $\delta$ < 30$^{\circ}$ \\
\hline
\end{tabular}
\caption{Adopted parameters for the LOW and MID SKA surveys \citep{stappers2018}. The columns show the bandwidth $\Delta \nu$, the antenna gain $G_{\rm A}$, the system temperature $T_{\rm rec}$, the sampling time $\tau_{\rm sampl}$, the integration time $t_{\rm int}$ and the sky coverages, respectively.}
\label{tab:surveysSKA}
\end{table}

\bsp	
\label{lastpage}
\end{document}